\pdfoutput=1
\documentclass[aip,jap,reprint,amsmath,amssymb,floatfix]{revtex4-2}
\usepackage{graphicx}
\usepackage{dcolumn}
\usepackage{bm}
\usepackage{booktabs}
\usepackage{longtable}
\usepackage[hidelinks]{hyperref}
\usepackage{url}

\begin{document}

\title{Designed atomic displacements for
finite-temperature phonons from few force calculations}

\author{Albert Linda}
\email{albert.hzbn@gmail.com}
\affiliation{Department of Materials Science and Engineering, Indian Institute
of Technology Kanpur, Kanpur 208016, India}

\date{\today}

\begin{abstract}
Finite-temperature phonons are usually fitted to first-principles forces in
randomly displaced supercells, and each force calculation can be expensive. We introduce special
optimal thermal cells (SOTCs), in which the displacements are designed rather
than drawn at random. Our design, SOTC-D, starts from a harmonic trial model.
It gives every supercell phonon mode its quantum-thermal amplitude, chooses the
sign of each mode so that the force constants are best determined, and pairs
each configuration with its inverse so that cubic forces cancel. Updating
the trial leads to the self-consistent harmonic solution. Cells that match
displacement correlations instead give good thermodynamics but not better
dispersions. For eleven crystals, eight force calculations reproduce the
effective dispersions within 0.03--0.31~THz, and SOTC-D is the most accurate of
five designs in 9 of them. After one trial update it beats random sampling in
all crystals at 4--16 calculations and two-iteration stochastic sampling (sTDEP)
in all at eight (supercell basis). On model potentials, the
self-consistent form needs 2--16 times fewer calculations than the stochastic
self-consistent harmonic approximation in 13 of 14 cases. With the HSE06 hybrid
functional it is ahead of pooled sTDEP at 4--16 calculations in Si and NaCl,
and it gives the thermal expansion of Al within 3\% of experiment at 300~K.
The method has limits: in Al and Si cells of more than 100 atoms, sTDEP is more
accurate beyond 4--16 calculations (not so in NaCl up to 16); for dynamically
stabilized bcc Zr the outcome depends on the seed; and above 400~K the free-energy
curvature is not resolved.
\end{abstract}

\maketitle

\section{Introduction}\label{sec:intro}

Finite-temperature phonons control phase stability, thermal transport,
electron--phonon renormalization, and vibrational thermodynamics
\cite{ref1,ref2,ref3}. At elevated temperature the atomic displacements reach a
few percent of the lattice constant, and anharmonic coupling between modes
shifts the phonon frequencies at fixed volume.

Finite-temperature lattice-dynamics methods differ in the ensemble over which
the forces are averaged and in the anharmonic terms they retain. The
quasiharmonic approximation (QHA) keeps harmonic force constants and lets them
depend on volume only, so it omits the explicit temperature dependence that
anharmonic coupling produces at fixed volume. Self-consistent harmonic
theories include this dependence through an effective harmonic Hamiltonian
with temperature-dependent second-order force constants (FC2). In the
self-consistent harmonic approximation (SCHA), the FC2 $\Phi$ parametrizes a
trial harmonic density matrix $\rho_\Phi$; the Gibbs--Bogoliubov upper bound
to the free energy is minimized with respect to $\Phi$ and the average atomic
positions, and at the minimum $\Phi$ equals the Hessian of the
Born--Oppenheimer surface averaged over the Gaussian displacement
distribution of $\rho_\Phi$. The stochastic SCHA (SSCHA) evaluates these
averages and the free-energy gradient on configurations drawn from
$\rho_\Phi$, and reuses each population over several minimization steps by
importance reweighting \cite{ref7,ref8,ref50}. The temperature-dependent
effective potential (TDEP) method is a force-constant fitting procedure
rather than a sampling scheme. It takes displacements and forces from a
sample of the anharmonic canonical ensemble, most commonly an \emph{ab
initio} molecular dynamics (AIMD) trajectory, which treats the nuclei
classically \cite{ref9}, and fits a model Hamiltonian of harmonic form,
optionally with third- and fourth-order terms, to these forces by least
squares \cite{ref5,ref6,ref52}. Mode-coupling theory provides the formal basis of
this construction: the force constants fitted in the canonical ensemble are
the static renormalized interactions of the phonon propagator, and the
higher-order constants determine its dynamical self-energy
\cite{ref51}. Stochastic TDEP (sTDEP) replaces the AIMD trajectory with
configurations drawn from the quantum canonical distribution of the current
harmonic model, in which each normal mode has its Bose--Einstein mean-square
amplitude. It then repeats fitting and sampling until the force constants are
self-consistent \cite{ref60}. When the average positions are fixed by
symmetry, the sTDEP fixed point is the SCHA auxiliary FC2
(Sec.~\ref{sec:effective}). The two methods therefore approach the same
effective FC2 by different numerical procedures, which differ in how samples
are generated, reused, and weighted. This self-consistent effective FC2 is
the target of the present work. It is the SCHA auxiliary FC2, not the
physical phonon spectrum of the anharmonic crystal, which in the SCHA follows
from the Hessian of the free energy and adds the cubic (bubble) correction
\cite{ref8}; we use ``effective dispersion'' for the spectrum of the auxiliary
FC2 throughout. The SCHA does not require sampling: the self-consistent
equations can also be solved with anharmonic force constants computed
beforehand, as in self-consistent phonon theory \cite{tadano2015}. When the forces are instead
taken from first-principles calculations on configurations, from a
molecular-dynamics trajectory or from Gaussian draws, these are statistical
samples of a canonical distribution, and the statistical error of the fitted
force constants decreases as the inverse square root of the number of
independent samples.
The number of first-principles force evaluations needed to converge the force
constants is therefore the dominant cost, and it is the subject of this
paper.

A deterministic alternative is suggested by the special quasirandom structure
(SQS) for substitutional alloys \cite{ref12}: instead of averaging many
disordered configurations, one constructs a single periodic structure whose pair
correlations reproduce those of the target ensemble within a cutoff. The
special displacement method of Zacharias and Giustino applies a related idea to
electron--phonon properties by choosing phonon-mode signs in one supercell
\cite{ref42}. Here we apply the SQS principle to thermal motion for the purpose
of effective-FC2 inference. A harmonic trial model defines the equal-time
displacement--displacement correlators at temperature $T$; a special optimal
thermal cell (SOTC) is a configuration of a fixed periodic supercell whose
shell-averaged correlators match those targets within the force-constant
range. First-principles forces on a few such cells are fitted to a constrained
FC2 model whose phonons define the next targets.

Whether designed configurations carry more information per force calculation
than samples of the canonical distribution is an empirical question. The present paper
answers it with controls that hold everything else fixed. For nine benchmark
crystals (fcc Al, Ag, Au, Cu; bcc Mo, W; rocksalt NaCl, MgO, PbTe) we
(i)~analyze the rank, conditioning, and prior sensitivity of correlator-matched
SOTC (SOTC-C) fits; (ii)~compare the resulting phonons with converged large-cell
finite-displacement (FD) references and with an effective FC2 fitted to a
40-snapshot canonical ensemble in the \emph{same} supercell at the same
temperature and electronic settings; and (iii)~compare SOTC with canonical
samples of exactly the same size, fitted by the same regression, on
held-out canonical forces, and also with special displacements, thermal
rattle, sTDEP, SSCHA (with density-functional theory (DFT) forces for Al and Si and on anharmonic model
potentials), TDEP fitted to \textit{ab initio}
molecular dynamics (Al and Si), and sparse and Bayesian regression. Diamond Si and hcp
Mg are added to test other structure classes. We further report the
non-analytic LO--TO correction for the polar crystals, Debye--Waller factors
against a tabulated reference, an independent test of constrained third-
and fourth-order fits, and the thermal expansion of Al and PbTe from self-consistent
SOTC-D on a volume grid.

\begin{figure*}[t]
\centering
\includegraphics[width=\textwidth]{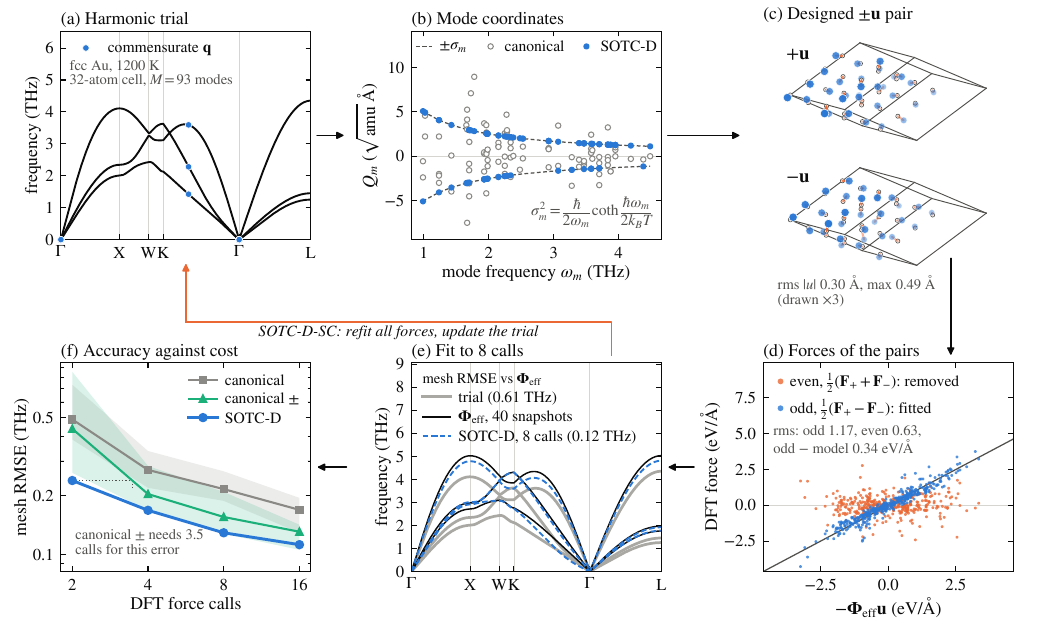}
\caption{Workflow of SOTC-D, with every panel computed from the DFT data of fcc Au at
1200~K in the 32-atom cell. (a)~The harmonic trial, initially the finite-displacement
FC2 of the cell, gives the commensurate modes (dots: wave vectors on this path that are
commensurate with the cell), with frequencies $\omega_m$, polarizations $\mathbf e_m$,
and quantum-thermal rms amplitudes $\sigma_m$. (b)~Mode coordinates $Q_m$ of one
configuration, recovered from its displacements: a canonical draw (open circles) is
Gaussian with variance $\sigma_m^2$, whereas SOTC-D (filled circles) has every mode at
exactly $\pm\sigma_m$ (dashed), with the signs chosen to maximize
$\det\mathbf X^{\rm T}\mathbf X$. (c)~The resulting configuration $\mathbf u$ and its
inverse $-\mathbf u$, which cost two force calls (open circles: ideal sites; orange:
displacements, enlarged threefold; symbol size and shade indicate depth; box: the cell).
(d)~DFT forces of the first eight calls against the force of the independent
40-snapshot effective FC2. The part odd under $\mathbf u\to-\mathbf u$, used in the
fit, follows this force (rms 1.17~eV/\AA, deviation 0.34~eV/\AA), whereas the even
part, which contains the cubic force (rms 0.63~eV/\AA), is removed by the pair.
(e)~The least-squares fit in the space-group adapted FC2 basis gives the effective
FC2 at temperature $T$, here strongly stiffened relative to the trial (legend: mesh
RMSE against the 40-snapshot effective FC2 $\mathbf\Phi_{\rm eff}$). (f)~Mesh RMSE
against a disjoint effective FC2 for SOTC-D and for random canonical subsets and
$\pm\mathbf u$ pairs (medians, bands 10--90\% over 20 draws; dotted: canonical pairs need
3.5 calls to reach the error of two SOTC-D calls); other crystals are shown in
Fig.~\ref{fig:counts}. In the self-consistent form SOTC-D-SC (orange arrow), the fitted
model becomes the trial for the next pair, all forces are refitted, and the iteration
has the SCHA and sTDEP fixed point.}
\label{fig:workflow}
\end{figure*}

This analysis identifies two weaknesses of correlator matching: its scalar
objective leaves the amplitude of individual modes, and hence the
information content of a configuration, uncontrolled, and its axial
regression prior biases the force constants. We therefore introduce SOTC-D, in
which each configuration excites every commensurate mode of the trial model at
exactly its quantum-thermal root-mean-square (rms) amplitude, the mode signs are chosen to
maximize the determinant of the FC2 information matrix (D-optimal design), and
every configuration is paired with its inverse (Fig.~\ref{fig:workflow}). An estimator theory
(Sec.~\ref{sec:theory}) identifies the target as the Gaussian-averaged
Hessian, gives the bias of the design exactly, and motivates a
self-consistent variant in which the trial is updated during the design.
At equal cost, SOTC-D gives lower dispersion errors than the competing
designs in most crystals at budgets of two to eight force calls. The gain is
largest at the smallest budgets and for crystals whose harmonic trial is far
from self-consistency; for a fixed trial it disappears at about 16 calls.

The paper is organized as follows. Section~\ref{sec:theory} defines the
effective force constants and their self-consistent fixed point, introduces
the correlator-matched and D-optimal designs, and derives the bias and
variance of the SOTC-D estimator. Section~\ref{sec:impl} describes the
regression and the iterative workflow, and Sec.~\ref{sec:comp} the
first-principles calculations and reference data. Section~\ref{sec:results}
compares the designs with each other and with canonical sampling, sTDEP,
SSCHA, special displacements, and thermal rattle, for DFT forces and for
anharmonic model potentials with known self-consistent solutions.
Section~\ref{sec:discussion} discusses the scope of the method, and
Sec.~\ref{sec:conclusions} summarizes the results.

\section{Theory}\label{sec:theory}

\subsection{Effective harmonic force constants}\label{sec:effective}

We consider nuclei moving on the Born--Oppenheimer surface $V(\mathbf R+\mathbf
u)$ about fixed reference positions $\mathbf R$, with displacements
$\mathbf u$ and forces $\mathbf F(\mathbf u)=-\nabla V$. In the
self-consistent harmonic description the thermal state at temperature $T$ is
represented by a harmonic Hamiltonian with force constants
$\boldsymbol\Phi$, whose quantum canonical displacement distribution is a
Gaussian $\rho_{\boldsymbol\Phi}(\mathbf u)$ with zero mean and covariance
$\mathbf C[\boldsymbol\Phi,T]$ given by Eq.~(\ref{eq:corr}) below. For a
given covariance $\mathbf C$, the effective FC2 is the harmonic model that best
reproduces the forces in this ensemble,
\begin{equation}
\boldsymbol\Phi_{\rm eff}[\mathbf C]
=\arg\min_{\boldsymbol\Phi}\big\langle|\mathbf F(\mathbf u)
+\boldsymbol\Phi\mathbf u|^2\big\rangle_{\mathbf C}
=-\langle\mathbf F\mathbf u^{\rm T}\rangle_{\mathbf C}\,\mathbf C^{-1}.
\label{eq:phieff}
\end{equation}
For a Gaussian distribution, Stein's lemma \cite{ref61} gives
$\langle\mathbf F\mathbf u^{\rm T}\rangle_{\mathbf C}
=-\langle\nabla\nabla V\rangle_{\mathbf C}\,\mathbf C$, so that
\begin{equation}
\boldsymbol\Phi_{\rm eff}[\mathbf C]=\langle\nabla\nabla V\rangle_{\mathbf C},
\label{eq:hessian}
\end{equation}
the Hessian averaged over the thermal displacement distribution. It
contains the renormalization of the harmonic force constants by the even
anharmonic terms, at leading order
$\tfrac12\sum_bV_{abbe}\sigma_b^2$ in normal coordinates. The
self-consistent effective FC2 is the fixed point
\begin{equation}
\boldsymbol\Phi_{\rm SC}=\boldsymbol\Phi_{\rm eff}\big[\mathbf C[\boldsymbol\Phi_{\rm SC},T]\big].
\label{eq:scfix}
\end{equation}
Equation~(\ref{eq:scfix}) is the stationarity condition of the SCHA free
energy with respect to the auxiliary force constants when the average
positions are fixed by symmetry \cite{ref7,ref50}, and it is the fixed point
of sTDEP \cite{ref60}, which alternates sampling from
$\rho_{\boldsymbol\Phi}$ and least-squares fitting. TDEP with molecular
dynamics evaluates the same least-squares problem with the average taken over
the classical anharmonic canonical ensemble instead of a Gaussian
\cite{ref5,ref6}. One evaluation of Eq.~(\ref{eq:phieff}) from a harmonic
trial gives $\boldsymbol\Phi_{\rm eff}[\mathbf C_{\rm harm}]$; we call it the
iteration-0 target, and $\boldsymbol\Phi_{\rm SC}$ the self-consistent target.
Equations~(\ref{eq:phieff})--(\ref{eq:scfix}) are standard
\cite{ref5,ref7,ref50,ref60}; the derivation of Eq.~(\ref{eq:hessian}) from
Stein's lemma and its relation to the SCHA free-energy gradient are given in
the Supplementary Material, Sec.~S4.

In every scheme the average in Eq.~(\ref{eq:phieff}) is replaced by a sum over
a finite set of configurations, each requiring one first-principles force
calculation. The estimate is a linear regression of forces on
displacements, and its accuracy for a given number of force calculations is
determined by how the configurations are chosen. Canonical (Gaussian) sampling, as in sTDEP and SSCHA, draws
them independently from $\rho_{\boldsymbol\Phi}$. The rest of this section
describes two deterministic alternatives and the statistical properties that
distinguish them.

\subsection{Thermal displacement correlators}

For a crystal with $n_b$ basis atoms and $N_c$ primitive cells, the
equal-time canonical correlator of a harmonic lattice is \cite{ref56}
\begin{equation}
\begin{split}
C_{ab}^{\alpha\beta}(\mathbf R,T)
 &=\frac{\hbar}{2N_c\sqrt{M_aM_b}}\sum_{\mathbf q s}
 \frac{e_a^{\alpha}(\mathbf q s)e_b^{\beta*}(\mathbf q s)}{\omega_{\mathbf q s}}\\
 &\quad\times\coth\!\frac{\hbar\omega_{\mathbf q s}}{2k_BT}\,
 e^{-i\mathbf q\cdot\mathbf R},
\end{split}
 \label{eq:corr}
\end{equation}
where $\mathbf q$ runs over the wave vectors commensurate with the
Born--von K\'arm\'an cell, $s$ labels branches, and $e$ are mass-weighted
polarization vectors in the gauge in which the Bloch phase carries only the
lattice vector $\mathbf R$; for polarization vectors in the gauge of the full
atomic position (as returned by Phonopy), $\mathbf R$ is replaced by
$\mathbf R+\boldsymbol\tau_b-\boldsymbol\tau_a$, and mixing the two gauges
leaves the on-site terms unchanged but corrupts the correlation between
different basis atoms (standard result \cite{ref56}; derivation in
Supplementary Material, Sec.~S8). The $\coth$ factor interpolates between zero-point motion
and classical equipartition. The SOTC target is the shell- and trace-averaged
scalar
\begin{equation}
\bar C_2^{ab}(R,T)=\frac{1}{3N_R^{ab}}\sum_{\mathbf R'\in\mathcal S_R^{ab}}
\sum_{\alpha}C_{ab}^{\alpha\alpha}(\mathbf R',T),
\label{eq:shell}
\end{equation}
where $\mathcal S_R^{ab}$ is the neighbor shell of radius $R$ for the ordered
basis pair $(a,b)$, with multiplicity $N_R^{ab}$. Shells are formed by binning
pair distances at 0.05~\AA\ separately for each species pair. The on-site value
$\bar C_2^{aa}(0,T)$ is the isotropic displacement parameter
$U_{\rm iso}=\langle|\mathbf u_a|^2\rangle/3$, related to the crystallographic
Debye--Waller factor by $B=8\pi^2U_{\rm iso}$.

\subsection{Correlator-matched cells (SOTC-C)}\label{sec:sotcc}

For a candidate displacement field $\mathbf u$ of an $n$-atom supercell, the
SOTC objective is
\begin{equation}
\mathcal Q(\mathbf u)=\sum_{a,b}\sum_{R\le r_{\rm max}}
e^{-R/\xi_w}\left[\bar C_{2,ab}^{\rm cell}(R;\mathbf u)
-\bar C_{2,ab}(R,T)\right]^2,
\label{eq:Q}
\end{equation}
where $\bar C_{2,ab}^{\rm cell}$ is evaluated from the single configuration
with the same shell, trace, and species averaging as Eq.~(\ref{eq:shell}), and
$\xi_w$ is a numerical weighting length ($\xi_w=r_{\rm cut}/2$; the gradient
is given in Supplementary Material, Sec.~S9). The
optimization starts from a superposition of commensurate trial phonon modes
with harmonic amplitudes
$A_{\mathbf qs}=[\hbar\coth(\hbar\omega_{\mathbf qs}/2k_BT)/2\omega_{\mathbf qs}]^{1/2}$
and random phases, and adjusts mode amplitudes and phases by gradient descent
with restarts; rigid translations are removed. Each iteration produces
$N_{\rm ens}$ such configurations with different random starts. The objective
constrains only the selected scalar moments: it does not by itself determine
the full tensor covariance, and the matched controls below are needed to assess
its value for force-constant inference. We refer to cells designed with this
objective as SOTC-C.

\subsection{D-optimal thermal-mode design (SOTC-D)}\label{sec:sotcd}

Equation~(\ref{eq:Q}) constrains shell sums of second moments, but not the
amplitude with which individual modes appear in a configuration; a mode that
happens to be weakly excited contributes little information to the force fit.
SOTC-D removes this freedom. Let $\omega_m$ and $\mathbf e_m$ be the
commensurate normal modes of the trial FC2 in the supercell (translations
excluded) and
$\sigma_m^2=\hbar\coth(\hbar\omega_m/2k_BT)/2\omega_m$ the quantum-thermal
variance of the mass-weighted mode coordinate. Configuration $k$ is
\begin{equation}
u_{i}^{(k)}=\frac{1}{\sqrt{M_i}}\sum_m s_{km}\,\sigma_m\,e_{im},\qquad
s_{km}\in\{-1,+1\},
\label{eq:sotcd}
\end{equation}
so that every configuration reproduces every mode variance exactly; the
off-diagonal products $s_{km}s_{km'}\sigma_m\sigma_{m'}$ vanish only on
average over sign patterns (Supplementary Material, Sec.~S8). The signs are
the only design variables. Equations~(\ref{eq:shell})--(\ref{eq:sotcd})
define the designs introduced in this work. Writing $\mathbf X$ for the force design matrix of the
space-group adapted FC2 basis (Sec.~\ref{sec:fit}), the signs of configuration
$k$ are chosen by greedy single-flip ascent of
$\log\det(\mathbf X^{\rm T}\mathbf X)$ given configurations $1,\dots,k-1$
(sequential D-optimal design \cite{ref57}), so that every prefix of the
sequence is itself a design. Each configuration is evaluated together with its
inverse $-\mathbf u^{(k)}$ (antithetic pairing \cite{ref58}). Because the third-order force is even in
$\mathbf u$, the pair cancels it exactly from the linear FC2 fit, whereas the
quartic renormalization that defines the effective FC2 is retained. A budget of
$n$ force calls thus corresponds to $n/2$ designed configurations. The
trial model enters only through $\omega_m$, $\mathbf e_m$, and $\sigma_m$, as
it does for canonical sampling.

\subsection{Bias and variance of the SOTC-D estimator}\label{sec:biasvar}

Writing the potential in normal
coordinates as $V=\tfrac12\sum\omega_a^2Q_a^2+\tfrac16\sum V_{abc}Q_aQ_bQ_c
+\tfrac1{24}\sum V_{abcd}Q_aQ_bQ_cQ_d$, the canonical target of Eq.~(\ref{eq:hessian}) is
$\Phi_{ae}=\omega_a^2\delta_{ae}+\tfrac12\sum_bV_{abbe}\sigma_b^2$ to fourth
order in the displacements. Averaging the regression coefficient of
Eq.~(\ref{eq:sotcd}) over independent sign patterns, the
only moment that differs from the Gaussian is $\langle Q_e^4\rangle=\sigma_e^4$
instead of $3\sigma_e^4$, so that
\begin{equation}
\Phi^{\rm SOTC\text{-}D}_{ae}-\Phi_{ae}=-\tfrac13V_{aeee}\sigma_e^2 .
\label{eq:bias}
\end{equation}
This is one diagonal term of the renormalization sum over the $M$
commensurate modes, so the bias is $O(1/M)$ relative to the quartic
renormalization and vanishes for large cells; the measured offsets of
Sec.~\ref{sec:scaling} contain further contributions and do not show this
decrease. D-optimal signs are not
independent, so Eq.~(\ref{eq:bias}) describes the design on average; the
residual is part of the variance discussed below. The bias is also removed exactly
in finite cells if the squared amplitudes of each mode are varied between
configurations with the Gaussian moments $\langle Q_e^2\rangle=\sigma_e^2$
and $\langle Q_e^4\rangle=3\sigma_e^4$ (moment-matched SOTC-D,
Sec.~\ref{sec:sc_surrogate}), at the price of a larger variance at small
budgets. Eq.~(\ref{eq:bias}) is derived
here (Supplementary Material, Sec.~S4) and was checked by exact enumeration
over sign patterns; the averaged-Hessian identity follows from Stein's lemma
\cite{ref61}, and its equivalence with the SCHA stationarity condition is
standard \cite{ref50}.

Under $\mathbf u\to-\mathbf u$ the cubic force is even and
the harmonic and quartic forces are odd. For a $\pm$ pair the cubic force
therefore lies in the null space of the linear fit and cancels exactly, for
any FC3; for unpaired canonical samples it enters as noise of order
$V_{abc}\sigma^2$. The remaining variance of the least-squares estimator is
$\propto(\mathbf X^{\rm T}\mathbf X)^{-1}$, and maximizing
$\det\mathbf X^{\rm T}\mathbf X$ minimizes the volume of its confidence
ellipsoid \cite{ref57}. Fixed amplitudes remove the amplitude variance of
$\mathbf X^{\rm T}\mathbf X$, which for canonical draws makes weakly excited
modes poorly determined at small budgets.

These statements can be made quantitative. For a single normal coordinate
with force $f=-\omega^2Q-cQ^2-dQ^3+\epsilon$, where $\epsilon$ (variance
$s^2$) is the part of the force that the FC2 model does not describe, the
least-squares estimate from $K$ calls has closed-form bias and variance
(Supplementary Sec.~S4). With $n$ independent Gaussian amplitudes ($n=K$
for canonical draws, $K/2$ for $\pm$ pairs) the noise variance is
$s^2n/[(n-2)\sigma^2K]$, against $s^2/(\sigma^2K)$ for fixed amplitudes;
unpaired draws add the cubic term $15c^2\sigma^2n/[(n+2)(n+4)]$, and
Gaussian amplitudes add the variance of $\sum Q^4/\sum Q^2$ and a
finite-budget bias $-6d\sigma^2/(n+2)$. Fixed amplitudes therefore have
the smaller mean-squared error up to a crossover budget beyond which
their bias, Eq.~(\ref{eq:bias}), dominates. In a cell with $P$ FC2
parameters and $M$ modes, each parameter averages over about $nM/P$
amplitudes, and the variance inflation of random designs becomes
$\approx1+2P/(nM)$. For the DFT design matrices of Al ($P=30$, $M=93$)
and NaCl ($P=53$, $M=189$), the A-criterion
$\mathrm{tr}(\mathbf X^{\rm T}\mathbf X)^{-1}$ of paired canonical
sampling exceeds that of SOTC-D by factors of 1.89 and 1.54 at two calls,
1.17 and 1.11 at eight, and 1.09 and 1.05 at sixteen, and for Si
($P=10$, $M=189$) by at most 1.13; D-optimal signs account for factors of
1.29 (two calls) to 1.02 (sixteen calls) of this relative to random signs.
The theory thus predicts the observations of Sec.~\ref{sec:results}: the
advantage of SOTC-D is largest at two to eight calls, vanishes near sixteen,
and is small for Si, whose few parameters each average over many modes.

Under $\mathbf u\to-\mathbf u$ the FC2 and FC4 features of the force
regression are odd and the FC3 features even. For $\pm$ pairs the
FC2--FC3 cross-information therefore vanishes exactly, and the difference
and the sum of a pair determine the odd and the even force constants
separately; per force call, the FC3 variance of paired designs is not
larger than that of unpaired ones (Supplementary Sec.~S4). Fixed
amplitudes match the second but not the fourth Gaussian moment, which
biases an FC2-only fit through the unfitted FC4 [Eq.~(\ref{eq:bias})]. When
FC4 is fitted jointly, the joint design is of full rank for fixed amplitudes
as well and this bias disappears; moment-matched amplitudes remove it
without fitting FC4.

\subsection{Self-consistent design (SOTC-D-SC)}\label{sec:sotcdsc}

The trial enters SOTC-D only
through $\omega_m$, $\mathbf e_m$, and $\sigma_m$, and can be updated during
the design. After each $\pm$ pair, all data so far are refitted; while the
data do not yet determine every coefficient, the update is the minimum-norm
correction of the trial coefficients consistent with the data. The modes and
amplitudes of the next pair are computed from the updated model, and its signs
maximize $\det(\mathbf X_{\rm prev}^{\rm T}\mathbf X_{\rm prev}
+\mathbf X^{\rm T}\mathbf X)$ given all previous configurations. This
converges to the self-consistent target without discarding the forces of
earlier iterations, unlike sTDEP, which redraws each iteration. It also
repairs a defective trial: modes that are unstable in the trial are omitted
from the first pair (as in canonical sampling) but acquire the correct
amplitude once the fitted model stabilizes them. In the DFT benchmark below we
use the simplest version, \emph{bootstrapped} SOTC-D, with a single update:
the first pair is designed from the harmonic trial, and the remaining pairs
are designed from the model refitted to that pair, with signs chosen
sequentially given all earlier pairs. The name SOTC-D-SC is reserved for the
version that updates the trial after every pair, which is used on the model
potentials of Sec.~\ref{sec:sc_surrogate}.

\section{Implementation}\label{sec:impl}

\subsection{Effective FC2 regression}\label{sec:fit}

Forces are evaluated for the undisplaced reference and for each designed
configuration in the same cell, and the response
$\Delta\mathbf F=\mathbf F(\mathbf u)-\mathbf F(\mathbf 0)$ is fitted to
the model below. At the ideal positions used here $\mathbf F(\mathbf 0)$
vanishes by site symmetry; the subtraction removes residual numerical forces
(rms $2\times10^{-6}$ to $1.4\times10^{-3}$~eV/\AA\ over the nine SOTC-C
data sets, at most $10^{-3}$ of the rms thermal force response, 0.48--2.3~eV/\AA),
while the effect of using a lattice parameter that is not the minimum of the
Perdew--Burke--Ernzerhof (PBE) functional is
assessed separately (Sec.~\ref{sec:comp}). The model is
\begin{equation}
\Delta F_i^\alpha=-\sum_{j\ne i,\beta}\Phi_{ij}^{\alpha\beta}
\left(u_j^\beta-u_i^\beta\right),
\label{eq:fit}
\end{equation}
with one $3\times3$ tensor per translationally distinct pair vector within the
cutoff $r_{\rm cut}$ and Newton's third law
$\boldsymbol\Phi(-\mathbf R)=\boldsymbol\Phi(\mathbf R)^{\rm T}$ built into the
parametrization. The tensors are obtained by ridge regression
\begin{equation}
\hat{\mathbf p}=-\left(\mathbf G^{\rm T}\mathbf G+\lambda\mathbf I\right)^{-1}
\mathbf G^{\rm T}\Delta\mathbf F,
\label{eq:ridge}
\end{equation}
with $\lambda=10^{-3}$~\AA$^2$. Equation~(\ref{eq:fit}) is the standard
pair-force model of TDEP-type fits \cite{ref5} and Eq.~(\ref{eq:ridge}) standard
ridge regression. Two physical priors are then applied.
First, each pair tensor is projected onto the axially symmetric form
$\boldsymbol\Phi=\alpha\,\hat{\mathbf R}\hat{\mathbf R}+\beta(\mathbf I-\hat{\mathbf R}\hat{\mathbf R})$
and $\alpha,\beta$ are averaged over each shell (the ``axial projection'').
Second, the on-site blocks are set by the acoustic sum rule,
$\boldsymbol\Phi_{ii}=-\sum_{j\ne i}\boldsymbol\Phi_{ij}$. The axial form is
exact for bonds along threefold or fourfold axes (bcc first neighbors, fcc and
rocksalt second neighbors, rocksalt first neighbors) but omits the third
independent constant allowed by symmetry for fcc first-neighbor $\langle110\rangle$
bonds. Because these priors can bias the result, Sec.~\ref{sec:audit} reports
the same data refitted without the projection, with different $\lambda$ and
cutoffs, and in a complete space-group, permutation, and sum-rule adapted FC2
basis constructed with symfc \cite{ref54}, solved by unregularized least
squares; this fit is used for SOTC-D and for all matched canonical baselines. It is the estimator of TDEP: ordinary least squares for the
second-order force constants in a basis that imposes the space-group,
permutation, and acoustic-sum-rule constraints \cite{ref5}. As a check,
the AIMD and canonical data of Al and Si were also fitted with the TDEP code
(\texttt{extract\_forceconstants}, Hellman \textit{et al.}) at the same pair
cutoff: in the same symmetry-adapted basis the two fits agree to
$10^{-11}$~THz on the phonon mesh. TDEP limits the pair cutoff to the
Wigner--Seitz radius of the supercell (2.87~\AA, the first shell, in the
32-atom Al cell, whose canonical fit is then 0.17~THz from the
self-consistent target against 0.056~THz with the 4.5~\AA\ cutoff of
the supercell basis) and solves the force-constant orders sequentially, so
its FC2 from MD data is not corrected for the cubic forces
(Sec.~\ref{sec:iter1}). The only differences from a TDEP fit are the basis (the
space group of the supercell rather than that of the crystal, compared in
Sec.~\ref{sec:design}) and the absence of TDEP's optional higher-order terms.
All protocols, including the canonical, sTDEP, and AIMD baselines, use this
same fit, so they differ only in the configurations.
For that basis, the rank and condition number of the column-normalized
design matrix are reported.

The symfc basis is adapted to the space group of the \emph{supercell}, which
for the low-symmetry SOTC cells is a subgroup of the crystal space group and
admits 30--90 parameters. We therefore also use the FC2 basis of the full
crystal space group, constructed with hiPhive \cite{ref59} from the primitive
cell and mapped to the supercell, with a pair cutoff limited to half the
shortest supercell image distance so that no pair interaction is aliased. It
has 3 (fcc), 7 (bcc), 8 (rocksalt) parameters at the same physical range
(Sec.~\ref{sec:design}). For the same data, fewer parameters at the same
physical range reduce the variance of every estimator.

\subsection{SOTC-C workflow and phonon properties}\label{sec:selfcons}

Iteration $n$ designs configurations for targets computed from the current
phonon model (a Debye model at $n=0$), evaluates forces, fits the FC2 model,
and recomputes $\bar C_2$ from Eq.~(\ref{eq:corr}) on an $8\times8\times8$ $q$
mesh. The targets are linearly mixed with weight 0.5; an update from a spectrum
with more than 1\% unstable modes is rejected. The loop stops when the
root-mean-square change in $\bar C_2$ falls below $3\times10^{-3}$~\AA$^2$ after
a material-specific minimum number of iterations, and the most stable iteration
is retained. Each iteration recomputes the reference, so the number of force
calculations is $N_{\rm iter}(N_{\rm ens}+1)$.

Dynamical matrices are Fourier interpolated from the real-space FC2.
Isochoric heat capacity $C_V$, vibrational entropy $S_{\rm vib}$, free energy
$F_{\rm vib}$, and zero-point energy follow from the harmonic partition
function of the fitted frequencies, omitting modes below 0.05~THz (including
imaginary ones); the spectral Debye temperature is the standard second-moment
value $T_D=(\hbar/k_B)\sqrt{5\langle\omega^2\rangle/3}$. These are quantities of the
effective harmonic model at the cell volume used. The electronic heat
capacity is not included. Thermal expansion, with the anharmonic correction
to the free energy, is computed for Al and PbTe (Sec.~\ref{sec:qha}).

\section{Computational details}\label{sec:comp}

\subsection{Electronic structure}

Forces were computed with Quantum ESPRESSO (QE) 7.5 \cite{ref46}, the PBE functional
\cite{ref17}, and pseudopotentials of the standard solid-state pseudopotential (SSSP)
efficiency library \cite{ref47} (optimized norm-conserving Vanderbilt (ONCV) for Ag,
Au, Mo; projector augmented-wave (PAW) for Al, Cu, Mg, O, Pb; ultrasoft for Na, Cl, Te, W). Wave-function
cutoffs were 40 (Al), 50 (Ag), 45 (Au, Cu), 40 (Mo), 30 (W), 60 (NaCl), 80
(MgO), and 44~Ry (PbTe); charge-density cutoffs were four (ONCV) or eight
(PAW, ultrasoft) times larger. Metals used 0.02~Ry Marzari--Vanderbilt
smearing \cite{ref18}; the rocksalt compounds used fixed occupations. The
self-consistent-field (SCF) threshold was $10^{-8}$~Ry for the primary SOTC runs and their same-cell
controls and $10^{-9}$~Ry for all other references.

\subsection{Correlator-matched SOTC runs}

The lattice parameters (Table~\ref{tab:runs}) are experimental values near the
design temperature $T_{\rm des}$; ideal positions were used. The supercells
were chosen by enumerating Hermite normal forms at the target size: 32 atoms
for the fcc metals, 64 for the bcc metals and NaCl and MgO, and 72 for PbTe.
Metals used a $2\times2\times2$ supercell $k$ mesh, rocksalt compounds
$\Gamma$ only. Table~\ref{tab:runs} lists the cutoffs, iteration counts, and
force-call accounting. The SOTC-C model of each crystal is the fit of
Eq.~(\ref{eq:fit}) to the configurations of the iteration selected by the
workflow, i.e., the most stable converged one (Sec.~\ref{sec:selfcons}); all SOTC-C results below are refits of the stored
force--displacement data.

\begin{table}[t]
\caption{Correlator-matched SOTC runs (SOTC-C). $a$: lattice parameter; $n$: supercell atoms; $N_{\rm iter}$, $N_{\rm ens}$: iterations and configurations per iteration; $N_{\rm force}=N_{\rm iter}(N_{\rm ens}+1)$ QE calls; $p$: parameters of the space-group adapted FC2 basis within $r_{\rm cut}$ in the SOTC cell.}
\label{tab:runs}
\begin{ruledtabular}
\begin{tabular}{lcccccccc}
 & $a$ (\AA) & $T_{\rm des}$ (K) & $n$ & $r_{\rm cut}$ (\AA) & $N_{\rm iter}$ & $N_{\rm ens}$ & $N_{\rm force}$ & $p$\\
\hline
Al & 4.05 & 300 & 32 & 4.5 & 2 & 5 & 12 & 30\\
Ag & 4.12 & 1000 & 32 & 4.5 & 4 & 5 & 24 & 30\\
Au & 4.13 & 1200 & 32 & 4.5 & 2 & 1 & 4 & 30\\
Cu & 3.67 & 1200 & 32 & 4.5 & 4 & 8 & 36 & 90\\
Mo & 3.16 & 1500 & 64 & 5.0 & 2 & 1 & 4 & 78\\
W & 3.18 & 1500 & 64 & 5.0 & 4 & 3 & 16 & 78\\
NaCl & 5.73 & 700 & 64 & 4.3 & 5 & 3 & 20 & 53\\
MgO & 4.26 & 1500 & 64 & 3.2 & 4 & 3 & 16 & 53\\
PbTe & 6.51 & 700 & 72 & 5.0 & 4 & 5 & 24 & 53\\
\end{tabular}
\end{ruledtabular}
\end{table}

\subsection{Reference and control calculations}

Three classes of additional calculations were performed.
\textit{Converged harmonic references} used central finite displacements
($\pm0.01$~\AA) with the open-source finite-displacement phonon code phonopy
\cite{ref53} in $5\times5\times5$ primitive
supercells (125 atoms) for the metals and $4\times4\times4$ (128 atoms) for the
rocksalt compounds, with $3\times3\times3$ and $2\times2\times2$ $k$ meshes,
respectively, at the production lattice parameters. The same calculations in
$3\times3\times3$ primitive supercells (27 or 54 atoms) were also performed,
together with references at the PBE equilibrium lattice parameter obtained from
third-order Birch--Murnaghan fits to primitive-cell energies at seven
lattice parameters within $\pm3$\%, and at the experimental
room-temperature lattice parameter for Al, Cu, and Au. The 128-atom PbTe
states used parallel-orbital diagonalization with local Thomas--Fermi
mixing, because Davidson diagonalization failed and plain mixing did not
converge in this cell.
\textit{Same-cell controls} used exactly the primary SOTC supercell and
electronic settings: a central FD harmonic FC2, and 48 quantum-canonical
configurations (24 $\pm\mathbf u$ pairs) drawn at $T_{\rm des}$ from that
harmonic model. Pairs 1--20 form a 40-snapshot training pool and pairs 21--24 a
held-out test set. The \textit{effective reference} is the symfc FC2 fitted to
all 40 training snapshots; \textit{matched canonical baselines} are the
space-group adapted least-squares fit (Sec.~\ref{sec:fit}) applied to 20 random subsets of the pool with exactly as
many configurations as the SOTC-C fit being compared, or of complete $\pm$
pairs. Because a canonical subset shares configurations with a reference
fitted to the whole pool, dispersion errors in the matched comparison are
measured against an effective FC2 fitted to the $40-n$ pool snapshots
\emph{disjoint} from the scored design (the complement of a canonical subset,
or a random $(40-n)$-subset for SOTC-D), and held-out forces are used as a
reference-free measure. Finally, \textit{SOTC-D designs} of eight
configurations (16 force calls with their inverses) were generated from the
same-cell harmonic FD model, i.e., from exactly the trial model used for the
canonical baselines, and evaluated in the same cell and settings.

Two further competitors were generated from the same trial and evaluated in
the same way (16 configurations each): \textit{special displacements} (ZG)
\cite{ref42}, in which the real, mass-normalized commensurate modes of the
half Brillouin zone are excited at the canonical amplitude with the
orthogonal Walsh sign patterns of Ref.~\onlinecite{ref42}, and a
\textit{thermal rattle} as used for hiPhive training sets \cite{ref59}, in
which every atom is displaced independently with the canonical per-atom
variance. To test the trial update (Sec.~\ref{sec:sotcdsc}), the effective FC2
of the 40 iteration-0 snapshots was used as a new trial (\textit{iteration
1}, one sTDEP step), from which 48 further canonical configurations (40
training, 8 test) and 16 SOTC-D configurations were computed; and a
bootstrapped SOTC-D set was computed whose first pair is that of the
iteration-0 SOTC-D and whose seven further pairs are designed from the model
refitted to the first pair (Sec.~\ref{sec:sotcdsc}). For a DFT-level sTDEP comparison, two-iteration sTDEP runs were computed
with our implementation of the protocol of Ref.~\onlinecite{ref60}, using the
regression of Sec.~\ref{sec:fit} rather than the TDEP code, for budgets of 8 and 16 calls (three
seeds each): half of the budget as canonical $\pm$ pairs from the harmonic
trial (taken from the iteration-0 pool), and half as new canonical $\pm$
pairs drawn from the model refitted to them (36 additional force calls per
crystal). For Al and Si, the self-consistent protocols were also run with
DFT forces in the SOTC cells, all from the same harmonic trial and up to 32
force calls: SOTC-D-SC (Sec.~\ref{sec:sotcdsc}, three seeds); sTDEP iterated
to self-consistency with four canonical configurations (two $\pm$ pairs) per
iteration, fitted to the last iteration as in Ref.~\onlinecite{ref60} or to
all iterations (three seeds); and a self-consistent form of special
displacements (ZG-SC), which adds one ZG $\pm$ pair designed from the
current model per iteration and refits all of them, in the spirit of the
anharmonic special-displacement method; its three replicates use the
first three Walsh sign patterns. SSCHA used python-sscha
\cite{ref50} with the whole cell as the unit cell ($\Gamma$ point only), the
same harmonic trial, and the minimizer settings of
Sec.~\ref{sec:sc_surrogate}, with populations of $N=16$, 32, 64, and 256
configurations; these runs stop after the first population, and further
runs with $N=16$, 32, and 64 were iterated over 6--10 populations with a
meaningful factor of $10^{-3}$ (at most 100 minimization steps per
population). SSCHA force constants, which are not restricted to $r_{\rm
cut}$, are projected onto the supercell basis in the metric of the thermal
displacement covariance, i.e., as a force regression on thermal
displacements would see them; a Euclidean projection of the same matrix
differs from it by 0.10 (Al) and 0.06~THz (Si). The reference for these
comparisons is the converged self-consistent effective FC2: sTDEP with
128 canonical configurations per iteration, four iterations, the last
iteration fitted alone, two independent seeds averaged (1024 force calls per
crystal). The two seeds agree within 0.029 (Al) and 0.013~THz (Si), and the
last two iterations of each change it by 0.04--0.05 (Al, the statistical
level of 128 samples) and 0.02~THz (Si). As an AIMD-TDEP baseline for Al and Si, four Born--Oppenheimer
molecular-dynamics trajectories per crystal were run with QE in the same cell and with the same
electronic settings (Verlet integration, 3.9~fs time step, stochastic velocity
rescaling at 300~K (Al) and 600~K (Si) with a 77~fs time constant), each started from an independent canonical
configuration and run for 1307--1372 (Al) and 791--820 steps (Si). After 250 equilibration steps, the
FC2 was fitted to the first $n$ production steps of each trajectory, with the
same basis and regression alone and jointly with an FC3 (symfc basis, cutoff
4.0~\AA); every MD step is counted as one force call. Iteration-1 results are scored against the iteration-1 effective
FC2. For the thermal expansion of Al, the 32-atom cell was scaled to seven
lattice parameters from 4.000 to 4.150~\AA\ in steps of 0.025~\AA. At each,
the harmonic FD trial (four displacements and the undisplaced cell) was
computed, and SOTC-D-SC was run with 32 force calls at 100, 300, 400, 500,
700, and 900~K, twice: once with design seeds that differ between volumes and
once with the same seeds at all volumes; the second grid was extended to
128 calls at 700 and 900~K, and a third grid with moment-matched amplitudes
(32 calls, common seeds) was computed at 700 and 900~K. The harmonic FD FC2 was also computed
on the same volumes in the 125-atom cell with the large-cell settings above.
For PbTe, the 72-atom cell was scaled to seven lattice parameters from 6.52
to 6.70~\AA\ in steps of 0.03~\AA, and SOTC-D-SC was run with 32 force
calls at 300, 500, and 700~K with the same seeds at all volumes.

The same protocol was applied to two further classes of crystal: covalent
diamond Si (600~K, 64-atom conventional $2\times2\times2$ cell, $r_{\rm
cut}=5.0$~\AA, ONCV pseudopotential \cite{ref47}, 30/120~Ry, $2\times2\times2$
$k$ mesh) and hexagonal close-packed Mg (600~K, $4\times4\times2$ 64-atom
cell, $r_{\rm cut}=4.6$~\AA, PAW, 40/320~Ry, $3\times3\times3$ $k$ mesh), with
converged references in 128-atom ($4\times4\times4$ primitive) and 150-atom
($5\times5\times3$) cells. For these crystals no SOTC-C runs were made, and
only the same-cell protocol above is used.

HSE06 calculations \cite{ref_hse06,ref_hse06b} used the 64-atom cells of Si and NaCl,
an exact-exchange $q$ mesh of $1\times1\times1$, and the adaptively
compressed exchange operator of QE. Si used the PBE settings above. For NaCl,
QE ignores a reduced Fock cutoff with ultrasoft pseudopotentials, and the
full Fock grid of the 60/480~Ry setting costs about three hours per call on
128 cores; HSE06 runs of NaCl therefore used SG15 ONCV pseudopotentials
\cite{ref_sg15} (80/320~Ry, Fock cutoff 160~Ry, $\Gamma$ only, symmetry
disabled), and the PBE runs entering the NaCl reference (harmonic FD,
self-consistent sTDEP with four iterations of 64 calls, SOTC-D-SC and sTDEP
with 16 calls) were repeated with the same pseudopotentials. Their harmonic
phonons agree with those of the ultrasoft setting within 0.015~THz. An HSE06
force call took about 15 minutes on 64 cores in both crystals.

Non-analytic corrections for NaCl, MgO, and PbTe used Born effective charges and
electronic dielectric tensors from QE density-functional perturbation theory (DFPT) in
the primitive cell at the production lattice parameters (60/480~Ry,
$8\times8\times8$ $k$ mesh), applied through the mixed-space formulation in
phonopy \cite{ref53}. Phonon densities of states and thermodynamic quantities
were evaluated on $12\times12\times12$ (comparisons) or $30\times30\times30$
(DOS) meshes; $U_{\rm iso}$ was evaluated on a shifted $24\times24\times24$
mesh, which changes the converged-reference value for Al by less than 1\%
relative to $60\times60\times60$.

\section{Results}\label{sec:results}

\subsection{Identifiability and priors of SOTC-C fits}\label{sec:audit}

Table~\ref{tab:audit} summarizes the identifiability of the nine SOTC-C data
sets. In the space-group adapted basis, the force design of every single
iteration has full column rank, with 30 (fcc metals), 53 (rocksalt), 78
(bcc), or 90 (Cu) FC2 parameters and normalized condition numbers of
2.9--17.1. This is possible because a single force calculation in an
$n$-atom cell gives $3n$ force components (96--216 here), each a linear
equation in the FC2 parameters, and because a collective displacement of all
atoms makes many of them linearly independent; the symmetry-adapted basis reduces
the unknowns to 30--90. A few collective SOTC configurations therefore
identify the FC2 model of these cells completely once the space-group
symmetry of the supercell and the acoustic sum rule are imposed, without any
further prior.
The parameter count is set by the number of neighbor shells within $r_{\rm
cut}$, not by the number of configurations. For Cu ($a=3.67$~\AA) the
4.5~\AA\ cutoff includes the third fcc shell at $a\sqrt{3/2}=4.49$~\AA, which
for Al, Ag, and Au lies beyond it (4.96--5.06~\AA); the Cu model therefore has
90 rather than 30 parameters in the same 32-atom cell; because this shell
lies only 0.005~\AA\ inside the cutoff, the Cu parameter count is sensitive
to the choice of $r_{\rm cut}$. The SOTC-C runs also
differ in the number of configurations per iteration, $N_{\rm ens}=1$ (Au,
Mo), 3 (W, NaCl, MgO), 5 (Al, Ag, PbTe), or 8 (Cu), as listed in
Table~\ref{tab:runs}; with $N_{\rm ens}=8$ each Cu iteration alone has full
rank 90.

The regression prior matters more than identifiability. Without the axial
projection, the ridge fits are unstable in every material (up to 33\% unstable
modes for Au), because the general $3\times3$ pair tensors of
Eq.~(\ref{eq:fit}) are not symmetry constrained. With the projection, eight of
the nine SOTC-C models are stable. The exception is Au, whose two iterations
each used a single configuration: 0.7\% and 15.9\% of the mesh modes are
unstable, and the selected first iteration therefore has 0.7\% unstable
modes (minimum $-0.95$~THz; Supplementary Table~S5). Pooling
the two Au configurations removes it (minimum frequency 0.27~THz), and the
converged harmonic FD reference at the same lattice parameter is stable
(minimum 0.36~THz), so the instability is an artifact of the one-configuration
fit rather than a property of PBE Au. The ridge parameter changes the results
by at most 0.09~THz between $\lambda=0$ and 0.1~\AA$^2$; the cutoff bracket
matters more, particularly for MgO, where lowering $r_{\rm cut}$ by 0.5~\AA\
to 2.7~\AA\ removes the second-neighbor shell at 3.0~\AA\ and changes the
root-mean-square error (RMSE) of the mesh frequencies by 0.83~THz (Table~\ref{tab:audit}).

\begin{table}[t]
\caption{Identifiability and priors of the SOTC-C force data. Rank/$p$ and the largest condition number over all iterations refer to the unregularized space-group adapted FC2 design of each single iteration. $f_{\rm unst}$ is the fraction of unstable $12^3$-mesh modes, with the axial prior and without the axial projection, in the worst iteration. $\Delta_\lambda$ and $\Delta_r$ are the ranges (THz) of the mesh RMSE against the converged FD reference over $\lambda\in[0,0.1]$ and $r_{\rm cut}\pm0.5$~\AA\ for the pooled data.}
\label{tab:audit}
\begin{ruledtabular}
\begin{tabular}{lccccc}
 & rank/$p$ & cond. & $f_{\rm unst}$ axial & $f_{\rm unst}$ no proj. & $\Delta_\lambda$, $\Delta_r$\\
\hline
Al & 30/30 & 5.2 & 0.000 & 0.010 & 0.04, 0.10\\
Ag & 30/30 & 5.7 & 0.000 & 0.063 & 0.00, 0.02\\
Au & 30/30 & 10.7 & 0.159 & 0.334 & 0.09, 0.04\\
Cu & 90/90 & 15.3 & 0.000 & 0.069 & 0.00, 0.09\\
Mo & 78/78 & 17.1 & 0.000 & 0.075 & 0.04, 0.05\\
W & 78/78 & 4.5 & 0.000 & 0.002 & 0.01, 0.06\\
NaCl & 53/53 & 3.9 & 0.000 & 0.106 & 0.00, 0.18\\
MgO & 53/53 & 2.9 & 0.000 & 0.033 & 0.01, 0.83\\
PbTe & 53/53 & 3.1 & 0.000 & 0.080 & 0.00, 0.03\\
\end{tabular}
\end{ruledtabular}
\end{table}

\subsection{Reference accuracy: cell size and temperature}\label{sec:accuracy}

Table~\ref{tab:accuracy} separates the error sources (overview in
Supplementary Fig.~S2). Against the converged
large-cell harmonic references, the harmonic FD FC2 in the 27--54-atom
($3\times3\times3$) cells deviates by 0.05--0.48~THz, and that in the SOTC
cells by 0.06--0.49~THz. Supercell truncation is therefore significant for Al
(0.49~THz in the 32-atom cell) and MgO (0.49~THz), and for PbTe, whose 54- and
72-atom FD models have 1.9\% unstable modes, whereas the 128-atom reference is
stable (lowest frequency 0.46~THz). The converged Al reference
gives 5.72 and 9.35~THz for the transverse and longitudinal acoustic (TA and
LA) modes at X and 4.25 and 9.66~THz at
L. The Born--von K\'arm\'an model fitted to the inelastic neutron
scattering spectrum of Al at 10~K (eight shells) \cite{kresch2008,kresch2009thesis}
gives 5.75 and 9.45~THz at X and 4.21 and 9.72~THz at L; the converged reference agrees with all four within 1.1\%
(see also the dispersion of Ref.~\onlinecite{ref34}). The 300~K model of the same
work, which includes thermal expansion, gives 5.39, 9.37, 4.16, and
9.43~THz. Its transverse branches are softer than both the converged
reference and the 300~K effective FC2 of the 32-atom cell (TA(X) 5.39 against
5.72 and 6.01--6.29~THz, the degeneracy being lifted by the cell). Most of the
stiffening of the effective FC2 is already present in the harmonic FD model of
the same cell (5.87--6.09~THz) and is therefore a truncation effect
(Supplementary Fig.~S4). The same comparison in real space is in Supplementary
Table~S10. The converged Al reference reproduces the 300~K neutron
nearest-neighbor constants within 4\%. For Cu at 1200~K, the effective and
SOTC-D first-shell constants are 16--18\% stiffer than the neutron value
interpolated to that temperature. This is the expected sign, because the
auxiliary FC2 omits the bubble softening of the measured spectrum
(Sec.~\ref{sec:intro}). For Al, by contrast,
the $3\times3\times3$ FD model places TA(L) at 4.91~THz, 15\% higher: in
small cells the truncation error is of the same size as the thermal effects
studied here. The effective FC2 at
$T_{\rm des}$ differs from the harmonic one by 0.16 (W, 1500~K) to 1.36~THz
(MgO, 1500~K) (last column of Table~\ref{tab:accuracy}); the effective
spectra are stiffer, with spectral Debye temperatures higher by 7~K (PbTe) to
54~K (MgO) for all materials except Al, Mo, and W, where the two agree within 4~K. Comparisons of high-temperature effective models with harmonic
references therefore conflate this physical
renormalization with fitting error. The effective references themselves are
converged to 0.02--0.20~THz between 20 and 40 snapshots.

Measured against the same-cell effective FC2, the SOTC-C models
deviate by 0.12--0.77~THz (0.12--0.44~THz excluding Au), and pooling all
SOTC-C iterations improves Au to 0.32~THz. For these nine crystals SOTC-D
with eight force calls deviates by 0.04--0.31~THz, a 1.3- to 9.8-fold
improvement over SOTC-C (1.3-fold for NaCl, 0.14 versus 0.18~THz) for
all materials except Cu. For Cu the SOTC-C model, fitted to the eight
configurations of the selected iteration of a 36-call workflow, is more
accurate (0.24 versus 0.31~THz).

\begin{figure*}[t]
\centering
\includegraphics[width=\textwidth]{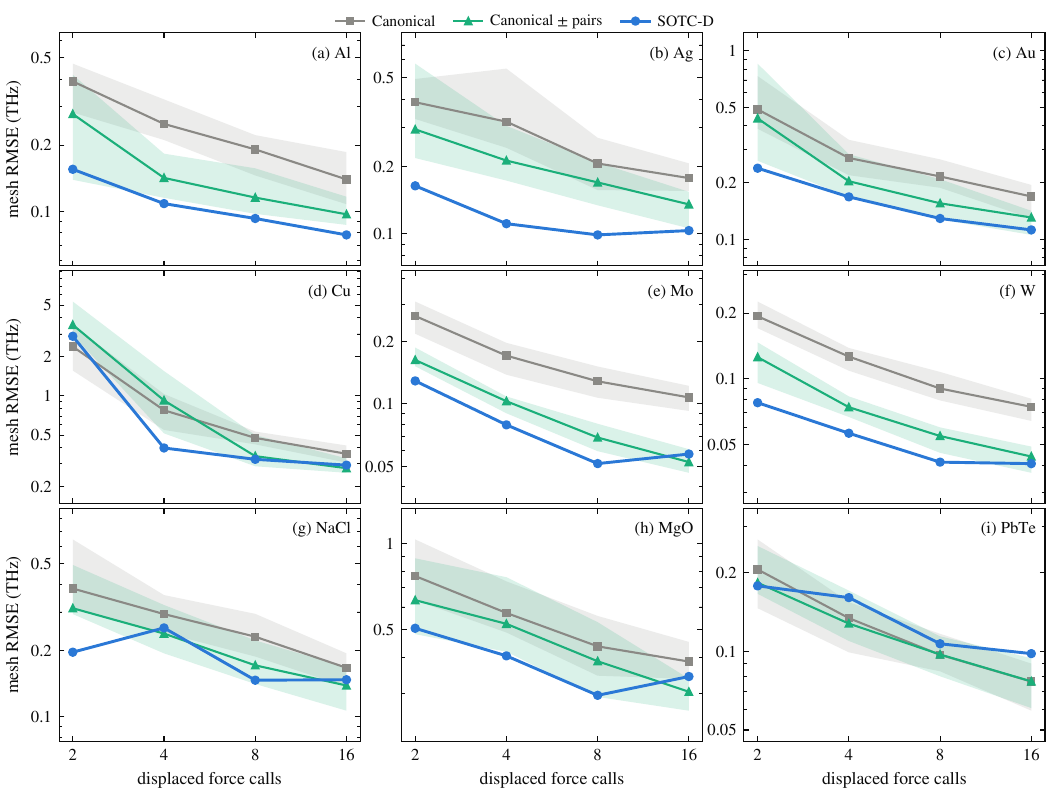}
\caption{Phonon-mesh RMSE against a disjoint same-cell effective FC2 versus
the number of displaced force calls, for SOTC-D (blue circles), random
canonical subsets (gray squares), and random canonical $\pm\mathbf u$ pairs
(green triangles), all designed from the same-cell harmonic FD trial
(iteration 0). Lines are medians and bands the 10--90\% range over 20 random
draws. All models use the space-group adapted fit in the SOTC cell with
identical electronic settings. Shown are the nine cubic crystals; Si and Mg
are included in Table~\ref{tab:matched}.}
\label{fig:counts}
\end{figure*}

\begin{table}[t]
\caption{Phonon-mesh RMSE (THz, $12^3$ mesh) of FC2 models for the eleven crystals (dashes: model not computed). Columns 2--4 against the converged large-cell harmonic FD reference, columns 5--7 against the effective FC2 fitted to 40 canonical snapshots at $T_{\rm des}$ in the SOTC cell. FD$_{3^3}$, FD$_{\rm cell}$: harmonic FD in the $3\times3\times3$ and in the SOTC cell; C: correlator-matched SOTC-C model (selected iteration, axial prior), C$_{\rm pool}$: SOTC-C pooled over iterations; D$_8$: SOTC-D with the space-group adapted fit at 8 force calls; $\Delta_T$: RMSE between the harmonic and effective models (finite-temperature renormalization). Because the reference here is the full 40-snapshot fit, SOTC-D values differ by up to 0.02~THz from those of Table~\ref{tab:matched}, whose reference is disjoint from the scored design.}
\label{tab:accuracy}
\begin{ruledtabular}
\begin{tabular}{lccccccc}
 & \multicolumn{3}{c}{vs harmonic FD} & \multicolumn{3}{c}{vs effective FC2} & \\
\cmidrule(lr){2-4}\cmidrule(lr){5-7}
 & FD$_{3^3}$ & FD$_{\rm cell}$ & C & C & C$_{\rm pool}$ & D$_8$ & $\Delta_T$\\
\hline
Al & 0.30 & 0.49 & 0.28 & 0.30 & 0.24 & 0.08 & 0.24\\
Ag & 0.05 & 0.06 & 0.71 & 0.29 & 0.25 & 0.09 & 0.46\\
Au & 0.09 & 0.12 & 0.91 & 0.77 & 0.32 & 0.12 & 0.58\\
Cu & 0.07 & 0.07 & 0.73 & 0.24 & 0.29 & 0.31 & 0.67\\
Mo & 0.38 & 0.27 & 0.44 & 0.44 & 0.11 & 0.04 & 0.28\\
W & 0.20 & 0.15 & 0.23 & 0.12 & 0.08 & 0.04 & 0.16\\
NaCl & 0.16 & 0.18 & 0.68 & 0.18 & 0.16 & 0.14 & 0.71\\
MgO & 0.48 & 0.49 & 1.38 & 0.37 & 0.43 & 0.28 & 1.36\\
PbTe & 0.22 & 0.24 & 0.47 & 0.25 & 0.30 & 0.10 & 0.32\\
Si & -- & 0.07 & -- & -- & -- & 0.04 & 0.26\\
Mg & -- & 0.05 & -- & -- & -- & 0.03 & 0.28\\
\end{tabular}
\end{ruledtabular}
\end{table}

\subsection{Matched comparison with canonical sampling and other designs}\label{sec:matched}

\begin{table*}[t]
\caption{Matched comparison in the SOTC cell at equal numbers of displaced force calls, all designed from the same-cell harmonic FD trial and fitted in the supercell space-group basis. D: SOTC-D; C: random canonical subsets; P: random canonical $\pm\mathbf u$ pairs (medians over 20 draws); Z: special displacements with Walsh sign patterns (ZG); R: thermal rattle (independent Gaussian atoms with the canonical per-atom variance). Upper block: held-out force RMSE on 8 canonical test snapshots relative to that of the 40-snapshot effective FC2. Lower block: mesh RMSE (THz) against an effective FC2 fitted to $40-n$ canonical snapshots disjoint from the scored design. The best entry is bold; the last row counts the materials in which D is best. Entries $>$10~THz ($>$100 in force) are singular fits of designs that do not identify the model.}
\label{tab:matched}
\begin{ruledtabular}
\begin{tabular}{lcccccccccc}
 & \multicolumn{5}{c}{2 force calls} & \multicolumn{5}{c}{8 force calls}\\
\cmidrule(lr){2-6}\cmidrule(lr){7-11}
 & D & C & P & Z & R & D & C & P & Z & R\\
\hline
\multicolumn{11}{l}{Held-out force RMSE / floor}\\
Al & \textbf{1.016} & 1.100 & 1.047 & 1.064 & 1.087 & \textbf{0.999} & 1.017 & 1.010 & 1.017 & 1.032\\
Ag & \textbf{1.008} & 1.103 & 1.071 & 1.123 & 1.444 & \textbf{0.999} & 1.028 & 1.016 & 1.011 & 1.042\\
Au & \textbf{1.007} & 1.099 & 1.094 & 1.124 & 1.302 & \textbf{1.009} & 1.026 & 1.013 & 1.024 & 1.083\\
Cu & \textbf{1.614} & 1.739 & 2.173 & $>$100 & 3.044 & \textbf{1.002} & 1.071 & 1.033 & 1.045 & 1.079\\
Mo & \textbf{1.022} & 1.152 & 1.069 & 1.250 & 1.120 & \textbf{0.999} & 1.028 & 1.008 & 1.012 & 1.021\\
W & \textbf{1.018} & 1.151 & 1.054 & 1.129 & 1.103 & \textbf{1.001} & 1.024 & 1.007 & 1.010 & 1.008\\
NaCl & \textbf{1.005} & 1.010 & 1.009 & 1.024 & 1.014 & 1.005 & 1.001 & 1.001 & 1.005 & \textbf{0.985}\\
MgO & \textbf{1.021} & 1.085 & 1.059 & 1.155 & 1.355 & 1.006 & 1.015 & 1.012 & \textbf{1.003} & 1.077\\
PbTe & 1.038 & 1.054 & 1.060 & 1.038 & \textbf{1.028} & 1.016 & 1.003 & \textbf{1.002} & 1.004 & 1.147\\
Si & \textbf{0.999} & 1.015 & 1.006 & 1.004 & 1.133 & \textbf{0.998} & 1.002 & 1.000 & 1.000 & 1.151\\
Mg & \textbf{1.002} & 1.016 & 1.005 & 1.003 & 1.033 & \textbf{1.000} & 1.003 & 1.001 & 1.000 & 1.021\\
D best & \multicolumn{5}{c}{10/11} & \multicolumn{5}{c}{8/11}\\
\hline
\multicolumn{11}{l}{Mesh RMSE vs disjoint effective FC2 (THz)}\\
Al & \textbf{0.16} & 0.39 & 0.28 & 0.30 & 0.34 & \textbf{0.09} & 0.19 & 0.12 & 0.13 & 0.22\\
Ag & \textbf{0.16} & 0.39 & 0.29 & 0.47 & 1.26 & \textbf{0.10} & 0.21 & 0.17 & 0.20 & 0.27\\
Au & \textbf{0.24} & 0.49 & 0.44 & 0.59 & 0.65 & \textbf{0.13} & 0.22 & 0.16 & 0.21 & 0.35\\
Cu & 2.88 & \textbf{2.40} & 3.52 & $>$10 & 4.43 & \textbf{0.33} & 0.48 & 0.34 & 0.54 & 0.64\\
Mo & \textbf{0.13} & 0.26 & 0.16 & 0.36 & 0.20 & \textbf{0.05} & 0.13 & 0.07 & 0.08 & 0.12\\
W & \textbf{0.08} & 0.19 & 0.13 & 0.23 & 0.15 & \textbf{0.04} & 0.09 & 0.05 & 0.05 & 0.06\\
NaCl & \textbf{0.20} & 0.38 & 0.31 & 0.80 & 0.97 & \textbf{0.15} & 0.23 & 0.17 & 0.19 & 0.89\\
MgO & \textbf{0.51} & 0.77 & 0.63 & 1.14 & 1.95 & \textbf{0.29} & 0.44 & 0.39 & 0.38 & 0.81\\
PbTe & \textbf{0.18} & 0.21 & 0.18 & 0.21 & 0.23 & 0.11 & \textbf{0.10} & 0.10 & 0.10 & 0.36\\
Si & 0.07 & 0.16 & 0.10 & \textbf{0.06} & 0.55 & 0.05 & 0.09 & 0.05 & \textbf{0.03} & 0.64\\
Mg & 0.07 & 0.14 & 0.08 & \textbf{0.05} & 0.23 & \textbf{0.04} & 0.07 & 0.04 & 0.04 & 0.20\\
D best & \multicolumn{5}{c}{8/11} & \multicolumn{5}{c}{9/11}\\
\end{tabular}
\end{ruledtabular}
\end{table*}

Figure~\ref{fig:counts} and Table~\ref{tab:matched} compare SOTC-D with
canonical sampling with least-squares fitting (as used in sTDEP), special
displacements, and thermal rattle at equal numbers of force calls, with
identical cells, electronic settings, trial model, and fit. The comparison
isolates the effect of how the configurations are chosen. The advantage of
SOTC-D is largest at the smallest budgets: with two force calls, a single
designed $\pm$ pair excites every commensurate mode at its thermal amplitude
and is free of cubic-force noise, whereas two canonical draws excite some
modes weakly and carry the full cubic noise. Accordingly, with two calls
SOTC-D gives the lowest dispersion and held-out force errors of all five
designs in dispersion for 8 of 11 crystals: the exceptions are Cu, whose
90-parameter model no two-call design determines, and Si and Mg, for which
special displacements are better (0.060 versus 0.066 and 0.046 versus
0.065~THz). In held-out force it is best for every crystal except PbTe,
whose trial is defective (see below). The ranking in dispersion holds in 8
of 11 crystals at four calls (Supplementary Table~S7): paired canonical
sampling is better for NaCl (0.239 versus 0.254~THz), and special
displacements are better for PbTe and Si (0.123 versus 0.161 and 0.032
versus 0.040~THz). At eight calls it holds in 9 of 11 crystals; the
exceptions are PbTe (unpaired canonical sampling, 0.097 versus 0.107~THz)
and Si (special displacements, 0.033 versus 0.047~THz). In held-out force
SOTC-D is best in 8 of 11 crystals at eight calls and within 2\% of the best
design in the others. At 16 calls SOTC-D is best in only 4 of 11 crystals: the sampling
error of every design approaches the noise of the disjoint reference, and
the designs can no longer be distinguished.

The special-displacement design, which also fixes the mode amplitudes but uses
orthogonal rather than information-optimal signs, is worse than SOTC-D in 9
of 11 crystals at eight calls, by up to a factor of 2.0 (Ag), and better for
Si and PbTe (factors 0.71 and 0.93); with two calls in Cu it does not
identify the model. Its Walsh patterns assign equal signs to modes that are
adjacent in the frequency ordering and are not chosen with respect to the
force-constant basis, so that in most crystals they leave the information
matrix less well conditioned than the D-optimal signs. For crystals with
more than one atom in the primitive cell, the real commensurate modes must
carry the Bloch phase of the full atomic position; with the phase of the
lattice vector alone, the configurations have the canonical mean-square
displacement of each atom but a nearest-neighbor bond-stretch variance up
to three times too large (Si), and the special-displacement errors are
then several times larger. Thermal rattle, a common training set of hiPhive workflows,
is worse than SOTC-D in dispersion in every crystal and the worst of all designs in 9 of
11 (0.06--0.89~THz at eight calls). Independent atomic displacements have
the canonical on-site variance but no interatomic correlations; the fitted
force constants are then the averaged Hessian of an uncorrelated ensemble,
which differs from the canonical target of Eq.~(\ref{eq:hessian}). The
held-out forces of NaCl, in contrast, are insensitive to the design: the
canonical, SOTC-D, and rattle designs lie within 2\% of each other at eight
calls.

In PbTe, one of the two crystals in which the iteration-0 SOTC-D is not the
best design at eight calls (0.11 versus 0.10~THz), the cause is the trial: the
same-cell harmonic FD model of PbTe has two unstable commensurate modes
($-1.72$ and $-0.49$~THz), which are omitted from SOTC-D and from canonical
draws alike, so that every iteration-0 design underexcites the soft optical
branch. This is a deficiency of the trial, not of the design, and is
removed by the trial update of the next subsection. The fixed amplitudes
also make the small-budget fits more robust: every SOTC-D model of Au with
four or more force calls is stable, and with two calls 0.2\% of the mesh
modes are unstable, whereas canonical sets of two calls produce unstable
modes in most draws.

\subsection{Trial update and equal-cost comparison}\label{sec:iter1}

\begin{figure*}[t]
\centering
\includegraphics[width=\textwidth]{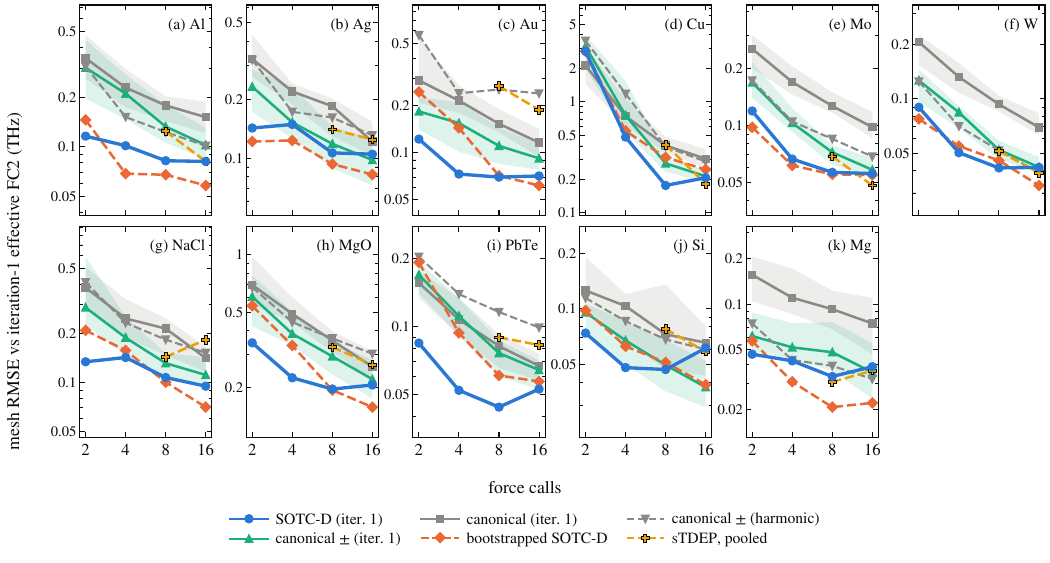}
\caption{Mesh RMSE against the iteration-1 effective FC2 versus force calls
for the eleven crystals. Solid lines: matched updated trial, SOTC-D (blue)
versus random canonical pairs (green) and subsets (gray), scored against the
disjoint effective FC2 (bands: 10--90\% over 20 draws). Dashed lines: equal
cost from the harmonic trial, bootstrapped SOTC-D (orange) versus one-shot
canonical pairs (gray) and two-iteration sTDEP fitted to both iterations
(yellow, 8 and 16 calls, median of three seeds), scored against the full
iteration-1 effective FC2, with which they share no configuration.}
\label{fig:selfconsistent}
\end{figure*}

\begin{table*}[t]
\caption{Updated trial (iteration 1, supercell basis). Mesh RMSE (THz) against the iteration-1 effective FC2. Matched trial: SOTC-D designed from the iteration-1 trial (D$_1$) versus the better of random canonical subsets and pairs drawn from the same trial (C$_1$), scored against the disjoint effective FC2. Equal cost: bootstrapped SOTC-D (B; first pair from the harmonic trial, later pairs from the model refitted to the first pair, no extra force calls) and one-shot SOTC-D (D$_0$) versus canonical pairs from the harmonic trial (C$_0$; a single sTDEP iteration), scored against the 40-snapshot iteration-1 effective FC2, which shares no configuration with them. B and D$_0$ coincide at 2 calls by construction. The best entry of each group is bold.}
\label{tab:selfconsistent}
\begin{ruledtabular}
\begin{tabular}{lcccccccccc}
 & \multicolumn{5}{c}{2 force calls} & \multicolumn{5}{c}{8 force calls}\\
\cmidrule(lr){2-6}\cmidrule(lr){7-11}
 & \multicolumn{2}{c}{matched trial} & \multicolumn{3}{c}{equal cost} & \multicolumn{2}{c}{matched trial} & \multicolumn{3}{c}{equal cost}\\
\cmidrule(lr){2-3}\cmidrule(lr){4-6}\cmidrule(lr){7-8}\cmidrule(lr){9-11}
 & D$_1$ & C$_1$ & B & D$_0$ & C$_0$ & D$_1$ & C$_1$ & B & D$_0$ & C$_0$\\
\hline
Al & \textbf{0.12} & 0.30 & \textbf{0.15} & 0.15 & 0.31 & \textbf{0.08} & 0.13 & \textbf{0.07} & 0.09 & 0.12\\
Ag & \textbf{0.14} & 0.23 & \textbf{0.12} & 0.12 & 0.32 & \textbf{0.11} & 0.12 & \textbf{0.09} & 0.13 & 0.16\\
Au & \textbf{0.12} & 0.18 & \textbf{0.24} & 0.24 & 0.56 & \textbf{0.07} & 0.11 & \textbf{0.07} & 0.20 & 0.25\\
Cu & 2.85 & \textbf{2.12} & \textbf{2.85} & 2.85 & 3.55 & \textbf{0.18} & 0.28 & 0.31 & \textbf{0.28} & 0.39\\
Mo & \textbf{0.12} & 0.17 & \textbf{0.10} & 0.10 & 0.17 & \textbf{0.06} & 0.07 & 0.06 & \textbf{0.04} & 0.08\\
W & \textbf{0.09} & 0.13 & \textbf{0.08} & 0.08 & 0.12 & \textbf{0.04} & 0.05 & 0.05 & \textbf{0.04} & 0.05\\
NaCl & \textbf{0.13} & 0.29 & \textbf{0.21} & 0.21 & 0.41 & \textbf{0.11} & 0.13 & \textbf{0.10} & 0.18 & 0.18\\
MgO & \textbf{0.34} & 0.60 & \textbf{0.54} & 0.54 & 0.68 & \textbf{0.20} & 0.29 & \textbf{0.19} & 0.38 & 0.36\\
PbTe & \textbf{0.08} & 0.16 & \textbf{0.19} & 0.19 & 0.20 & \textbf{0.04} & 0.08 & \textbf{0.06} & 0.13 & 0.12\\
Si & \textbf{0.07} & 0.10 & \textbf{0.10} & 0.10 & 0.11 & \textbf{0.05} & 0.05 & \textbf{0.05} & 0.09 & 0.07\\
Mg & \textbf{0.05} & 0.06 & \textbf{0.06} & 0.06 & 0.07 & \textbf{0.03} & 0.05 & \textbf{0.02} & 0.06 & 0.04\\
\hline
D$_1<$C$_1$ / B$<$C$_0$ & \multicolumn{2}{c}{10/11} & \multicolumn{3}{c}{11/11} & \multicolumn{2}{c}{11/11} & \multicolumn{3}{c}{11/11}\\
held-out force D$_1<$C$_1$ & \multicolumn{2}{c}{11/11} & \multicolumn{3}{c}{} & \multicolumn{2}{c}{10/11} & \multicolumn{3}{c}{}\\
\end{tabular}
\end{ruledtabular}
\end{table*}

Updating the trial from the harmonic FC2 to the iteration-1 effective FC2
changes the target by 0.03--0.28~THz and, more importantly, changes the
amplitudes at which the modes are excited. Table~\ref{tab:selfconsistent} and
Fig.~\ref{fig:selfconsistent} repeat the comparison with this updated trial.
With a trial close to the effective FC2, the advantage of the design becomes
uniform: SOTC-D has the lower dispersion error at four and eight calls in
every crystal, and at two calls in all except Cu, for the reason given
above; its held-out force error is the lower one in 11, 10, and 10 of the 11
crystals at two, four, and eight calls (Table~\ref{tab:selfconsistent} and
Supplementary Table~S9). PbTe changes from the only crystal without a gain to the one with the
largest gain (0.08 versus 0.16~THz with two calls, 0.04 versus 0.08~THz with
eight). The updated trial is dynamically stable, so the soft transverse
optical branch, which the harmonic trial could not excite, is now sampled at
its thermal amplitude.

Because the iteration-1 trial itself requires 40 force calls, a comparison
at equal total cost must start every protocol from the harmonic trial.
Bootstrapped SOTC-D (Sec.~\ref{sec:sotcdsc}) requires no force calls beyond
the budget: its first pair is the iteration-0 SOTC-D pair, and the model
refitted to this pair defines the modes and amplitudes of the remaining
pairs. Measured against
the iteration-1 effective FC2, it is more accurate than one-shot canonical
pairs from the harmonic trial (a single sTDEP iteration) in all 11 crystals at
every budget of 2, 4, 8, and 16 calls; at two calls neither design identifies
the 90-parameter Cu model (2.85 against 3.55~THz). At eight calls its error is 0.02--0.31~THz
against 0.04--0.39~THz, and the largest gains are for the crystals with the
largest trial update: Au (0.07 versus 0.25~THz), MgO (0.19 versus 0.36),
NaCl (0.10 versus 0.18), and PbTe (0.06 versus 0.12). It is also better than
one-shot SOTC-D for the same crystals, so that the update, and not only the
design, is responsible; for Cu, Mo, and W, whose trial changes least relative
to the cell error, one-shot SOTC-D is slightly more accurate at eight calls
(0.28, 0.044, and 0.041 versus 0.31, 0.055, and 0.045~THz). Starting from the harmonic trial alone, eight
bootstrapped SOTC-D calls are more accurate than sixteen one-shot canonical
calls for 9 of the 11 crystals (all except Cu and W).

\begin{table*}[t]
\caption{Self-consistent protocols with DFT forces, equal cost. Mesh RMSE (THz) against the 40-snapshot iteration-1 effective FC2 (supercell basis). B: bootstrapped SOTC-D. sTDEP: half of the calls as canonical $\pm$ pairs from the harmonic trial, half as canonical $\pm$ pairs from the model refitted to them, fitted to the second iteration only (S$_{\rm L}$) or to both (S$_{\rm P}$); median of three seeds. C$_0$: one-shot canonical pairs from the harmonic trial (median of 20 subsets). No configuration of any protocol enters the reference. The best entry of each group is bold.}
\label{tab:stdep}
\begin{ruledtabular}
\begin{tabular}{lcccccccc}
 & \multicolumn{4}{c}{8 calls} & \multicolumn{4}{c}{16 calls}\\
\cmidrule(lr){2-5}\cmidrule(lr){6-9}
 & B & S$_{\rm L}$ & S$_{\rm P}$ & C$_0$ & B & S$_{\rm L}$ & S$_{\rm P}$ & C$_0$\\
\hline
Al & \textbf{0.07} & 0.15 & 0.12 & 0.12 & \textbf{0.06} & 0.10 & 0.08 & 0.10\\
Ag & \textbf{0.09} & 0.12 & 0.14 & 0.16 & \textbf{0.08} & 0.10 & 0.12 & 0.13\\
Au & \textbf{0.07} & 0.13 & 0.27 & 0.25 & \textbf{0.06} & 0.09 & 0.19 & 0.24\\
Cu & \textbf{0.31} & 0.60 & 0.41 & 0.39 & 0.25 & 0.28 & \textbf{0.18} & 0.29\\
Mo & \textbf{0.06} & 0.12 & 0.07 & 0.08 & 0.05 & 0.07 & \textbf{0.05} & 0.07\\
W & \textbf{0.05} & 0.08 & 0.05 & 0.05 & \textbf{0.03} & 0.05 & 0.04 & 0.04\\
NaCl & \textbf{0.10} & 0.20 & 0.14 & 0.18 & \textbf{0.07} & 0.16 & 0.18 & 0.15\\
MgO & \textbf{0.19} & 0.32 & 0.33 & 0.36 & \textbf{0.16} & 0.30 & 0.26 & 0.30\\
PbTe & \textbf{0.06} & 0.13 & 0.09 & 0.12 & \textbf{0.06} & 0.09 & 0.08 & 0.10\\
Si & \textbf{0.05} & 0.08 & 0.08 & 0.07 & \textbf{0.04} & 0.05 & 0.06 & 0.06\\
Mg & \textbf{0.02} & 0.05 & 0.03 & 0.04 & \textbf{0.02} & 0.04 & 0.04 & 0.03\\
\hline
B best & \multicolumn{4}{c}{11/11} & \multicolumn{4}{c}{9/11}\\
\end{tabular}
\end{ruledtabular}
\end{table*}

The iterated canonical protocol, sTDEP \cite{ref60}, was run with DFT forces
in all eleven crystals at the same budgets (Table~\ref{tab:stdep}), in our
implementation with the regression of Sec.~\ref{sec:fit}. With
eight calls, bootstrapped SOTC-D has the lowest error in all 11 crystals,
with errors lower by up to a factor of 2.3 than sTDEP fitted to its last
iteration and up to 3.8 (Au) than sTDEP fitted to both iterations (pooled);
with 16 calls it is best in 9 of 11,
the exceptions being Cu (0.25 versus 0.18~THz for pooled sTDEP) and Mo (0.055
versus 0.048). Eight bootstrapped SOTC-D calls are more accurate than sixteen
sTDEP calls with either fit in 7 of 11 crystals (not Cu, Mo, W, Si). In the
crystal-symmetry basis, where all errors are smaller, the margins shrink:
bootstrapped SOTC-D is best in 8 of 11 crystals at eight calls and in 6 of
11 at sixteen, and pooled sTDEP is better for Cu at both budgets
(Supplementary Table~S3). The advantage of the design is thus largest at the
smallest budgets and for the crystals whose harmonic trial is furthest from
the effective FC2 (Au, NaCl, MgO, PbTe). Both iterated protocols aim at
the self-consistent FC2, of which the iteration-1 FC2 used for scoring is
only the first approximation; the model potentials of
Sec.~\ref{sec:sc_surrogate}, for which the self-consistent target is known
exactly, remove this limitation.

\begin{table*}[t]
\caption{Protocols with DFT forces in the SOTC cells of Al (300~K) and Si (600~K). Mesh RMSE (THz) against the converged self-consistent effective FC2 at 4/8/16/32 force calls; medians over seeds where several were run (sTDEP and SOTC-D-SC: 3; ZG-SC: 3 Walsh sign patterns; canonical: 20 random subsets of pairs). sTDEP draws 4 canonical configurations per iteration and fits the last iteration or all of them; ZG-SC adds one special-displacement $\pm$ pair per iteration and refits all of them. SSCHA (python-sscha) with its default stopping criterion stops after the first population; iterated runs use a meaningful factor of $10^{-3}$ and 10 ($N=16$) or 8 ($N=32$) populations of $N$ configurations. SSCHA force constants are projected onto the supercell basis in the metric of the thermal displacement covariance. ``Canonical, SSCHA population'': the same SSCHA configurations and DFT forces refitted by the regression of Sec.~\ref{sec:fit}.}
\label{tab:dft_sc}
\begin{ruledtabular}
\begin{tabular}{lcccccccc}
 & \multicolumn{4}{c}{Al} & \multicolumn{4}{c}{Si}\\
\cmidrule(lr){2-5}\cmidrule(lr){6-9}
Force calls & 4 & 8 & 16 & 32 & 4 & 8 & 16 & 32\\
\hline
\textit{One shot from the harmonic trial} &  &  &  &  &  &  &  & \\
SOTC-D & .088 & .076 & .059 & -- & .072 & .076 & .076 & --\\
Canonical $\pm$ (TDEP) & .161 & .117 & .083 & -- & .079 & .051 & .051 & --\\
Canonical, SSCHA population & -- & -- & .072 & .064 & -- & -- & .061 & .047\\
Special displacements (ZG) & .158 & .137 & .125 & -- & .053 & .059 & .052 & --\\
Thermal rattle & .269 & .207 & .155 & -- & .629 & .642 & .555 & --\\
\textit{Self-consistent} &  &  &  &  &  &  &  & \\
sTDEP, last iteration & .171 & .150 & .200 & .151 & .091 & .092 & .073 & .097\\
sTDEP, pooled & .171 & .117 & .068 & .057 & .091 & .064 & .040 & .035\\
ZG-SC (pooled) & .222 & .216 & .179 & .214 & .044 & .027 & .022 & .014\\
SSCHA, default stopping & -- & -- & .250 & .223 & -- & -- & .089 & .079\\
SSCHA, iterated ($N=16$) & -- & -- & .227 & .163 & -- & -- & .054 & .046\\
SSCHA, iterated ($N=32$) & -- & -- & -- & .128 & -- & -- & -- & .049\\
SOTC-D-SC & .087 & .076 & .052 & .047 & .079 & .065 & .058 & .051\\
\hline
Harmonic trial & \multicolumn{4}{c}{0.308} & \multicolumn{4}{c}{0.095}\\
Iteration-0 effective FC2 (256 calls) & \multicolumn{4}{c}{0.037} & \multicolumn{4}{c}{0.048}\\
Reference: seed-to-seed & \multicolumn{4}{c}{0.029} & \multicolumn{4}{c}{0.013}\\
\end{tabular}
\end{ruledtabular}
\end{table*}

\begin{figure*}[t]
\centering
\includegraphics[width=\textwidth]{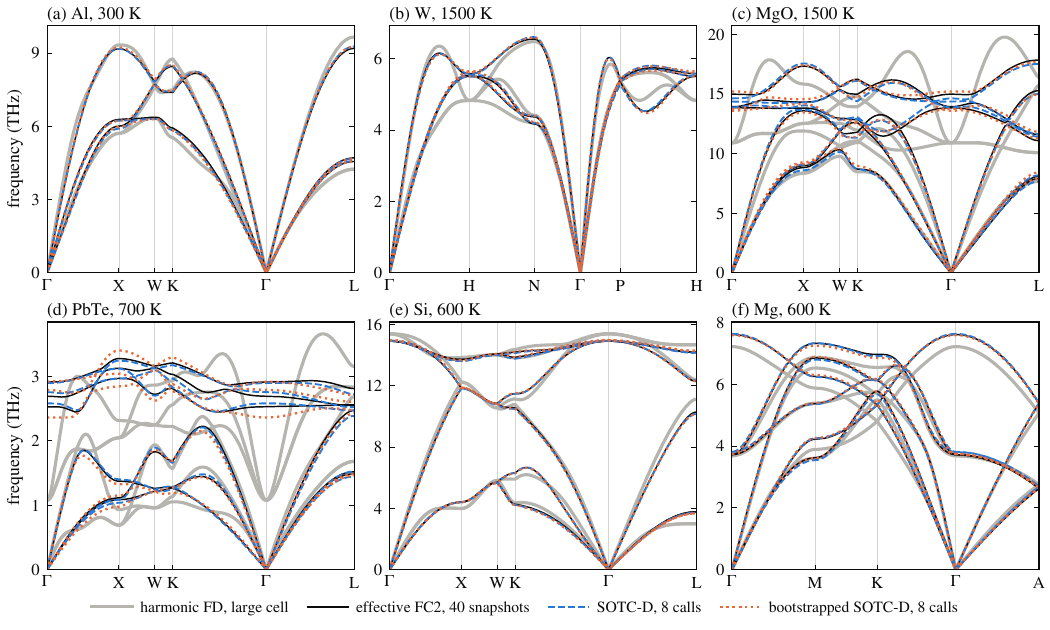}
\caption{Dispersions of the converged large-cell harmonic FD reference
(gray), the 40-snapshot iteration-1 effective FC2 (black), and SOTC-D with
eight force calls from the iteration-1 trial (blue dashed) and bootstrapped
from the harmonic trial (orange dotted), for six representative crystals; effective models are Fourier interpolated from the SOTC cell, and the
non-analytic correction is omitted for MgO and PbTe.}
\label{fig:dispersions}
\end{figure*}

Table~\ref{tab:dft_sc} compares the self-consistent protocols, and the
one-shot designs from the harmonic trial, with DFT forces in Al and Si
against the converged self-consistent effective FC2. The two crystals differ
in how far this target lies from the trial: 0.31~THz for Al at 300~K and
0.095~THz for Si at 600~K, with both models in the supercell basis of the fit.
These distances differ from the $\Delta_T$ column of
Table~\ref{tab:accuracy} (0.24 and 0.26~THz), whose effective FC2 is fitted to the
ensemble of the harmonic trial and is not self-consistent. In Al, SOTC-D-SC reaches 0.086--0.095~THz with
four force calls (three seeds), where iterated sTDEP reaches 0.16--0.19~THz
and ZG-SC 0.18--0.24~THz. With 32 calls SOTC-D-SC (0.043--0.050~THz) and sTDEP
pooled over its iterations (0.052--0.063~THz) both approach the 0.03~THz
precision of the reference; fitted to its last iteration alone, as in the
original protocol, sTDEP stays at 0.15~THz, because every refit then rests on
four configurations. In Si, where the anharmonic correction is small, the
ranking changes: with 32 calls ZG-SC (three sign patterns, 0.014--0.019~THz) and pooled sTDEP
(0.031--0.037~THz) are closer to the target than SOTC-D-SC
(0.046--0.059~THz). The advantage of the D-optimal design is thus largest
at small budgets and for strong renormalization; for weakly anharmonic Si
with DFT forces it is not present.

One-shot designs from the harmonic trial cannot reach the self-consistent
target; their limit is the iteration-0 effective FC2, 0.037 (Al) and
0.048~THz (Si) from it. Within this limit, SOTC-D is the best one-shot design
for Al at every budget (four calls: 0.088~THz, against 0.16 for canonical
pairs and for special displacements and 0.27 for rattle), and special
displacements are the best for Si at four calls (0.053~THz, against 0.072 for
SOTC-D and 0.079 for canonical pairs). Thermal rattle, which ignores the
correlation of neighboring displacements, is the worst design in both
crystals (Si: 0.55--0.64~THz at every budget).

With its default stopping criterion, python-sscha stops after its first
population in both crystals, because the statistical error of its
free-energy gradient exceeds the gradient itself; its result then stays
near the effective FC2 of the harmonic-trial ensemble, 0.25 and 0.22~THz
(Al) and 0.089 and 0.079~THz (Si) from the target with 16 and 32
configurations. With a meaningful factor of $10^{-3}$, so that every
population is minimized until the Kong--Liu ratio drops below 0.5, and 8--10
populations ($N=16$ and 32; six for $N=64$ in Al), SSCHA iterates. In Al it then reaches 0.10~THz after 64 calls
($N=16$) and 0.055--0.066 after 80--96 calls ($N=16$ and 32), and it
fluctuates between 0.05 and 0.08~THz up to 384 calls, the noise level of
its populations; SOTC-D-SC reaches 0.087~THz with four calls and
0.047~THz with 32. In Si the iterated SSCHA lies 0.054~THz from the target
after its first population of 16 calls and 0.043--0.066~THz thereafter,
which is as accurate as SOTC-D-SC at 16--32 calls (0.056--0.063 and
0.046--0.059~THz) and less accurate than pooled sTDEP and ZG-SC. For
$N=16$ in the 64-atom Si cell the free-energy gradient is about $10^{-3}$
of its own statistical error, so the minimization is a random walk within
each population. Refitting all configurations of an SSCHA run by the
regression of Sec.~\ref{sec:fit} gives 0.037--0.059~THz (Al) and
0.042--0.048~THz (Si).

AIMD-TDEP samples unpaired configurations, whose odd displacement moments
do not vanish, so a fit of the FC2 alone absorbs part of the cubic forces;
the $\pm$ pairs of the Gaussian designs cancel these forces exactly. Fitted
alone, the FC2 of all production steps lies 0.29 (Al) and 0.27~THz (Si) from
the converged self-consistent FC2 and 0.22 and 0.23~THz from the joint
FC2+FC3 fit to the same steps. In Al its mean mesh frequency, 5.84~THz,
equals that of the harmonic trial, against 6.11~THz for the
self-consistent FC2. The joint fit raises the mean to 6.05~THz (Si: 9.72,
against 9.77~THz) and lies 0.081 (Al) and 0.055~THz (Si) from the converged
self-consistent FC2. To separate the classical--quantum and non-Gaussian
contributions to this difference, 40 configurations (20
$\pm$ pairs) were drawn from the same harmonic trial with classical
amplitudes $\sigma_e^2=k_BT/\omega_e^2$ and computed with DFT. In Al their
effective FC2 lies 0.081~THz from the quantum one (iteration-0 pool; mean
frequency 6.05 against 6.11~THz; rms displacements 0.153 and 0.158~\AA) and
0.056~THz from the joint AIMD-TDEP fit (mean frequency 6.05~THz, rms
displacement 0.154~\AA). In Si at 600~K the
three ensembles agree more closely: the classical and quantum Gaussian FC2
differ by 0.033~THz, and the joint AIMD-TDEP fit lies 0.019 and 0.028~THz
from them, comparable to the statistical error of 40 configurations. The
non-Gaussian part of the distribution sampled by the MD therefore changes
the FC2 by at most 0.06~THz in both crystals; in Al the zero-point
amplitudes contribute a similar amount.

Measured against its own limit, the joint fit to a single trajectory has a
median error of 0.14~THz after 282 steps (including the 250 equilibration
steps), 0.10~THz after 314, 0.080~THz after 506, and 0.047~THz after 1274 in
Al, and 0.054~THz after 282 and 0.046~THz after 762 steps in Si. Against the
converged self-consistent FC2, it levels off at 0.09--0.11~THz in Al and
0.07--0.09~THz in Si, the difference between the two limits. SOTC-D-SC reaches
0.087~THz in Al with four force calls and 0.065~THz in Si with eight
(Table~\ref{tab:dft_sc}). The 250 equilibration steps account for most of
the cost of AIMD-TDEP, and successive MD steps are correlated.

Supplementary Fig.~S4 shows, branch by branch, how the SOTC-D dispersions of Al and Au
converge with 2--16 force calls; for Au, whose single-configuration SOTC-C fits
are unstable, SOTC-D from the updated trial is stable at every budget. Figure~\ref{fig:dispersions} shows the resulting dispersions for six
representative crystals. The difference between the effective FC2 and the
large-cell harmonic reference combines the thermal renormalization with the
truncation of the SOTC cell, which Table~\ref{tab:accuracy} separates. The
renormalization is largest for the optical branches of the polar crystals.
In PbTe at 700~K, the zone-center transverse optical mode, which is soft in
the harmonic model (1.1~THz in the 128-atom cell), is stiffened to about
2.5~THz, the anharmonic hardening of the soft mode of this incipient
ferroelectric. In MgO at 1500~K the optical branches differ by up to 3~THz,
of which 1.4~THz (rms) is renormalization in the same cell. Si softens
slightly at 600~K (the zone-center optical mode from 15.4 to 15.0~THz),
whereas Mg stiffens (the highest zone-center mode from 7.2 to 7.6~THz). The
metals Al and W change little. In all six crystals, SOTC-D with eight force
calls follows the 40-snapshot effective dispersion across the Brillouin zone,
for the iteration-1 design and for the bootstrapped design started from the
harmonic trial, including the soft optical branch of PbTe; the largest
deviations are in the optical branches of MgO and PbTe near X and W.

\subsection{Force-constant basis and regression frameworks}\label{sec:design}

Because the information matrix is fixed by the data and the basis, the
basis is the other lever on the variance. Table~\ref{tab:basis} compares the
supercell basis with the crystal-symmetry basis. The latter has 3 (fcc), 7
(bcc), or 8 (rocksalt) parameters instead of 30--90, yet the held-out force
errors of the two 40-snapshot effective models agree within 0.007~eV/\AA,
and their dispersions within 0.04--0.18~THz; for Si and Mg the SOTC cells
carry the full point symmetry and the two bases coincide. With fewer
parameters every design converges faster, so the absolute errors of
Table~S2 are lower, and the ranking of the designs is preserved: in the
crystal basis, SOTC-D with the updated trial is better than canonical
sampling in 8, 9, and 9 of 11 crystals at two, four, and eight calls, and
bootstrapped SOTC-D beats one-shot canonical pairs in 7, 9, 9, and 10 of 11
at 2--16 calls (Tables~S2 and S8). The margins are smaller than in the supercell
basis, because with three to eight parameters even canonical designs are well
conditioned.

The regression method matters much less than either. Fitting the same data
with the hiPhive cluster space \cite{ref59} and nine methods of the trainstation
library distributed with hiPhive
(least squares, ridge, Bayesian ridge, automatic relevance determination,
recursive feature elimination, orthogonal matching pursuit, LASSO, adaptive
LASSO, elastic net) changes the dispersion error at eight calls by at most
0.03~THz for any material and design (Table~S4), and least squares is never
significantly worse than a sparse or Bayesian method; with SOTC-D data the
error is lower than with canonical data for every method in 9 of the 11
crystals. Regularization and sparsity are therefore no substitute for the
design in this regime, in which the parameter count is small and the
designs are full rank.

\begin{table}[t]
\caption{Supercell versus crystal space-group FC2 basis: number of parameters $p$, alias-free crystal-basis cutoff $r_c$, held-out force RMSE (eV/\AA) of the 40-snapshot iteration-1 effective FC2, and mesh RMSE (THz) between the two effective FC2s.}
\label{tab:basis}
\begin{ruledtabular}
\begin{tabular}{lcccccc}
 & $p_{\rm sc}$ & $p_{\rm cr}$ & $r_c$ (\AA) & $F_{\rm sc}$ & $F_{\rm cr}$ & $\Delta\omega$\\
\hline
Al & 30 & 3 & 3.78 & 0.089 & 0.092 & 0.18\\
Ag & 30 & 3 & 3.84 & 0.255 & 0.257 & 0.07\\
Au & 30 & 3 & 3.85 & 0.340 & 0.347 & 0.11\\
Cu & 90 & 3 & 3.42 & 0.320 & 0.319 & 0.13\\
Mo & 78 & 7 & 4.66 & 0.274 & 0.276 & 0.07\\
W & 78 & 7 & 4.69 & 0.283 & 0.283 & 0.04\\
NaCl & 53 & 8 & 4.31 & 0.223 & 0.223 & 0.05\\
MgO & 53 & 8 & 3.21 & 0.504 & 0.504 & 0.14\\
PbTe & 53 & 8 & 5.01 & 0.259 & 0.259 & 0.04\\
Si & 10 & 10 & 5.01 & 0.191 & 0.191 & 0.00\\
Mg & 12 & 12 & 4.61 & 0.146 & 0.146 & 0.00\\
\end{tabular}
\end{ruledtabular}
\end{table}

\subsection{Other structure classes: diamond Si and hcp Mg}

To test transferability beyond the cubic metals and rocksalt compounds,
the complete protocol was applied without changes to covalent diamond Si
and to hexagonal Mg. The same-cell harmonic FD models lie within 0.07 (Si)
and 0.05~THz (Mg) of the 128- and 150-atom references, and thermal
renormalization at 600~K changes the dispersions by 0.26 and 0.28~THz
(Table~\ref{tab:accuracy}), softening Si and stiffening Mg. SOTC-D with eight
force calls reproduces the effective FC2 within 0.04 (Si) and 0.03~THz (Mg)
(Table~\ref{tab:accuracy}),
the same accuracy as for the bcc metals. At two calls it is better than canonical
sampling for both crystals (Si: 0.07 versus 0.10~THz for
paired canonical sampling; Mg: 0.07 versus 0.08) and equal to paired
canonical sampling at eight; special displacements are better than SOTC-D
for both at two calls and for Si at eight (0.03 versus 0.05~THz against the
disjoint reference of Table~\ref{tab:matched}), and
rattle is again the worst design (Table~\ref{tab:matched}),
and bootstrapped SOTC-D reaches 0.05 (Si) and 0.02~THz (Mg) at eight calls
(Table~\ref{tab:selfconsistent}). The behavior of the method thus does not
depend on the structure class.

\subsection{Hybrid functional: HSE06}\label{sec:hse}

\begin{figure}[t]
\centering
\includegraphics[width=\columnwidth]{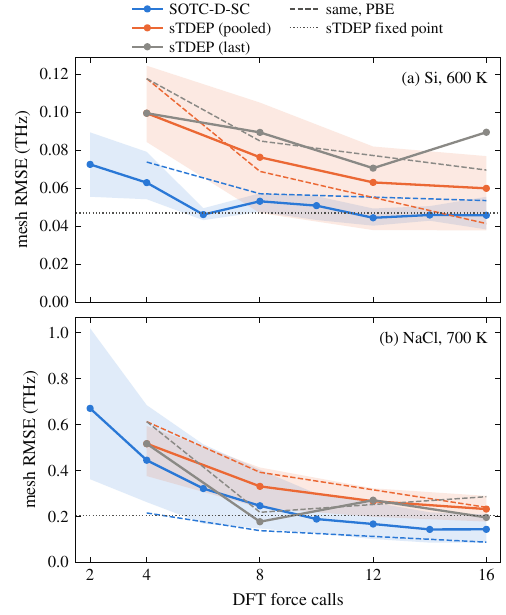}
\caption{HSE06: mesh RMSE against the composite HSE06 reference
(self-consistent PBE FC2 plus the harmonic HSE06$-$PBE difference) versus
force calls in the 64-atom cells of Si (600~K) and NaCl (700~K). Lines: seed
means of SOTC-D-SC (blue), pooled sTDEP (orange), and sTDEP fitted to the
last iteration (gray); bands: range over three seeds for SOTC-D-SC and pooled
sTDEP. Dashed: the same
protocols with PBE against the PBE reference. Dotted: distance of the
pooled-sTDEP HSE06 fixed point from the composite reference.}
\label{fig:hse}
\end{figure}

The case for few force calls is strongest when each call is expensive. With
HSE06 a call costs about 16 core-hours in the 64-atom cells of Si and NaCl
(Sec.~\ref{sec:comp}), so a self-consistent HSE06 reference of the kind used
above (128--1024 calls) is out of reach. The reference is therefore composed
as the converged self-consistent PBE FC2 plus the harmonic HSE06$-$PBE
difference from finite displacements in the same cell with the same
pseudopotentials. This assumes that the anharmonic renormalization is the
same in both functionals; as a check that does not use this assumption,
every run is also scored against the pooled-sTDEP HSE06 fixed point.
SOTC-D-SC and sTDEP (four calls per iteration) were run with HSE06 from the
HSE06 harmonic trial with 16 calls and three seeds each
(Fig.~\ref{fig:hse}).

In Si the hybrid functional stiffens the harmonic phonons by 0.40~THz (mesh
RMSE), more than the difference between the harmonic FD model and the
self-consistent PBE FC2 at 600~K (0.23~THz, without projection onto the fit
basis).
SOTC-D-SC lies 0.063, 0.053, and 0.046~THz from the composite reference after
4, 8, and 16 calls (seed means), against 0.099, 0.076, and 0.060~THz for
pooled sTDEP and 0.089~THz for sTDEP fitted to its last iteration at 16
calls; all three SOTC-D-SC seeds are below 0.1~THz after two calls. At 4
and 8 calls the ranking and the errors are those of the PBE runs in the same
cell (dashed lines in Fig.~\ref{fig:hse}); at 16 calls pooled sTDEP is ahead
with PBE (0.041 against 0.053~THz) but not with HSE06. The fixed points of both protocols lie
0.046~THz from the composite reference, which is thus consistent with them at
the level of the PBE comparison. Against the pooled-sTDEP fixed point, which
favors sTDEP, pooled sTDEP is closer from eight calls on (0.050 and
0.038~THz at 8 and 16 calls, against 0.067 and 0.058~THz for SOTC-D-SC).

NaCl at 700~K is the harder case. Its renormalization in the 64-atom cell is
0.60~THz in PBE, and 16 calls do not converge either protocol: SOTC-D-SC
reaches 0.44, 0.25, and 0.14~THz at 4, 8, and 16 calls, pooled sTDEP 0.52,
0.33, and 0.23~THz, and sTDEP fitted to the last iteration 0.18 and 0.20~THz
at 8 and 16 calls, with seed spreads of 0.05--0.18~THz. The PBE runs in the
same cell show the same order at 16 calls (0.088 against 0.24~THz). The
HSE06 correction of the NaCl harmonic phonons, 0.05~THz, is smaller than
these errors, so NaCl demonstrates that a hybrid-functional run is
affordable rather than testing hybrid-specific physics. In both crystals the
cost of SOTC-D-SC with HSE06 is that of the same number of single-point
calculations: about 260 core-hours for 16 calls.

\subsection{Larger cells}\label{sec:scaling}

Since each configuration of an $N$-atom cell gives $3N$ force components and
the number of FC2 parameters $P$ does not grow once the cutoff fits in the
cell ($P=5$--10 here, against 30 in the 32-atom Al cell), few calls should
suffice in large cells. We repeated the self-consistent comparison in
108- and 256-atom Al (300~K), 128- and 216-atom Si (600~K), and 128- and
216-atom NaCl (700~K), at 3--24 core-hours per call: SOTC-D-SC and sTDEP with
32 calls and three seeds each, and one-shot SOTC-D (16 calls) and canonical pairs
(32 calls, not run for 216-atom NaCl), all from the FD harmonic trial of the same
cell. The reference is the mean of the fits of iterations 2--4 of an sTDEP run
with four iterations of 64 calls (32 for 216-atom NaCl), whose last iteration
changes the dispersion by 0.003--0.033~THz. Since this
reference is an sTDEP estimate, it favors sTDEP at large budgets.

\begin{figure*}[t]
\centering
\includegraphics[width=\textwidth]{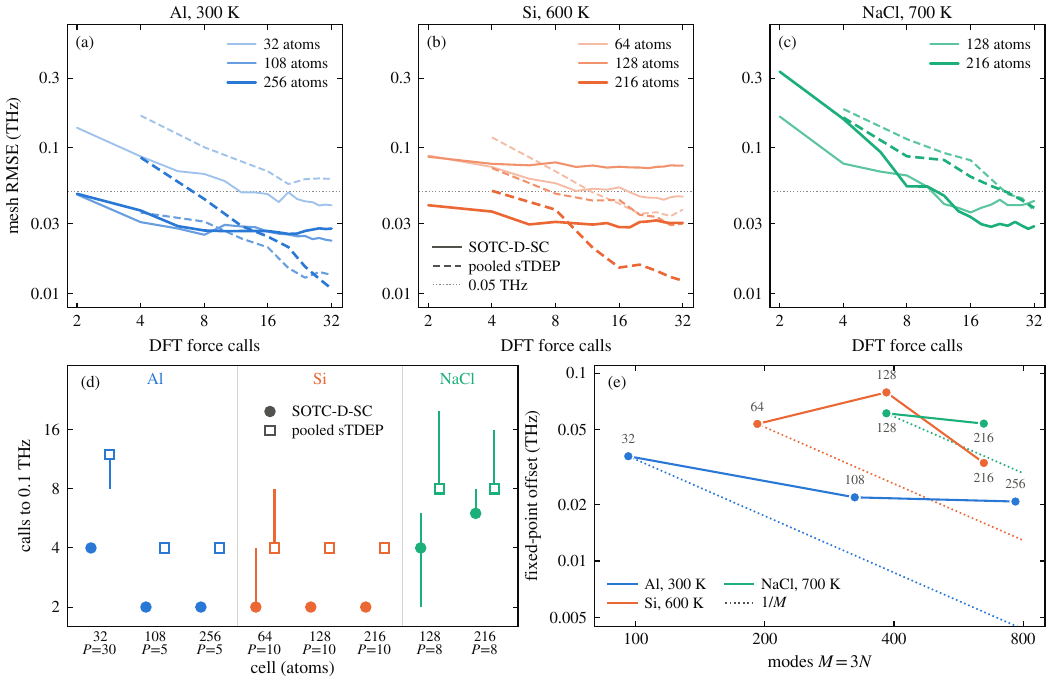}
\caption{Cell-size scaling. (a)--(c) Mesh RMSE against the self-consistent
reference of each cell versus force calls (seed means) for SOTC-D-SC (solid)
and pooled sTDEP (dashed); dotted: 0.05~THz. (d) Calls after which the error
stays below 0.1~THz for each cell (atom count and number of FC2 parameters $P$
below the axis): median (marker) and range (bar) over three seeds. sTDEP has
its first fit at four calls and SOTC-D-SC at two. (e) Mesh RMSE between the
SOTC-D-SC and pooled-sTDEP fixed points versus the number of modes, labeled with
the atom count; dotted: $1/M$ through the smallest cell of each material.}
\label{fig:scaling}
\end{figure*}

Figure~\ref{fig:scaling} shows two regimes. At very few calls SOTC-D-SC is
the more accurate and the cost of reaching a coarse target does not grow with
the cell: at or above 108 atoms, every SOTC-D-SC seed of Al and Si is
within 0.1~THz after two calls, the first fit available; sTDEP, with four
calls per iteration, is there after four, its first fit. For the 0.05~THz target
SOTC-D-SC needs 2--6 calls in 108- and 256-atom Al and two in 216-atom Si,
against 4--12 and 4--8 for pooled sTDEP.

NaCl, where the renormalization is largest, differs in two ways. The coarse
target does depend on the cell: the first fit is 0.08--0.24~THz from the
reference at 128 atoms and 0.26--0.40~THz at 216, and SOTC-D-SC reaches 0.1~THz
after 2--6 and 6--8 calls, against 8--20 and 8--16 for pooled sTDEP. And
SOTC-D-SC stays ahead up to 16 calls in both cells (0.036 against 0.082~THz and
0.034 against 0.063~THz). At 216 atoms it reaches 0.05~THz after 8--14 calls,
against 20--32, and it is still ahead at 32 calls (0.029 against 0.039~THz),
although the seeds overlap and this reference has half the calls; at 128 atoms
pooled sTDEP is slightly ahead at 32 calls (0.037 against 0.043~THz).

At larger budgets, SOTC-D-SC loses in the Al and Si cells of 108 atoms or more.
Its error saturates at 0.023--0.028~THz in Al and 0.031 and 0.075~THz in
216- and 128-atom Si, while pooled sTDEP keeps improving, to 0.011--0.013~THz in
Al and 0.012 and 0.030~THz in Si at 32 calls, and is ahead from 4--16 calls on
(4 in 128-atom Si, 16 in 256-atom Al). In 128-atom Si SOTC-D-SC never reaches
0.05~THz. The saturation level is the distance between the fixed points of the
two protocols, panel (e), and it does not follow the $1/M$ estimate of
Sec.~\ref{sec:biasvar}: from the smallest to the largest cell it changes
from 0.036 to 0.021~THz in Al, with no decrease from 108 to 256 atoms, and
from 0.054 through 0.079 to 0.033~THz in Si, and from 0.061 to 0.054~THz in NaCl,
where $1/M$ would give 0.036~THz. That estimate covers only the leading
quartic term, and the offset also contains the reference error; we do not
assign it a mechanism, but it means
that SOTC-D-SC is not the method of choice for convergence below about
0.05~THz in large Al and Si cells. In NaCl we see no such crossover up to 32 calls
at 216 atoms, but we cannot exclude one at larger budgets.

The one-shot design gains less from large cells than from small ones. Against
the median canonical pair run, SOTC-D at two calls is ahead in 108-atom Al
(0.029 against 0.047~THz) and in 128-atom NaCl (0.155 against 0.181~THz), but
within 0.01~THz in the other cells (behind in 256-atom Al, 0.064 against
0.056~THz), and at 16 calls behind in all Al and Si cells of 108 atoms or more
(0.033--0.083 against 0.020--0.055~THz). In 216-atom NaCl it stays at
0.12--0.17~THz from 2 to 16 calls, far from the 0.034~THz of SOTC-D-SC at 16 calls,
so the trial update matters more than the design there. A plausible reason is that a single
configuration of a large cell already averages over many modes, which leaves
less for the design to control.

\subsection{Dynamically stabilized bcc Zr}\label{sec:zr}

bcc Zr at 1188~K is the hardest test in this work: it is
stable only through anharmonicity, the harmonic trial has imaginary modes at N
and P (down to $-2.67$ and $-1.80$~THz in the 64-atom cell), and the soft
N-point mode is overdamped \cite{ref_heiming_zr}. SOTC-D-SC and sTDEP were run
with three seeds in the 54-atom (64 calls) and 64-atom (32 calls) cells
[PBE, GBRV ultrasoft pseudopotential, 30/240~Ry cutoffs, $a=3.606$~\AA,
$3\times3\times3$ $k$ mesh, 0.02~Ry Marzari--Vanderbilt smearing, 21--34~core-hours
per call in the 64-atom cell]. The reference in the 54-atom cell is the pooled
fit of a 4$\times$64-call sTDEP run, whose iterations still differ by
0.11--0.28~THz. The same protocol with 4$\times$32 calls failed in the 64-atom
cell. Because modes that are unstable in the trial are given zero amplitude, the
42 of 48 unstable supercell modes that the first fit left imaginary were
never sampled, and the N mode stayed between $-2.0$ and $-2.4$~THz in all four
iterations, without any close contact. In the 54-atom cell, which does not contain
N, the first fit removed all 28 unstable modes. We do not use this run as a
reference: the 64-atom runs are instead
compared with the mean of the two stable pooled-sTDEP seeds, which agree
within 0.08~THz, and with the inelastic neutron energies
(Table~\ref{tab:zr}).

The results depend on the seed, for both protocols. In the 54-atom cell one
SOTC-D-SC seed is stable from two calls and ends 0.074~THz from the
reference, whereas the other two remain on a separate, unstable branch with
the N mode near $-0.67$~THz through all 64 calls. In the 64-atom cell all three
seeds are stable from four calls, but only one of them cleanly, ending 0.046~THz from
the stable sTDEP result. The other two end 0.45 and 0.17~THz away,
and each contains a configuration with an interatomic distance of 1.52 or
1.82~\AA\ (nearest-neighbor distance 3.1~\AA). In these seeds the close pair
coincides with the stabilization: a refit of the same forces without it,
which is not self-consistent because the later trials were updated with
it, returns the N mode to $-1.38$ and $-0.63$~THz. They therefore cannot be
counted as successes.

sTDEP is exposed to close contacts more severely, since Gaussian draws are
not bounded. Fitted to all data, pooled
sTDEP is stable in only three of six seeds, with distances as short as
0.85~\AA\ and dispersion errors of up to 4.9~THz in the others, and sTDEP
fitted to the last iteration, which forgets such events, is stable at the end
of four of six runs. Excluding pairs with distances below 2~\AA\ after the
fact gives a stable N mode (1.28--1.73~THz) in five of six seeds (the sixth stays
at $-2.25$~THz), against two of six for SOTC-D-SC, again without
self-consistency. A distance screen thus appears to be a cheap safeguard for
sTDEP, and it does not repair SOTC-D-SC. SSCHA in the 54-atom cell is stable
from 64 calls with 32 configurations per population, 0.24~THz from the reference
after 192 calls, and is still unstable after 96 calls with 16.

Where a run converges, it agrees with experiment as well as the reference does
(Table~\ref{tab:zr}): the clean SOTC-D-SC seeds reproduce H and the upper N modes
within 0.13--0.35~THz and give the soft N mode at 1.27 and 1.38~THz, against
1.08~THz measured and 1.47--1.55~THz for sTDEP. A clean seed needs only 2--4
calls for this. The difficulty is that a single run does not reveal whether it
is on the correct branch: the unstable branch shows in the frequencies, but the
contact-stiffened seeds appear stable and are found only by checking the
shortest distance. We therefore
regard SOTC-D-SC as unreliable for dynamically stabilized crystals: any such
application requires several seeds and a distance check, and for this
class of crystal screened sTDEP with frequent trial updates appears the more
robust choice, although this screening was tested only after the fact and the
trapped 64-atom reference shows that sTDEP with long iterations can fail too.

\begin{table}[t]
\caption{bcc Zr at 1188~K: frequencies (THz) from inelastic neutron scattering
(INS, 895--960~$^\circ$C \cite{ref_heiming_zr}), the harmonic trial and the
reference of the 54-atom cell, the SOTC-D-SC seeds free of close contacts and stable
(D$_{54}$: 54 atoms, 64 calls; D$_{64}$: 64 atoms, 32 calls), and the mean of the
stable pooled-sTDEP seeds in the 64-atom cell (S$_{64}$, 32 calls). L$_{2/3}$:
longitudinal $\frac23(1,1,1)$, not commensurate with the 64-atom cell.}
\label{tab:zr}
\begin{ruledtabular}
\begin{tabular}{lcccccc}
 & INS & Harm. & Ref. & D$_{54}$ & D$_{64}$ & S$_{64}$\\
\hline
H & 4.79(7) & 4.91 & 4.86 & 4.92 & 4.97 & 4.88\\
N$_{T1}$ & 1.08(3) & $-2.41$ & 1.55 & 1.27 & 1.38 & 1.47\\
N$_{T2}$ & 3.39(12) & 3.17 & 2.99 & 3.04 & 3.05 & 3.11\\
N$_{L}$ & 4.72(5) & 3.58 & 4.90 & 4.87 & 4.85 & 4.81\\
P & & $-1.04$ & 3.44 & 3.35 & 3.35 & 3.39\\
L$_{2/3}$ & 1.81(12) & 1.25 & 2.27 & 2.17 & 2.22 & 2.29\\
\end{tabular}
\end{ruledtabular}
\end{table}

\subsection{Higher-order force constants}\label{sec:ho}

The constrained regression also extends to anharmonic tensors when enough
forces are available. Joint FC2+FC3+FC4 fits to 80 canonical snapshots in
$3\times3\times3$ cells (Al 300~K, W 1500~K, NaCl 700~K) lower the held-out
force RMSE relative to the harmonic FD FC2 by 93.7\%, 87.5\%, and 72.3\%, with
full-rank designs (106, 126, and 48 parameters) and held-out $R^2$ of 0.9997,
0.9994, and 0.969 (Fig.~S1). These tensors define the anharmonic model
potentials (surrogates) of the next subsection.

For the lattice thermal conductivity $\kappa_L$, the D-optimal criterion was
applied to the FC3 block of the information matrix (an FC3 design), and the
FC2 and FC3 of Si at 300~K were fitted jointly to designs of 4--40 force calls
in the 64-atom cell; $\kappa_L$ was computed with phono3py \cite{ref53} in the
relaxation-time approximation on a $17\times17\times17$ mesh. The result is
insensitive to the design and the budget: 113--120~W/(m\,K) for SOTC-D (4--16
calls), 118--121 for canonical pairs (4--40 calls), and 124--125 for the FC3
design (4--16 calls). The FC3 design therefore brings no measurable advantage
here. A finite-displacement FC3 in the same cell (51.5~W/(m\,K) with the
large-cell FC2) is dominated by numerical noise: it is nonzero for every
atom triplet of the cell, and restricted to the triplets of the fitted FC3 it
gives 37~W/(m\,K). It is therefore not a usable reference, and the accuracy
of the fitted values is not established.

\subsection{Anharmonic model potentials with known self-consistent solutions}\label{sec:sc_surrogate}

With DFT forces the self-consistent effective FC2 of Eq.~(\ref{eq:scfix}) is
not known exactly, and the comparisons above are made against finite-sample
estimates. Anharmonic model potentials fitted to DFT data remove this
limitation: their forces are noise-free and cheap, so the self-consistent
solution can be converged to any precision and each protocol can be repeated
many times. To separate the ingredients of SOTC-D, the FC2+FC3+FC4 models fitted to 80
QE snapshots of Al, W, and NaCl (Sec.~\ref{sec:ho}) were used as noise-free
anharmonic surrogates, with the effective FC2 converged on 6000 canonical
samples (0.001--0.006~THz between 3000 and 6000). Thirty repetitions of each
design were scored (Table~S1 in the Supplementary Material). Antithetic
pairing is the largest single ingredient: it lowers the median mesh error of
two-configuration canonical designs from 0.199 to 0.066~THz (Al), 0.091 to
0.035~THz (W), and 0.224 to 0.176~THz (NaCl), because the cubic force, which
is pure noise for an FC2 fit, cancels exactly. Exact rms amplitudes
improve the larger budgets further (16 calls: 0.025 versus 0.032~THz for Al and
0.063 versus 0.078~THz for NaCl). D-optimal signs add little in these small
surrogate cells, which have only 7--8 FC2 parameters. In the
30--90-parameter SOTC cells the signs matter more: special displacements,
which fix the amplitudes but choose the signs by orthogonal Walsh patterns
rather than by D-optimality, are worse than SOTC-D in 9 of 11 crystals at
eight calls (Sec.~\ref{sec:matched}). On the surrogates their accuracy
depends strongly on the crystal: they are the best design for Si
(0.022--0.036~THz at 2--16 calls) but become worse with more sign patterns
for NaCl (0.133~THz at four calls, 0.174 at sixteen; Table~S1). In these one-shot designs with up to 16 calls,
stratified Gaussian-quantile amplitudes, which match the fourth moment of
the canonical distribution, give no further improvement, and no amplitude
bias of the fixed-amplitude design is detectable, as expected from
Eq.~(\ref{eq:bias}).

\textit{Self-consistent protocols and SSCHA.} The surrogates also give the
exact self-consistent target, obtained by iterating the canonical fit with
6000 samples per iteration to a mesh change below 0.01~THz, so that complete
self-consistent workflows can be compared at equal numbers of force calls
(Table~\ref{tab:sc_surrogate}): one-shot canonical pairs and SOTC-D from the
harmonic trial; sTDEP with three iterations of equal size, fitted to the last
iteration or to all of them; SOTC-D-SC (Sec.~\ref{sec:sotcdsc}); and SSCHA,
run with python-sscha \cite{ref50} with the surrogate as an exact calculator,
starting from the same harmonic trial (minimization step 0.1, threshold 0.5 on the
Kong--Liu effective sample size, stopping when the gradient is below 0.2
times its statistical error, fixed centroids, at most 30
populations of $N=8$--128 configurations, 5 repetitions against 12 for the
other protocols). Because the smallest SSCHA population has eight
configurations, a factor of 2 between SSCHA and SOTC-D-SC at a threshold
reached with four calls is limited by this grid. For Al and W, whose harmonic trials are
within 0.15--0.22~THz of the self-consistent target, all protocols are
comparable; SOTC-D-SC is best at 16 and 32 calls and within 0.003~THz of
the best at 8 calls. NaCl is the stringent case: the harmonic trial is
0.53~THz from the self-consistent target, so one-shot designs stall at
0.27--0.33~THz irrespective of the budget. SOTC-D-SC reaches 0.096, 0.054,
0.035, and 0.029~THz with 8, 16, 32, and 64 calls, whereas sTDEP fluctuates
between 0.036 and 0.18~THz because each iteration redraws a small sample, and
SSCHA needs 96 calls for 0.041~THz and 192 for 0.027~THz. At equal accuracy,
SOTC-D-SC therefore needs about three times fewer force calls than SSCHA for
NaCl, and it is the most accurate protocol at 32 and 64 calls; only pooled
sTDEP is marginally better at 16 calls (0.048 versus 0.054~THz).

The same study was repeated for diamond Si and hcp Mg at 600~K, with
FC2+FC3+FC4 surrogates fitted to all 194 same-cell QE configurations of each
crystal (cutoffs 5.0/4.0/2.5~\AA\ for Si and 4.6/3.5/3.3~\AA\ for Mg; held-out
force RMSE 0.075 and 0.040~eV/\AA, $R^2=0.993$ and 0.992, against 0.95 and 0.88
for FC2 alone). Their harmonic trials are 0.12 and 0.27~THz from the
self-consistent target. SOTC-D-SC reaches 0.037, 0.024, 0.014, and
0.013~THz for Si and 0.026, 0.021, 0.016, and 0.014~THz for Mg with 8, 16,
32, and 64 calls; pooled sTDEP is within 0.002~THz for Si and slightly better for Mg at
32 and 64 calls (0.015 and 0.011~THz), as is one-shot canonical sampling for
Mg at 64 calls (0.010~THz), because the Mg trial changes little between
iterations. SSCHA needs 64 (Si) and 16 calls (Mg) for 0.05~THz and 128 and
64 calls for 0.03~THz, against 4 and 16 (Si) and 4 and 8 calls (Mg) for
SOTC-D-SC (Table~\ref{tab:calls}).

\textit{Amplitude bias.} The flattening of the SOTC-D-SC error between 32
and 64 calls for Si and Mg is consistent with the fixed-amplitude bias of
Eq.~(\ref{eq:bias}). It can be removed exactly at fourth order: if, for
every mode, the squared amplitudes $q_e^2/\sigma_e^2$ are drawn without
replacement within blocks of four pairs from the equal-weight values
$\{0,0,2-\sqrt2,2+\sqrt2\}$, independently for each mode, they have the
Gaussian moments $\langle q_e^2\rangle=\sigma_e^2$ and $\langle
q_e^4\rangle=3\sigma_e^4$, so that the $V_{aeee}$ term of Eq.~(\ref{eq:bias})
vanishes while the signs remain D-optimal. On the five surrogates with 128
calls (12 repetitions), this moment-matched variant lowers the error from
0.0098 to 0.0080 (Al), 0.0080 to 0.0047 (W), and 0.0122 to 0.0092~THz
(Mg) and raises it from 0.0074 to 0.0092 (Si) and 0.0170 to 0.0181~THz
(NaCl); with 16--32 calls the two variants differ by 0.001--0.015~THz in
either direction. The fixed-amplitude bias is therefore of order
0.01~THz, an order of magnitude below the supercell and reference errors of
Sec.~\ref{sec:accuracy}, and fixed amplitudes, which minimize the variance
at small budgets, remain the default; the moment-matched variant is the
appropriate choice for budgets above about 100 calls.

\begin{table*}[t]
\caption{Self-consistent protocols on the anharmonic surrogates. Median mesh RMSE (THz) against the self-consistent effective FC2 (= SCHA auxiliary FC2) at 8, 16, 32, and 64 force calls (12 repetitions). SSCHA: python-sscha with the surrogate as calculator, populations of $N=8$--128 configurations until convergence (5 repetitions), listed at the median total number of force calls (the better population size where two need the same number). Other SSCHA budgets: Al 128 calls, 0.008; W 128 calls, 0.009; NaCl 96 calls, 0.041; 192 calls, 0.027; 384 calls, 0.025; Si 128 calls, 0.024; 256 calls, 0.020; Mg 128 calls, 0.015 THz.}
\label{tab:sc_surrogate}
\begin{ruledtabular}
\footnotesize
\setlength{\tabcolsep}{2pt}
\begin{tabular}{lcccc@{\hspace{9pt}}cccc@{\hspace{9pt}}cccc@{\hspace{9pt}}cccc@{\hspace{9pt}}cccc}
 & \multicolumn{4}{c}{Al} & \multicolumn{4}{c}{W} & \multicolumn{4}{c}{NaCl} & \multicolumn{4}{c}{Si} & \multicolumn{4}{c}{Mg}\\
\cmidrule(lr){2-5}\cmidrule(lr){6-9}\cmidrule(lr){10-13}\cmidrule(lr){14-17}\cmidrule(lr){18-21}
Force calls & 8 & 16 & 32 & 64 & 8 & 16 & 32 & 64 & 8 & 16 & 32 & 64 & 8 & 16 & 32 & 64 & 8 & 16 & 32 & 64\\
\hline
Canonical $\pm$, one shot & .042 & .030 & .026 & .019 & .029 & .020 & .020 & .017 & .271 & .278 & .299 & .331 & .036 & .023 & .025 & .021 & .026 & .019 & .018 & .010\\
SOTC-D, one shot & .031 & .028 & .017 & .013 & .023 & .020 & .018 & .013 & .273 & .308 & .312 & .309 & .032 & .025 & .018 & .023 & .027 & .023 & .021 & .018\\
sTDEP (last iteration) & .051 & .038 & .039 & .026 & .030 & .026 & .021 & .013 & .150 & .055 & .091 & .036 & .049 & .035 & .029 & .019 & .040 & .028 & .024 & .017\\
sTDEP (pooled) & .044 & .040 & .019 & .013 & .024 & .017 & .014 & .008 & .181 & .048 & .060 & .081 & .038 & .024 & .016 & .014 & .033 & .022 & .015 & .011\\
SOTC-D-SC & .034 & .022 & .016 & .012 & .020 & .014 & .011 & .008 & .096 & .054 & .035 & .029 & .037 & .024 & .014 & .013 & .026 & .021 & .016 & .014\\
SSCHA & .034 & .030 & .021 & .010 & .023 & .017 & .012 & .016 & -- & .092 & .069 & -- & .098 & .070 & -- & .042 & .055 & .041 & -- & .019\\
\hline
Harmonic vs self-consistent & \multicolumn{4}{c}{0.215} & \multicolumn{4}{c}{0.151} & \multicolumn{4}{c}{0.528} & \multicolumn{4}{c}{0.121} & \multicolumn{4}{c}{0.268}\\
\end{tabular}
\end{ruledtabular}
\end{table*}

\subsection{Cost}

Table~\ref{tab:cost} converts the comparison into computing time. A force
call in the SOTC cell costs 0.3--6.9 core-hours; the harmonic FD trial costs
three to nine calls. To reach the iteration-1 dispersion error that SOTC-D
attains with two calls, paired canonical sampling needs 4--16 calls, and to
reach its eight-call error, more than 16 calls for 9 of the 11 crystals.

For complete self-consistent workflows, Table~\ref{tab:calls} gives the
number of force calls that each complete protocol, started from the same
harmonic trial, needs to reach and keep a target accuracy on the model
potentials of Sec.~\ref{sec:sc_surrogate}. SOTC-D-SC needs the fewest calls, or is tied
for the fewest, in 13 of the 14 attainable material--threshold combinations
of the five surrogates (the exception is Mg at 0.02~THz, 32 versus 16 calls
for one-shot canonical pairs), and strictly the fewest for W at 0.02~THz and
for NaCl at 0.05 and 0.03~THz, where it needs 32 and 64 calls against 96 and
192 for SSCHA, 64 or more for sTDEP, and more than 64 for every one-shot protocol. It
needs a factor of 2--16 fewer calls than SSCHA in 13 of the 14 combinations and as
many in the remaining one (Al, 0.03~THz). Since all protocols evaluate forces
in the same cell, the cost in core-hours is proportional to these numbers,
with the per-call costs of Table~\ref{tab:cost}.

\begin{table}[t]
\caption{Cost. $t$: measured cost per force call in the SOTC cell (core-hours, 32 cores, mean over all same-cell calls); $N_{\rm FD}$: calls of the harmonic FD trial. Columns 4--5: number of random canonical pairs (force calls) from the same trial needed to reach the iteration-1 mesh error that SOTC-D attains with 2 and 8 calls.}
\label{tab:cost}
\begin{ruledtabular}
\begin{tabular}{lcccc}
 & $t$ (core-h) & $N_{\rm FD}$ & match D$_1$(2) & match D$_1$(8)\\
\hline
Al & 0.28 & 5 & 16 & $>$16\\
Ag & 1.06 & 5 & 8 & 16\\
Au & 1.09 & 5 & 8 & $>$16\\
Cu & 0.58 & 7 & 4 & $>$16\\
Mo & 6.75 & 7 & 4 & $>$16\\
W & 6.92 & 7 & 4 & $>$16\\
NaCl & 1.27 & 9 & 8 & $>$16\\
MgO & 0.47 & 9 & 8 & $>$16\\
PbTe & 3.84 & 9 & 8 & $>$16\\
Si & 0.40 & 3 & 4 & 16\\
Mg & 1.13 & 3 & 16 & $>$16\\
\end{tabular}
\end{ruledtabular}
\end{table}

\begin{table*}[t]
\caption{End-to-end cost on the anharmonic surrogates: force calls needed by each complete protocol, started from the same harmonic trial, to reach and keep a median mesh error of 0.05, 0.03, 0.02~THz against the self-consistent effective FC2 (tested budgets 4--64 calls; SSCHA 8--384).}
\label{tab:calls}
\begin{ruledtabular}
\begin{tabular}{lccccc}
Protocol & Al & W & NaCl & Si & Mg\\
 & \multicolumn{5}{c}{calls to reach 0.05/0.03/0.02~THz in each column}\\
\hline
Canonical $\pm$, one shot & 4/32/64 & 4/8/16 & $>$64/$>$64/$>$64 & 4/16/$>$64 & 4/8/16\\
SOTC-D, one shot & 8/16/32 & 4/4/16 & $>$64/$>$64/$>$64 & 4/16/$>$64 & 4/8/64\\
sTDEP (last) & 16/64/$>$64 & 8/8/64 & 64/$>$64/$>$64 & 8/32/64 & 8/16/64\\
sTDEP (pooled) & 4/32/32 & 4/4/16 & $>$64/$>$64/$>$64 & 4/16/32 & 4/16/32\\
SOTC-D-SC & 4/16/32 & 4/4/8 & 32/64/$>$64 & 4/16/32 & 4/8/32\\
SSCHA & 8/16/64 & 8/8/16 & 96/192/$>$384 & 64/128/$>$256 & 16/64/64\\
\end{tabular}
\end{ruledtabular}
\end{table*}

\subsection{Thermodynamics and displacement parameters}

Because $C_V$ is close to the Dulong--Petit limit at $T_{\rm des}$ for all
nine materials, entropy, free energy, and Debye temperature are the more
discriminating thermodynamic quantities (Table~\ref{tab:thermo}). The
SOTC-C models reproduce the effective $S_{\rm vib}$ to within
0.2--1.4~J\,mol$^{-1}$K$^{-1}$ for the metals, NaCl, and MgO (2.3 for PbTe)
and the effective $T_D$ to within 19~K, whereas the harmonic references of
Ag, Au, Cu, NaCl, MgO, and PbTe are higher in $S_{\rm vib}$ by up to
5.3~J\,mol$^{-1}$K$^{-1}$ (Au) at the same volume; for Al, Mo, and W the
harmonic and effective values agree within 0.4~J\,mol$^{-1}$K$^{-1}$.
DOS-integrated quantities are thus considerably less sensitive to the design
than branch-resolved frequencies. Using the PBE equilibrium lattice
parameters (0.01--1.1\% from the production values) changes the harmonic
$S_{\rm vib}$ at $T_{\rm des}$ by up to 1.7~J\,mol$^{-1}$K$^{-1}$ (Cu, PbTe) and the
mesh frequencies by 0.00--0.35~THz (Supplementary Table~S6).

For the Debye--Waller comparison the force constants must correspond to the
measurement temperature and volume. The converged harmonic models at the
room-temperature lattice parameters give $U_{\rm iso}(300\,{\rm K})$ of
0.0104 (Al), 0.0066 (Cu), and 0.0081~\AA$^2$ (Au), within 6\% of the values
0.0099, 0.0070, and 0.0078~\AA$^2$ tabulated by Peng \textit{et al.}\ at
293~K \cite{ref55}. For Al, whose production lattice parameter is the
room-temperature one, the SOTC-cell models can be compared directly
(Fourier-interpolated on a $20^3$ mesh). The 300~K effective FC2 gives
0.0104~\AA$^2$, and the SOTC-D models with eight force calls give 0.0101
(harmonic trial), 0.0104 (bootstrapped), and 0.0106~\AA$^2$ (updated trial),
2--7\% above the tabulated value. The harmonic FD model of the same 32-atom cell
gives 0.0150~\AA$^2$, because its interpolated spectrum has small regions of
imaginary and near-zero frequencies (minimum $-0.67$~THz). For Cu and Au,
the same 32-atom cell was scaled to the room-temperature lattice parameters
(3.6149 and 4.0782~\AA), and SOTC-D-SC was run at 300~K with eight force
calls after the harmonic FD trial. It gives 0.0062 (Cu) and
0.0074~\AA$^2$ (Au), 11\% and 5\% below the tabulated values, against 0.0064
and 0.0088~\AA$^2$ for the harmonic FD model of that cell. The result
changes by less than 3\% between two and eight calls. For Au the
self-consistent model removes most of the overestimate of the small-cell
harmonic model; for Cu the remaining deficit is that of the cell, since the
harmonic model of the same cell is also 9\% low. At the production lattice parameters, Ag, Mo, and W give
0.0096, 0.0029, and 0.0022~\AA$^2$, 3--10\% above the tabulated 0.0093,
0.0027, and 0.0020~\AA$^2$, as expected because the production lattice
parameters of these crystals are those near their design temperatures
(1000--1500~K) and thus larger than at room temperature. Debye--Waller factors at the design temperatures
require force constants and lattice parameters at those temperatures; they
are not compared with room-temperature data.

\begin{table*}[t]
\caption{Vibrational thermodynamics per atom at $T_{\rm des}$ and displacement parameters. $S_{\rm vib}$ (J\,mol$^{-1}$K$^{-1}$), $F_{\rm vib}$ (kJ\,mol$^{-1}$), and $T_D$ (K) from the converged harmonic FD (FD), the same-cell effective FC2 (eff.), and the SOTC-C model. $U_{\rm iso}(300\,{\rm K})$ (\AA$^2$) from the converged harmonic model at the room-temperature (Al, Cu, Au) or production lattice, compared with Peng \textit{et al.}\ \cite{ref55} at 293~K.}
\label{tab:thermo}
\begin{ruledtabular}
\begin{tabular}{lccccccccccc}
 & \multicolumn{3}{c}{$S_{\rm vib}$} & \multicolumn{3}{c}{$F_{\rm vib}$} & \multicolumn{3}{c}{$T_D$} & \multicolumn{2}{c}{$U_{\rm iso}(300\,{\rm K})$}\\
\cmidrule(lr){2-4}\cmidrule(lr){5-7}\cmidrule(lr){8-10}\cmidrule(lr){11-12}
 & FD & eff. & SOTC-C & FD & eff. & SOTC-C & FD & eff. & SOTC-C & calc. & Ref.~\cite{ref55}\\
\hline
Al & 27.8 & 27.8 & 28.2 & -0.2 & -0.2 & -0.3 & 397 & 394 & 396 & 0.0104 & 0.0099\\
Ag & 72.9 & 69.8 & 68.9 & -47.9 & -44.8 & -43.9 & 210 & 238 & 252 & 0.0096 & 0.0093\\
Au & 84.1 & 78.8 & 78.0 & -70.9 & -64.6 & -63.8 & 168 & 204 & 204 & 0.0081 & 0.0078\\
Cu & 68.7 & 65.9 & 65.7 & -52.5 & -49.0 & -48.8 & 292 & 332 & 335 & 0.0066 & 0.0070\\
Mo & 68.4 & 68.8 & 67.4 & -65.1 & -65.7 & -63.6 & 358 & 354 & 373 & 0.0029 & 0.0027\\
W & 72.8 & 72.7 & 73.1 & -71.7 & -71.6 & -72.2 & 300 & 303 & 300 & 0.0022 & 0.0020\\
NaCl & 58.8 & 55.9 & 56.7 & -23.6 & -21.5 & -22.1 & 254 & 290 & 286 & 0.0224 & --\\
MgO & 51.8 & 49.8 & 50.1 & -39.8 & -36.7 & -37.3 & 726 & 780 & 774 & 0.0044 & --\\
PbTe & 76.4 & 75.2 & 72.9 & -36.0 & -35.2 & -33.5 & 133 & 140 & 155 & 0.0159 & --\\
\end{tabular}
\end{ruledtabular}
\end{table*}

\subsection{Thermal expansion}\label{sec:qha}

Because the SCHA free energy is stationary with respect to the auxiliary
force constants, its temperature derivative at fixed volume is minus the
harmonic entropy of the self-consistent FC2, $S=S_{\rm h}[\boldsymbol\Phi_{\rm
SC}](T)$. The free energy per atom on the volume grid is
\begin{align}
F(V,T)={}&E_{\rm st}(V)+F_{\rm h}[\boldsymbol\Phi_{\rm SC}(V,T)](T)\nonumber\\
&+\big\langle E-E_0-\tfrac12\mathbf u\cdot\boldsymbol\Phi_{\rm SC}\mathbf u\big\rangle ,
\label{eq:fqha}
\end{align}
in which all quantities are per atom. $E_{\rm st}$ is the static energy
(third-order Birch--Murnaghan fit of the primitive-cell PBE energies on a
$16^3$ $k$ mesh), $F_{\rm h}$ is the quantum harmonic free energy of the
self-consistent FC2 (Fourier-interpolated from the 32-atom cell on a $20^3$
mesh), and the last term, with $E$ and $E_0$ the DFT energies of a designed
configuration $\mathbf u$ and of the undisplaced cell, is the anharmonic
energy correction of the SCHA \cite{ref7,ref50}. It is averaged over the
second half of the designed configurations at each volume and temperature
and fitted linearly in $V$. Because $E$ and $E_0$ are computed in the same
32-atom cell with the same $k$ mesh, their difference does not mix the two
$k$ meshes. The linear thermal expansion coefficient follows as
$\alpha=(\partial S/\partial V)_T/3B_T$, with the isothermal bulk modulus
$B_T=V\partial^2F/\partial V^2$, both at the volume that minimizes $F$.
$F(V)$ is fitted by a cubic and $S(V)$ by a quadratic polynomial over the
seven volumes. For the QHA, $F_{\rm h}$ and $S$ are those of the harmonic FD
FC2 and the correction term is omitted. Neither route requires derivatives
of the FC2 with respect to temperature.

Table~\ref{tab:qha} gives the results; Supplementary Table~S11 adds $B_T$
and $C_P$ and a Gr\"uneisen cross-check. Evaluated at $V(T)$ with $B_T$, the
Gr\"uneisen formula reproduces the minimization within
$0.1\times10^{-6}$~K$^{-1}$, whereas the usual shortcut with the static $B_0$
and $V_0$ is 13--44\% low at 300--900~K. The converged QHA gives $C_P$
within 1~J~mol$^{-1}$K$^{-1}$ of the NIST-JANAF values \cite{chase1998}
between 300 and 900~K. The QHA with the converged 125-atom
harmonic FC2 gives 24.1 and $26.2\times10^{-6}$~K$^{-1}$ at 300 and 400~K,
4--5\% above the measured 23.2 and 25.1, and 40.2 at 900~K, against 36 at
923~K, of which vacancies contribute about $2\times10^{-6}$~K$^{-1}$
\cite{nakashima2020}.
With the harmonic FD FC2 of the 32-atom cell, the QHA overestimates $\alpha$
by 13--15\% at 300--400~K and diverges above 500~K; at 900~K its free-energy
minimum lies outside the volume grid. The cause is the truncated cell: its
interpolated harmonic spectrum has imaginary frequencies at all but the
smallest volume, with the minimum frequency (full $20^3$ mesh) falling from
$-0.45$ to $-1.25$~THz between 4.025 and 4.150~\AA. The self-consistent
FC2 is stable at every volume and temperature (minimum frequency
0.30--0.41~THz).

Two independent self-consistent grids were computed, one with design seeds
that differ between volumes and one with the same seeds at all volumes,
each with 32 force calls per volume and temperature. Their statistical
error was estimated by resampling the 16 $\pm$ pairs of every volume with
replacement (40 resamplings) and repeating the whole analysis. Up to 400~K
the bootstrap noise of $F(V)$ is below 0.35~meV per atom, less than 15\% of
the second difference of $F$ between neighboring volumes (2.4--2.7~meV),
and $\alpha$ is resolved: 13.2(2) and 13.0(2) at 100~K, 22.5(8) and 23.0(9)
at 300~K, and 23.1(13) and 23.4(10)$\times10^{-6}$~K$^{-1}$ at 400~K for the
two grids, so the grids agree within one standard deviation. At 300~K this
is 1--3\% below experiment, and at 400~K 7--8\% below it and 11\% below the
converged QHA. The second-order cumulant of free-energy perturbation,
$-\mathrm{Var}(\Delta U)/2k_BT$ with
$\Delta U=E-E_0-\tfrac12\mathbf u\cdot\boldsymbol\Phi_{\rm SC}\mathbf u$,
changes $\alpha$ by at most $0.5\times10^{-6}$~K$^{-1}$ in this range, and
omitting the first-order energy term by less than
$1\times10^{-6}$~K$^{-1}$.

Above 400~K the bootstrap noise of $F$ grows to 0.5--1.3~meV per atom,
17--37\% of the second difference, and the cubic fit of $F(V)$ becomes
unstable: with 32 calls the grids give 22--25 at 500--700~K and differ by
$11\times10^{-6}$~K$^{-1}$ at 900~K, and the bootstrap standard deviation
reaches 15--100 at 700~K. Extending the common grid to 128 calls per volume
at 700 and 900~K halves the noise of $F$ and raises $\alpha$ to 30.4(79) and
$29.8(146)\times10^{-6}$~K$^{-1}$, consistent with the converged QHA (32.9
and 40.2) within their uncertainty but still not resolved. The apparent
underestimate of the rise of $\alpha$ with 32 calls is therefore statistical.
Moment-matched amplitudes (Sec.~\ref{sec:sc_surrogate}), which remove the
fixed-amplitude bias of the FC2 and of the energy term, do not reduce the
noise at 32 calls (0.9--1.4~meV per atom at 700--900~K). Because the
standard deviation falls only as the inverse square root of the number of
calls, resolving $\alpha$ to $\pm1.5\times10^{-6}$~K$^{-1}$ at 700~K would
require several thousand calls per volume and temperature, far beyond the
budgets for which SOTC-D is designed. We therefore report the
self-consistent thermal expansion of Al only up to 400~K.

The same construction was applied to PbTe, whose harmonic FC2 in the
72-atom cell is unstable at every volume of the grid ($a=6.52$--6.70~\AA,
minimum frequency $-1.7$ to $-2.0$~THz), with 32 force calls per volume at
300, 500, and 700~K and design seeds common to all volumes. The
self-consistent FC2 is stable everywhere (minimum frequency 0.09--0.11~THz).
At 300~K it gives $\alpha=16.4(15)\times10^{-6}$~K$^{-1}$ (14.2 and 15.7
from the first 8 and 16 calls, 15.8 with a quartic $F(V)$), 20\% below the
measured $20.4\times10^{-6}$~K$^{-1}$ \cite{houston1968,madelung2004}, a
difference larger than the statistical error. The QHA in the same cell
gives 19.9, but only because the unstable modes are excluded from
$F_{\rm h}$, so its agreement is not meaningful. At 500 and 700~K the noise
of $F$ (2.2 and 4.6~meV per atom) is comparable to its curvature, and
$\alpha$ is not resolved.

In summary, the self-consistent design removes the instability of the
small-cell QHA and gives a reproducible thermal expansion of Al between 100
and 400~K, within 3\% of experiment at 300~K and 7--8\% below it at 400~K,
and a stable self-consistent free energy for PbTe, whose small-cell harmonic
FC2 is unstable, with $\alpha$ 20\% below experiment at 300~K. At higher
temperatures the curvature of $F(V)$ is not resolved with 32--128 calls per
volume; for a crystal whose large-cell harmonic FC2 is stable, such as Al,
the converged harmonic QHA is then the more accurate route. The calculation
used 4515 force calls in the 32-atom cell (two grids of seven volumes and
six temperatures with 32 calls each, the 128-call extension and the
moment-matched grid at 700 and 900~K, and the FD trials) and 21 in the
125-atom cell.

\begin{table}[t]
\caption{Linear thermal expansion coefficient $\alpha$ ($10^{-6}$~K$^{-1}$) of Al (PBE), from $(\partial S/\partial V)_T/3B_T$ on seven volumes. SC: self-consistent SOTC-D in the 32-atom cell, two grids with 32 force calls per volume and temperature (design seeds independent across volumes, and common to all volumes) and the common grid extended to 128 calls at 700 and 900~K (column 128); in parentheses the standard deviation over 40 bootstrap resamplings of the $\pm$ pairs. QHA: harmonic FD FC2 in the same cell and in the converged 125-atom cell. Experiment: averages of measured values at 298, 400, and 923~K (300, 400, and 900~K rows) compiled in Ref.~\onlinecite{nakashima2020}.}
\label{tab:qha}
\footnotesize
\begin{ruledtabular}
\begin{tabular}{lcccccc}
 & \multicolumn{3}{c}{SC} & \multicolumn{2}{c}{QHA} & \\
\cmidrule(lr){2-4}\cmidrule(lr){5-6}
$T$ (K) & indep. & common & 128 & 32 at. & 125 at. & expt.\\
\hline
100 & 13.2(0.2) & 13.0(0.2) & -- & 14.6 & 12.9 & --\\
300 & 22.5(0.8) & 23.0(0.9) & -- & 26.3 & 24.1 & 23.2\\
400 & 23.1(1.3) & 23.4(1.0) & -- & 28.9 & 26.2 & 25.1\\
500 & 23.0(1.7)$^b$ & 25.1(1.3)$^b$ & -- & 31.8 & 28.2 & --\\
700 & 22.0(104)$^b$ & 23.4(15)$^b$ & 30.4(7.9)$^b$ & 41.1 & 32.9 & --\\
900 & 21.5(26)$^b$ & 10.5(3.0)$^b$ & 29.8(15)$^b$ & 79.4$^a$ & 40.2 & 36.0\\
\end{tabular}
\end{ruledtabular}
$^a$Minimum outside the volume grid. $^b$Not resolved: the bootstrap noise of $F(V)$ exceeds 15\% of its second difference between neighboring volumes.
\end{table}

\subsection{Polar crystals}

\begin{figure*}[t]
\centering
\includegraphics[width=\textwidth]{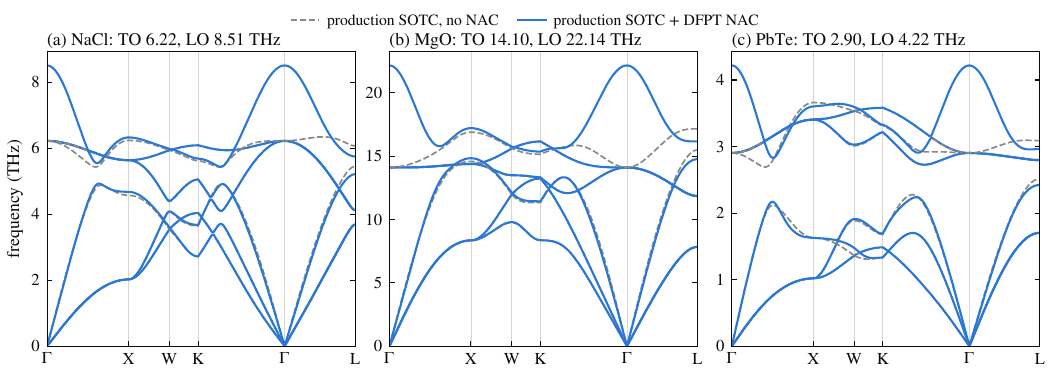}
\caption{SOTC-C FC2 dispersions of NaCl, MgO, and PbTe without (gray dashed) and
with (blue) the non-analytic correction from same-volume QE-DFPT Born charges
and dielectric tensors. The zone-center TO and LO frequencies are given above
each panel.}
\label{fig:polar}
\end{figure*}

Born effective charges and dielectric constants from DFPT at the production
volumes are $(Z^*_{\rm cation},\varepsilon_\infty)=(1.111, 2.471)$ for NaCl,
$(1.973, 3.167)$ for MgO, and $(5.767, 28.83)$ for PbTe. Applied to the
selected SOTC-C refits, they open zone-center LO--TO splittings of 2.29, 8.05,
and 1.32~THz (Fig.~\ref{fig:polar}). For NaCl, the same correction applied to
the $3\times3\times3$ harmonic FD model gives TO and LO limits of 4.37 and
7.26~THz, 11\% and 10\% below the infrared values of 4.92 and 8.03~THz
\cite{ref48}; the LO/TO ratio, 1.66, is within 2\% of experiment. The
non-analytic correction is independent of the displacement design and can be
combined with any of the FC2 models above (NaCl comparison with neutron
and infrared data in Supplementary Fig.~S3).
\section{Discussion}\label{sec:discussion}

The results allow a precise statement of what correlator design contributes.
The correlator objective of SOTC-C matches shell-averaged second moments,
which by construction the canonical distribution also matches on average; it
does not control the per-mode amplitudes or the conditioning of the force fit,
and its regression adds an axial prior that is inexact for fcc first-neighbor
bonds. Consequently the SOTC-C models are good effective-FC2
estimates---their thermodynamics agree closely with the effective
references---but their dispersions are not systematically better than those
of canonical sampling. SOTC-D keeps the deterministic, temperature-aware
character of SOTC but optimizes what matters for regression: exact thermal
mode variances in every configuration, a maximal information determinant, and
cancellation of the cubic force by antithetic pairs.

Within the class of finite-temperature effective-harmonic methods that fit
force constants to sampled forces, the comparisons above cover the relevant
alternatives at equal cost and with identical data: canonical sampling,
unpaired and paired, fitted by least squares as in TDEP \cite{ref5,ref6},
and its iterated form sTDEP \cite{ref60}; SSCHA \cite{ref7,ref50}, whose fixed point is the same
self-consistent effective FC2; special displacements \cite{ref42}; thermal
rattle with hiPhive regression \cite{ref59}; and nine sparse, Bayesian, and
least-squares fit methods. Three results stand out. First, at the small
budgets typical of high-throughput or expensive-functional calculations,
SOTC-D is the best design with the same trial in 8, 8, and 9 of 11 crystals
at two, four, and eight force calls, and its bootstrapped form beats
one-shot canonical sampling in all 11 crystals at every budget from 4 to 16
calls and two-iteration sTDEP with DFT forces in all 11 at eight calls and
9 of 11 at sixteen. Second, on model potentials with an exactly known
self-consistent target, SOTC-D-SC is at least as accurate as sTDEP and SSCHA
in 50 of 56 comparisons at equal budget; in the other six, pooled sTDEP (W
and NaCl at 64 and 16 calls, respectively, Si at 16, Mg at 32 and 64) or SSCHA (Al at 64)
is better by at most 0.006~THz. It needs a factor of 2--16 fewer force calls
than SSCHA for a given accuracy in 13 of 14 cases. With DFT forces, against
a converged self-consistent reference, it reaches for Al with four calls
the accuracy that iterated SSCHA reaches with about 64 and halves the error
of iterated sTDEP and self-consistent special displacements at four calls;
for weakly anharmonic Si, pooled sTDEP and self-consistent special
displacements are more accurate at 32 calls, and iterated SSCHA is as
accurate from 16 calls on. Third, the regression method, including sparsity-promoting
and Bayesian fits, changes the result by at most 0.03~THz, whereas the
design changes it by up to a factor of 2.6 relative to random canonical
subsets at two and eight calls in the supercell basis
(Table~\ref{tab:matched}), and by less in the crystal basis; the crystal-symmetry basis
reduces the number of parameters by an order of magnitude at no loss in
held-out accuracy and should be used by any method. The gain of SOTC-D
diminishes with budget, as expected once the sampling error falls below the
supercell and reference errors.

The estimator theory explains these observations. The effective FC2 is the
Gaussian-averaged Hessian; SOTC-D estimates it with a bias that is a single
diagonal term of the $M$-mode renormalization sum, $O(1/M)$, whereas
antithetic pairs remove the cubic force, the dominant noise of every
unpaired design, exactly; fixed amplitudes and D-optimal signs then
minimize the remaining variance for the given number of calls. The theory
also identifies the one failure mode found in the iteration-0 benchmark: a
trial with unstable commensurate modes (PbTe) gives a defective target and
underexcites the soft modes for any design. Updating the trial during the
design repairs it at no extra cost, and the same update converges to the
SCHA fixed point.

SOTC-D changes only how the configurations are chosen; the target, the
regression, and the self-consistency condition are those of sTDEP and the
SCHA. It is therefore complementary to these methods rather than an
alternative to them. Within sTDEP it replaces the Gaussian draws at each
iteration, and the bootstrapped and self-consistent variants are sTDEP
iterations in which all earlier forces are retained. Its use within SSCHA is
less direct, because the SSCHA reweighting of a population to a new trial
requires the probability density of the configurations, which the
deterministic design does not provide; the free energy and its gradients
would have to be evaluated from the fitted effective model instead. AIMD-based
TDEP samples the classical anharmonic ensemble, including non-Gaussian
displacement statistics, and is not replaced by a Gaussian design; SOTC-D
addresses the self-consistent harmonic level, at which quantum zero-point
motion enters through the mode amplitudes. Because MD samples are not
paired, the FC2 must be fitted jointly with the FC3; fitted alone it is
0.27--0.29~THz from the self-consistent FC2 in Al and Si. Fitted jointly, it
lies 0.08 (Al) and 0.06~THz (Si) from it; zero-point motion and
non-Gaussian statistics each change the FC2 by at most about 0.06--0.08~THz
(Sec.~\ref{sec:iter1}). With DFT forces, a single trajectory needed
about 310 MD steps in Al and at most 282 in Si (the shortest length
tested), including 250 equilibration steps, to come within 0.1~THz of its
own limit, whereas SOTC-D-SC needed four (Al)
and eight (Si) force calls to come within 0.09 and 0.07~THz of the
self-consistent target (Sec.~\ref{sec:iter1}).

SOTC-D gives the effective FC2 at one volume. Repeated on a volume grid, it
gives the free energy of the self-consistent harmonic model, because the
designed configurations also provide the energies for its anharmonic
correction. For Al, with 32 force calls per volume and temperature, this
removes the divergence of the QHA of the same 32-atom cell and gives a
reproducible thermal expansion between 100 and 400~K, within 3\% of
experiment at 300~K and 7--8\% below it at 400~K (Sec.~\ref{sec:qha}). Above
400~K the statistical noise of the free energy is comparable to its
curvature with respect to volume, and even 128 calls per volume do not
resolve $\alpha$. For PbTe, whose small-cell harmonic FC2 is
unstable at every volume, the self-consistent FC2 is stable and gives
$\alpha$ about 20\% below experiment at 300~K. Thermal expansion is therefore an application in
which the small-budget advantage of the design is limited: it needs
free-energy differences between volumes to better than 1~meV per atom, and
for crystals with a stable harmonic FC2 in a large cell the converged QHA
is the cheaper and more accurate route.

The comparison also shows where the remaining errors are. In 32--72-atom cells
the supercell truncation error (up to 0.5~THz) is comparable to or larger than
the sampling error of a few SOTC-D configurations; larger cells and the
non-analytic correction for polar crystals are therefore needed for
quantitative dispersions, independently of the sampling scheme. The method
assumes a unimodal displacement distribution; double-well systems near
displacive transitions are outside its scope, dynamically stabilized crystals
are unreliable (Sec.~\ref{sec:zr}), and also outside its scope are anharmonic free
energies, lifetimes, and thermal conductivity, which require the
higher-order tensors of Sec.~\ref{sec:ho} together with appropriate
many-body theory.

\section{Conclusions}\label{sec:conclusions}

We have introduced special optimal thermal cells for finite-temperature
effective-harmonic phonons and compared two design objectives. Correlator-matched
cells (SOTC-C) of nine crystals identify their FC2 models completely in a
space-group adapted basis and reproduce effective thermodynamics, but their
dispersion accuracy is limited by an inexact regression prior and by the lack
of control over mode amplitudes; for Au, single-configuration iterations
give an unstable model. The D-optimal antithetic design SOTC-D, which
excites every mode at its exact thermal amplitude, chooses signs that maximize
the FC2 information, and cancels cubic forces by inversion pairs, has an
amplitude bias of order $1/M$ at fourth order and no cubic noise. With eight force calls
it reproduces same-cell effective FC2 models within 0.03--0.31~THz for eleven
crystals of five structure classes (fcc and bcc metals, rocksalt, diamond,
and hcp), and with an updated trial it is more accurate than canonical
sampling with the same trial in all eleven at four and eight calls. Its
bootstrapped form needs no additional calculations and, in the supercell
basis, outperforms one-shot
canonical sampling in every crystal at every budget from 4 to 16 calls and
two-iteration sTDEP in every crystal at eight calls; on anharmonic model
potentials of five crystals, the fully self-consistent form reaches a given
accuracy with a factor of 2--16 fewer force calls than SSCHA in 13 of 14
cases. With DFT forces and a converged self-consistent reference, it halves
the error of iterated sTDEP and of self-consistent special displacements at
four calls for Al and needs about 16 times fewer calls than iterated SSCHA
for 0.09--0.10~THz, whereas for weakly anharmonic Si those two protocols
are more accurate at 32 calls and iterated SSCHA is as accurate.
Special displacements are competitive for Si and Mg but not in
general, and thermal rattle and alternative regression methods do not close
this gap. Converged references reveal that
finite-temperature renormalization (up to 1.4~THz) and supercell truncation
(up to 0.5~THz) are as large as or larger than the sampling errors, and
room-temperature Debye--Waller factors of Al, Cu, and Au are recovered within
6\% of tabulated values by converged harmonic references and within 2--11\%
by SOTC-D with eight force calls, when force constants at the
room-temperature lattice parameter are used. On a volume grid, the self-consistent
form gives a reproducible thermal expansion of Al between 100 and 400~K,
within 3\% of experiment at 300~K and 7--8\% below it at 400~K, whereas the
QHA of the same small cell diverges above 500~K; at higher temperatures the
statistical noise of the free energy is comparable to its curvature even
with 128 force calls per volume, and for Al the converged harmonic QHA is
then more accurate. For PbTe, whose
small-cell harmonic FC2 is unstable, the self-consistent form gives
$\alpha$ about 20\% below experiment at 300~K. With HSE06 the self-consistent
form is ahead of pooled sTDEP at 4--16 calls in Si and NaCl, against a composite
reference. In Al and Si cells of 108--256 atoms it is within 0.1~THz after two calls
but saturates at 0.02--0.08~THz, so that sTDEP is more accurate beyond about 4--16
calls, and the offset does not follow the $1/M$ estimate; in 128- and 216-atom NaCl
it stays ahead up to 16 calls. For dynamically
stabilized bcc Zr only two of six seeds converge cleanly, and an sTDEP run with
long iterations stayed trapped in the unstable modes of its first fit. Because the gain is
largest at budgets of a few force calls, the design is most useful where each
calculation is expensive---hybrid functionals or
high-throughput screening of finite-temperature phonons, and large cells when
a coarse accuracy of about 0.1~THz suffices (beyond some 4--16 calls sTDEP is more
accurate there)---and it can be used
as the sampling step of existing self-consistent workflows. The
implementation, inputs, and all comparison data are openly available.

\begin{acknowledgments}
No external funding was received.
\end{acknowledgments}

\section*{Author Declarations}
\subsection*{Conflict of Interest}
The author has no conflicts to disclose.
\subsection*{Author Contributions}
\textbf{Albert Linda:} Conceptualization; Methodology; Software; Investigation;
Formal analysis; Writing -- original draft; Writing -- review \& editing.
\subsection*{Use of Generative AI}
During the preparation of this work the author used Claude (Anthropic,
Claude Opus 5.5, through Claude Code) to assist with writing and testing
the analysis code, setting up and monitoring the calculations, and editing
the manuscript text. The author reviewed and edited all such output and
takes full responsibility for the content of the publication.

\section*{Supplementary Material}
See the Supplementary Material for the derivations of Eqs.~(\ref{eq:corr})
and (\ref{eq:bias}) and of the averaged-Hessian identity, the gradient of
Eq.~(\ref{eq:Q}), the surrogate design study, the design comparison in the
crystal-symmetry basis, the regression-framework comparison, the
individual SOTC-C iterations, the lattice-parameter sensitivity, an
overview of the reference accuracy, the electronic-convergence and cutoff
tests of the higher-order fits, the NaCl comparison with neutron data, the
branch-resolved convergence of SOTC-D for Al and Au with the Al neutron
model, the design comparisons at four and sixteen force calls, and the
nearest-neighbor force constants compared with neutron Born--von K\'arm\'an
fits.

\section*{Data Availability}
The data that support the findings of this study, including all QE inputs and
outputs, force--displacement data, fitted force constants, and the
scripts that generate every table and figure, are openly available at
\url{https://github.com/albert-hzbn/SOTC}.

\bibliography{ms}

\clearpage
\onecolumngrid
\begin{center}
{\large\bfseries Supplementary Material for ``Designed atomic displacements for
finite-temperature phonons from few force calculations''\par}
\vspace{0.6em}
Albert Linda\par
\vspace{1.2em}
\end{center}
\twocolumngrid
\setcounter{section}{0}\setcounter{figure}{0}\setcounter{table}{0}\setcounter{equation}{0}
\renewcommand{\thesection}{S\arabic{section}}
\renewcommand{\thefigure}{S\arabic{figure}}
\renewcommand{\thetable}{S\arabic{table}}
\renewcommand{\theequation}{S\arabic{equation}}
\renewcommand{\theHsection}{S\arabic{section}}
\renewcommand{\theHfigure}{S\arabic{figure}}
\renewcommand{\theHtable}{S\arabic{table}}
\renewcommand{\theHequation}{S\arabic{equation}}

\section{Surrogate design study}

Table~\ref{tab:surrogate} lists the median phonon-mesh errors of six
displacement designs on the DFT-fitted FC2+FC3+FC4 surrogates of Al, W, and
NaCl. The surrogate effective FC2 was fitted to 6000 paired canonical
samples; fitting half of them changes the mesh frequencies by 0.002 (Al),
0.001 (W), and 0.006~THz (NaCl). All designs were drawn from the harmonic FD
model of the same $3\times3\times3$ cell and fitted in the space-group adapted
FC2 basis with the cutoffs of the main text. Held-out forces on 200 canonical
configurations follow the same ordering.

\begin{table*}[t]
\caption{Surrogate design study (Al 300~K, W 1500~K, NaCl 700~K, Si 600~K; DFT-fitted FC2+FC3+FC4 surrogates in $3\times3\times3$ cells). Median mesh RMSE (THz) against the surrogate effective FC2 over 30 repetitions, at 2/4/8/16 force calls.}
\label{tab:surrogate}
\begin{ruledtabular}
\footnotesize
\setlength{\tabcolsep}{3pt}
\begin{tabular}{lcccc}
Design & Al & W & NaCl & Si\\
 & \multicolumn{4}{c}{2/4/8/16 force calls in each column}\\
\hline
Canonical & .199/.123/.088/.064 & .091/.068/.048/.033 & .224/.185/.116/.089 & .191/.122/.079/.064\\
Canonical $\pm$ & .066/.054/.038/.032 & .035/.027/.021/.018 & .176/.127/.078/.078 & .060/.045/.033/.027\\
Exact ampl., random signs & .066/.050/.038/.025 & .035/.028/.017/.016 & .166/.131/.093/.063 & .052/.042/.032/.018\\
SOTC-D & .063/.056/.040/.025 & .039/.024/.021/.015 & .150/.131/.090/.062 & .054/.040/.031/.021\\
Stratified ampl. + D-opt. & .073/.046/.038/.028 & .035/.031/.020/.015 & .151/.134/.088/.089 & .065/.039/.030/.023\\
Gaussian ampl. + D-opt. & .074/.053/.046/.030 & .048/.032/.024/.019 & .184/.124/.114/.077 & .064/.043/.032/.025\\
Special displ. (ZG) & .084/.067/.048/.041 & .042/.033/.017/.033 & .163/.133/.161/.174 & .036/.033/.022/.025\\
Thermal rattle & .089/.074/.069/.076 & .053/.038/.032/.030 & .300/.295/.305/.323 & .313/.300/.296/.309\\
\hline
Harmonic vs effective & 0.229 & 0.149 & 0.765 & 0.225\\
\end{tabular}
\end{ruledtabular}
\end{table*}

\section{Designs in the crystal-symmetry basis}

Table~\ref{tab:selfconsistent_crystal} repeats Table~V of the main text with
all fits in the FC2 basis of the full crystal space group (hiPhive cluster
space, 3--12 parameters, cutoff limited to half the shortest image distance).
For Si and Mg the two bases coincide.

\begin{table*}[t]
\caption{Updated trial (iteration 1, crystal basis). Mesh RMSE (THz) against the iteration-1 effective FC2. Matched trial: SOTC-D designed from the iteration-1 trial (D$_1$) versus the better of random canonical subsets and pairs drawn from the same trial (C$_1$), scored against the disjoint effective FC2. Equal cost: bootstrapped SOTC-D (B; first pair from the harmonic trial, later pairs from the model refitted to the first pair, no extra force calls) and one-shot SOTC-D (D$_0$) versus canonical pairs from the harmonic trial (C$_0$; a single sTDEP iteration), scored against the 40-snapshot iteration-1 effective FC2, which shares no configuration with them. B and D$_0$ coincide at 2 calls by construction. The best entry of each group is bold.}
\label{tab:selfconsistent_crystal}
\begin{ruledtabular}
\begin{tabular}{lcccccccccc}
 & \multicolumn{5}{c}{2 force calls} & \multicolumn{5}{c}{8 force calls}\\
\cmidrule(lr){2-6}\cmidrule(lr){7-11}
 & \multicolumn{2}{c}{matched trial} & \multicolumn{3}{c}{equal cost} & \multicolumn{2}{c}{matched trial} & \multicolumn{3}{c}{equal cost}\\
\cmidrule(lr){2-3}\cmidrule(lr){4-6}\cmidrule(lr){7-8}\cmidrule(lr){9-11}
 & D$_1$ & C$_1$ & B & D$_0$ & C$_0$ & D$_1$ & C$_1$ & B & D$_0$ & C$_0$\\
\hline
Al & \textbf{0.06} & 0.08 & \textbf{0.07} & 0.07 & 0.10 & \textbf{0.03} & 0.05 & \textbf{0.03} & 0.05 & 0.04\\
Ag & \textbf{0.04} & 0.11 & \textbf{0.02} & 0.02 & 0.13 & 0.06 & \textbf{0.05} & \textbf{0.06} & 0.09 & 0.10\\
Au & 0.08 & \textbf{0.08} & \textbf{0.15} & 0.15 & 0.24 & \textbf{0.05} & 0.05 & \textbf{0.02} & 0.17 & 0.22\\
Cu & \textbf{0.08} & 0.15 & 0.20 & 0.20 & \textbf{0.07} & \textbf{0.04} & 0.08 & 0.17 & \textbf{0.10} & 0.14\\
Mo & 0.06 & \textbf{0.06} & \textbf{0.05} & 0.05 & 0.07 & \textbf{0.02} & 0.03 & \textbf{0.02} & 0.03 & 0.06\\
W & 0.05 & \textbf{0.04} & 0.05 & 0.05 & \textbf{0.04} & \textbf{0.02} & 0.02 & 0.02 & \textbf{0.02} & 0.02\\
NaCl & \textbf{0.05} & 0.11 & \textbf{0.11} & 0.11 & 0.14 & 0.09 & \textbf{0.07} & \textbf{0.05} & 0.15 & 0.12\\
MgO & \textbf{0.22} & 0.32 & 0.42 & 0.42 & \textbf{0.34} & \textbf{0.09} & 0.15 & \textbf{0.11} & 0.32 & 0.26\\
PbTe & \textbf{0.02} & 0.07 & 0.19 & 0.19 & \textbf{0.13} & \textbf{0.02} & 0.03 & \textbf{0.03} & 0.12 & 0.09\\
Si & \textbf{0.07} & 0.10 & \textbf{0.10} & 0.10 & 0.11 & \textbf{0.05} & 0.05 & \textbf{0.05} & 0.09 & 0.07\\
Mg & \textbf{0.05} & 0.06 & \textbf{0.06} & 0.06 & 0.07 & \textbf{0.03} & 0.05 & \textbf{0.02} & 0.06 & 0.04\\
\hline
D$_1<$C$_1$ / B$<$C$_0$ & \multicolumn{2}{c}{8/11} & \multicolumn{3}{c}{7/11} & \multicolumn{2}{c}{9/11} & \multicolumn{3}{c}{9/11}\\
held-out force D$_1<$C$_1$ & \multicolumn{2}{c}{9/11} & \multicolumn{3}{c}{} & \multicolumn{2}{c}{8/11} & \multicolumn{3}{c}{}\\
\end{tabular}
\end{ruledtabular}
\end{table*}

Table~\ref{tab:stdep_crystal} repeats the DFT sTDEP comparison of the main
text in the crystal-symmetry basis.

\begin{table}[t]
\caption{Self-consistent protocols with DFT forces, equal cost. Mesh RMSE (THz) against the 40-snapshot iteration-1 effective FC2 (crystal basis). B: bootstrapped SOTC-D. sTDEP: half of the calls as canonical $\pm$ pairs from the harmonic trial, half as canonical $\pm$ pairs from the model refitted to them, fitted to the second iteration only (S$_{\rm L}$) or to both (S$_{\rm P}$); median of three seeds. C$_0$: one-shot canonical pairs from the harmonic trial (median of 20 subsets). No configuration of any protocol enters the reference. The best entry of each group is bold.}
\label{tab:stdep_crystal}
\begin{ruledtabular}
\begin{tabular}{lcccccccc}
 & \multicolumn{4}{c}{8 calls} & \multicolumn{4}{c}{16 calls}\\
\cmidrule(lr){2-5}\cmidrule(lr){6-9}
 & B & S$_{\rm L}$ & S$_{\rm P}$ & C$_0$ & B & S$_{\rm L}$ & S$_{\rm P}$ & C$_0$\\
\hline
Al & \textbf{0.03} & 0.05 & 0.05 & 0.04 & 0.02 & 0.03 & \textbf{0.01} & 0.03\\
Ag & 0.06 & \textbf{0.04} & 0.07 & 0.10 & 0.04 & \textbf{0.03} & 0.09 & 0.10\\
Au & \textbf{0.02} & 0.06 & 0.17 & 0.22 & \textbf{0.01} & 0.04 & 0.15 & 0.21\\
Cu & 0.17 & 0.18 & \textbf{0.10} & 0.14 & 0.16 & 0.08 & \textbf{0.06} & 0.14\\
Mo & \textbf{0.02} & 0.05 & 0.03 & 0.06 & 0.02 & 0.03 & \textbf{0.02} & 0.05\\
W & 0.02 & 0.05 & 0.03 & \textbf{0.02} & 0.02 & 0.03 & \textbf{0.02} & 0.02\\
NaCl & \textbf{0.05} & 0.13 & 0.09 & 0.12 & \textbf{0.04} & 0.10 & 0.13 & 0.12\\
MgO & \textbf{0.11} & 0.22 & 0.18 & 0.26 & \textbf{0.10} & 0.19 & 0.22 & 0.25\\
PbTe & \textbf{0.03} & 0.05 & 0.07 & 0.09 & \textbf{0.04} & 0.04 & 0.06 & 0.08\\
Si & \textbf{0.05} & 0.08 & 0.08 & 0.07 & \textbf{0.04} & 0.05 & 0.06 & 0.06\\
Mg & \textbf{0.02} & 0.05 & 0.03 & 0.04 & \textbf{0.02} & 0.04 & 0.04 & 0.03\\
\hline
B best & \multicolumn{4}{c}{8/11} & \multicolumn{4}{c}{6/11}\\
\end{tabular}
\end{ruledtabular}
\end{table}

\section{Regression frameworks}

Table~\ref{tab:fitters} compares nine regression methods of the trainstation
package, as used in hiPhive workflows, on the same DFT data and the same
crystal-symmetry cluster space. Hyperparameters were selected by the
package's internal cross-validation. All methods agree to within 0.03~THz.

\begin{table*}[t]
\caption{Regression frameworks (hiPhive cluster space with crystal symmetry, trainstation fit methods) on the same DFT data, 8 force calls. Entries: mesh RMSE (THz) against the disjoint effective FC2 (supercell basis, so the entries include the basis difference listed in the main text) for random canonical subsets (median of 10) / SOTC-D.}
\label{tab:fitters}
\footnotesize
\begin{ruledtabular}
\begin{tabular}{lccccccccccc}
Method & Al & Ag & Au & Cu & Mo & W & NaCl & MgO & PbTe & Si & Mg\\
 & \multicolumn{11}{c}{canonical / SOTC-D in each column}\\
\hline
least-squares & 0.20/0.18 & 0.11/0.09 & 0.12/0.13 & 0.23/0.20 & 0.09/0.08 & 0.06/0.05 & 0.13/0.11 & 0.31/0.26 & 0.06/0.11 & 0.10/0.05 & 0.08/0.04\\
ridge & 0.20/0.18 & 0.11/0.09 & 0.12/0.13 & 0.24/0.20 & 0.09/0.08 & 0.06/0.05 & 0.13/0.12 & 0.31/0.26 & 0.06/0.12 & 0.10/0.05 & 0.08/0.04\\
bayesian-ridge & 0.20/0.18 & 0.11/0.09 & 0.12/0.14 & 0.24/0.20 & 0.09/0.08 & 0.06/0.05 & 0.13/0.12 & 0.31/0.26 & 0.06/0.12 & 0.10/0.05 & 0.08/0.03\\
ardr & 0.20/0.18 & 0.11/0.09 & 0.12/0.14 & 0.24/0.20 & 0.09/0.08 & 0.06/0.05 & 0.14/0.11 & 0.32/0.27 & 0.07/0.12 & 0.08/0.06 & 0.07/0.04\\
rfe & 0.20/0.18 & 0.11/0.09 & 0.12/0.13 & 0.23/0.20 & 0.09/0.08 & 0.06/0.05 & 0.13/0.11 & 0.33/0.26 & 0.06/0.12 & 0.10/0.07 & 0.07/0.04\\
omp & 0.20/0.18 & 0.11/0.09 & 0.12/0.13 & 0.23/0.20 & 0.09/0.08 & 0.06/0.05 & 0.14/0.11 & 0.31/0.26 & 0.06/0.11 & 0.10/0.08 & 0.07/0.04\\
lasso & 0.20/0.18 & 0.11/0.09 & 0.12/0.13 & 0.23/0.20 & 0.09/0.08 & 0.06/0.05 & 0.14/0.11 & 0.31/0.26 & 0.06/0.11 & 0.09/0.06 & 0.07/0.04\\
adaptive-lasso & 0.20/0.18 & 0.13/0.12 & 0.12/0.16 & 0.23/0.20 & 0.09/0.08 & 0.06/0.05 & 0.14/0.11 & 0.31/0.26 & 0.06/0.11 & 0.09/0.05 & 0.08/0.06\\
elasticnet & 0.20/0.18 & 0.11/0.09 & 0.12/0.13 & 0.23/0.20 & 0.09/0.08 & 0.06/0.05 & 0.14/0.11 & 0.31/0.26 & 0.06/0.11 & 0.09/0.06 & 0.07/0.04\\
\end{tabular}
\end{ruledtabular}
\end{table*}

\section{Derivations for the estimator theory}

\textit{Effective FC2 as averaged Hessian.} Let $\mathbf u\sim
N(\mathbf 0,\mathbf C)$ and let $V$ be smooth. Stein's lemma states
\begin{equation}
\langle\mathbf u\,g(\mathbf u)\rangle=\mathbf C\,\langle\nabla g\rangle .
\end{equation}
With $g=\mathbf F^{\rm T}=-\nabla V^{\rm T}$ this gives
\begin{equation}
\langle\mathbf u\,\mathbf F^{\rm T}\rangle=-\mathbf C\,\langle\nabla\nabla V\rangle ,
\end{equation}
so the population least-squares coefficient of $\mathbf F$ on $\mathbf u$ is
\begin{equation}
-\langle\mathbf F\,\mathbf u^{\rm T}\rangle\,\mathbf C^{-1}
=\langle\nabla\nabla V\rangle .
\end{equation}
The SCHA free-energy gradient with respect to the auxiliary FC2 vanishes
when
\begin{equation}
\langle\nabla\nabla V\rangle_{\mathbf C(\boldsymbol\Phi)}=\boldsymbol\Phi
\end{equation}
for a fixed centroid, which is the self-consistent fixed point of the main
text.

\textit{Fixed-amplitude bias.} In mass-weighted normal coordinates the
forces are
\begin{equation}
f_a=-\omega_a^2Q_a-\tfrac12V_{abc}\,Q_bQ_c-\tfrac16V_{abcd}\,Q_bQ_cQ_d .
\end{equation}
For a design with independent, symmetrically distributed coordinates the
regression coefficient is
\begin{equation}
\Phi_{ae}=-\frac{\langle f_aQ_e\rangle}{\langle Q_e^2\rangle}.
\end{equation}
The cubic term vanishes by symmetry. The quartic term needs the fourth
moment $\langle Q_bQ_cQ_dQ_e\rangle$:
\begin{itemize}
\item for distinct indices paired as $(bb)(ee)$ it equals
$\sigma_b^2\sigma_e^2$, for both Gaussian and $\pm\sigma$ coordinates;
\item for $b=c=d=e$ it equals $3\sigma_e^4$ (Gaussian) or $\sigma_e^4$
(fixed amplitude).
\end{itemize}
Collecting the three pairings,
\begin{equation}
\Phi_{ae}=\omega_a^2\delta_{ae}
+\frac12\sum_{b}V_{abbe}\,\sigma_b^2
-\frac16\,(3-\kappa_e)\,V_{aeee}\,\sigma_e^2 ,
\end{equation}
with the kurtosis $\kappa_e=\langle Q_e^4\rangle/\sigma_e^4$. The last term
vanishes for Gaussian coordinates ($\kappa_e=3$); for the fixed-amplitude
design ($\kappa_e=1$) it equals $-\tfrac13V_{aeee}\sigma_e^2$.

This result is verified numerically in the test suite
(\texttt{test\_optimal\_design.py}) by exact enumeration of all $2^6$ sign
patterns of a random six-mode quartic model. The same tests verify:
\begin{itemize}
\item the exact cancellation of arbitrary cubic forces by $\pm$ pairs;
\item the thermal mean-square displacements of the ZG and rattle designs;
\item for a crystal with a two-atom basis, the canonical nearest-neighbor
bond-stretch variance of the ZG configurations, which requires the Bloch
phase of the full atomic position.
\end{itemize}

\textit{Exact single-coordinate statistics.} Consider one coordinate with
force
\begin{equation}
f=-\omega^2Q-cQ^2-dQ^3+\epsilon .
\end{equation}
Over $K$ calls the fitted slope is
\begin{equation}
\hat\phi=\omega^2
+c\,\frac{\sum Q^3}{\sum Q^2}
+d\,\frac{\sum Q^4}{\sum Q^2}
-\frac{\sum\epsilon Q}{\sum Q^2},
\end{equation}
and the target is $\omega^2+3d\sigma^2$.

For $n$ independent Gaussian amplitudes, $x_i=Q_i^2/\sigma^2$ are $\chi^2_1$
variables. Their sum $T\sim\chi^2_n$ is independent of the weights
$w_i=x_i/T$, which are Dirichlet($\tfrac12$) distributed. The Dirichlet
moments give
\begin{align}
E\!\left[\frac{\sum Q^4}{\sum Q^2}\right]&=\frac{3\sigma^2n}{n+2},\\
E\!\left[\left(\frac{\sum Q^4}{\sum Q^2}\right)^{\!2}\right]
&=\frac{\sigma^4n\,(9n+96)}{(n+4)(n+6)},\\
E\!\left[\left(\frac{\sum Q^3}{\sum Q^2}\right)^{\!2}\right]
&=\frac{15\sigma^2n}{(n+2)(n+4)},
\end{align}
and $E[1/T]=1/(n-2)$ gives the noise variance
\begin{equation}
\mathrm{Var}_\epsilon=\frac{s^2\,n}{(n-2)\,\sigma^2K}.
\end{equation}
Pairs have $\sum Q^3=0$ and $n=K/2$. Fixed amplitudes have
$\sum Q^4/\sum Q^2=\sigma^2$, no amplitude variance, and the bias
$-2d\sigma^2$.

Monte Carlo with $4\times10^5$ repetitions reproduces all variances within
2\% and all biases within the sampling error for $K=4$--64
(\texttt{sotc/design\_theory.py}, \texttt{theory\_checks.py}, and unit
tests). For $\sigma^2=1$, $c=0.6$, $d=0.2$, and $s^2=0.5$, fixed amplitudes
have the smaller mean-squared error up to 12 calls against paired and up
to 38 calls against unpaired canonical sampling.

\textit{Bias against the number of modes.} Consider the quartic model
\begin{equation}
V_4=\frac18\sum_{a\neq b}W_{ab}\,Q_a^2Q_b^2+\frac1{24}\sum_aW_{aa}\,Q_a^4
\end{equation}
with random positive $W$. Exact averaging over all $2^M$ sign patterns gives
a relative bias of the diagonal FC2 of
\begin{equation}
\frac{W_{aa}/3}{\tfrac12\sum_bW_{ab}},
\end{equation}
equal to Eq.~(8) of the main text to $10^{-15}$. Its product with $M$ is
0.65--0.70 for $M=2$--12, i.e., the bias falls as $2/(3M)$.

\textit{Higher orders.} In the Al surrogate cell ($3\times3\times3$; hiPhive
cluster space with 5 FC2, 10 FC3, and 27 FC4 parameters, cutoffs 4.5, 3.5,
and 3.0~\AA), designs generated from the thermal modes at 300~K were
compared by the relative FC2--FC3 cross-information
\begin{equation}
\frac{\|\mathbf X_2^{\rm T}\mathbf X_3\|}
{\left(\|\mathbf X_2^{\rm T}\mathbf X_2\|\,
\|\mathbf X_3^{\rm T}\mathbf X_3\|\right)^{1/2}} .
\end{equation}
It is zero (to machine precision) for every paired design and 0.06--0.22
for unpaired ones. One configuration
already determines the 10 FC3 parameters, so pairs do not lose FC3 rank in
cells of this size, and the FC3 variance per call of paired designs lies
within the range of unpaired ones. The joint FC2+FC4 design has full rank
(32) with similar smallest singular values for fixed, moment-matched, and
Gaussian amplitudes. Two plausible expectations are therefore not borne
out: pairs do not halve the FC3
information, and fixed amplitudes do not make FC2 and FC4 collinear in a
Cartesian cluster basis; the amplitude bias arises only when FC4 is left
out of the fit.

\section{Higher-order force-constant validation}

\begin{figure*}[t]
\centering
\includegraphics[width=0.95\textwidth]{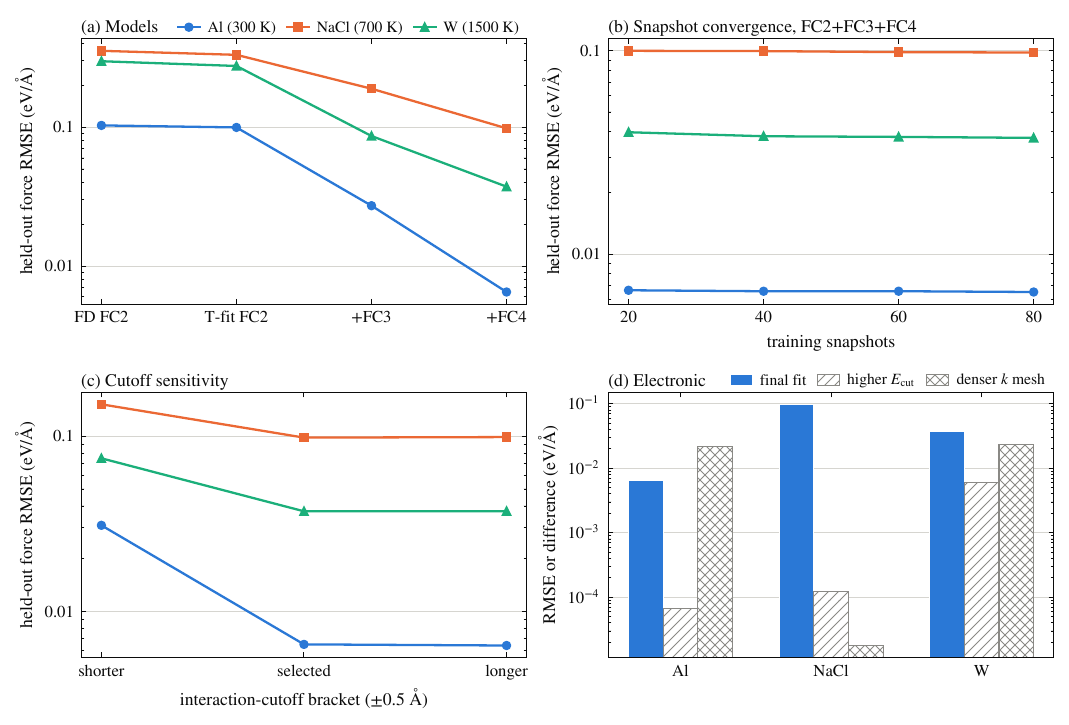}
\caption{Validation of the FC2+FC3+FC4 fits to 80 canonical QE snapshots
(Al 300~K, W 1500~K, NaCl 700~K; $3\times3\times3$ cells). (a) Held-out force
RMSE of harmonic FD FC2 and of temperature-fitted FC2, FC2+FC3, and
FC2+FC3+FC4 models. (b) Convergence with training snapshots. (c) Cutoff
sensitivity ($\pm0.5$~\AA). (d) Fit error compared with electronic-setting
force differences.}
\label{fig:ifc34}
\end{figure*}

Figure~\ref{fig:ifc34} documents the joint FC2+FC3+FC4 fits used as
surrogates: held-out errors of the nested models, convergence with the number
of training snapshots, sensitivity to the interaction cutoffs, and comparison
with the force changes produced by higher plane-wave cutoffs and denser $k$
meshes. For Al the denser-$k$-mesh force change exceeds the final fit error,
so the Al tensors are used here only as model potentials for the
sampling study, not as converged inputs for transport calculations.

\section{Individual iterations of the SOTC-C runs}

\begin{table*}[t]
\caption{Individual iterations of the SOTC-C runs: configurations, rms displacement, unstable mesh fraction and minimum frequency of the axial-prior fit, and normalized condition number of the space-group adapted design. An asterisk marks the iteration selected by the workflow.}
\label{tab:iterations}
\begin{ruledtabular}
\renewcommand{\arraystretch}{1.12}
\begin{tabular}{lcccccc}
Material & iter. & $N_{\rm ens}$ & $u_{\rm rms}$ (\AA) & $f_{\rm unst}$ & $\omega_{\min}$ (THz) & cond.\\
\hline
Al & 1 & 5 & 0.106 & 0.000 & 0.60 & 5.2\\
Al & 2* & 5 & 0.106 & 0.000 & 0.48 & 3.4\\
\noalign{\smallskip}
Ag & 1 & 5 & 0.167 & 0.000 & 0.26 & 5.7\\
Ag & 2 & 5 & 0.167 & 0.000 & 0.42 & 3.2\\
Ag & 3 & 5 & 0.167 & 0.000 & 0.44 & 3.2\\
Ag & 4* & 5 & 0.167 & 0.000 & 0.32 & 2.8\\
\noalign{\smallskip}
Au & 1* & 1 & 0.186 & 0.007 & -0.95 & 10.7\\
Au & 2 & 1 & 0.186 & 0.159 & -4.58 & 7.5\\
\noalign{\smallskip}
Cu & 1 & 8 & 0.169 & 0.000 & 0.51 & 15.3\\
Cu & 2 & 8 & 0.169 & 0.000 & 0.45 & 7.9\\
Cu & 3* & 8 & 0.169 & 0.000 & 0.45 & 7.5\\
Cu & 4 & 8 & 0.169 & 0.000 & 0.52 & 8.2\\
\noalign{\smallskip}
Mo & 1 & 1 & 0.111 & 0.000 & 0.52 & 17.1\\
Mo & 2* & 1 & 0.111 & 0.000 & 0.99 & 8.0\\
\noalign{\smallskip}
W & 1 & 3 & 0.115 & 0.000 & 0.76 & 4.5\\
W & 2* & 3 & 0.115 & 0.000 & 0.73 & 2.9\\
W & 3 & 3 & 0.115 & 0.000 & 0.75 & 3.5\\
W & 4 & 3 & 0.115 & 0.000 & 0.74 & 3.3\\
\noalign{\smallskip}
NaCl & 1 & 3 & 0.191 & 0.000 & 0.31 & 3.9\\
NaCl & 2 & 3 & 0.191 & 0.000 & 0.34 & 2.8\\
NaCl & 3 & 3 & 0.191 & 0.000 & 0.32 & 2.9\\
NaCl & 4 & 3 & 0.191 & 0.000 & 0.39 & 3.1\\
NaCl & 5* & 3 & 0.191 & 0.000 & 0.31 & 2.4\\
\noalign{\smallskip}
MgO & 1 & 3 & 0.141 & 0.000 & 0.96 & 2.9\\
MgO & 2 & 3 & 0.141 & 0.000 & 0.73 & 2.1\\
MgO & 3 & 3 & 0.141 & 0.000 & 0.96 & 2.6\\
MgO & 4* & 3 & 0.141 & 0.000 & 0.88 & 2.2\\
\noalign{\smallskip}
PbTe & 1 & 5 & 0.190 & 0.000 & 0.21 & 3.1\\
PbTe & 2* & 5 & 0.190 & 0.000 & 0.22 & 2.2\\
PbTe & 3 & 5 & 0.190 & 0.000 & 0.19 & 2.1\\
PbTe & 4 & 5 & 0.190 & 0.000 & 0.22 & 2.4\\
\end{tabular}
\end{ruledtabular}
\end{table*}

Table~\ref{tab:iterations} lists every iteration of the SOTC-C runs. The two
Au iterations each contain a single configuration; their
axial-prior fits are unstable, whereas the pooled fit is stable.

\section{Lattice-parameter sensitivity}

Table~\ref{tab:volume} compares the production lattice parameters with the
PBE equilibrium values and gives the change of the harmonic FD model of the
$3\times3\times3$ cell when the PBE value is used: the mesh RMSE between the
two models and the change of $S_{\rm vib}$ at the design temperature.

\begin{table*}[t]
\caption{PBE equilibrium lattice parameters from Birch--Murnaghan fits (primitive cells) and the effect of using them instead of the production values on the harmonic $3\times3\times3$ FD model: mesh RMSE and change in $S_{\rm vib}(T_{\rm des})$.}
\label{tab:volume}
\begin{ruledtabular}
\begin{tabular}{lcccccc}
 & $a_{\rm prod}$ (\AA) & $a_{\rm PBE}$ (\AA) & $\Delta a/a$ (\%) & $B_0$ (GPa) & RMSE (THz) & $\Delta S_{\rm vib}$\\
\hline
Al & 4.050 & 4.039 & -0.26 & 78 & 0.10 & -0.38\\
Ag & 4.120 & 4.146 & +0.63 & 91 & 0.16 & +1.17\\
Au & 4.130 & 4.150 & +0.49 & 142 & 0.12 & +1.17\\
Cu & 3.670 & 3.631 & -1.06 & 142 & 0.35 & -1.71\\
Mo & 3.160 & 3.160 & -0.01 & 259 & 0.00 & -0.01\\
W & 3.180 & 3.184 & +0.14 & 305 & 0.03 & +0.18\\
NaCl & 5.730 & 5.704 & -0.45 & 24 & 0.12 & -0.64\\
MgO & 4.260 & 4.249 & -0.25 & 153 & 0.14 & -0.30\\
PbTe & 6.510 & 6.567 & +0.88 & 39 & 0.14 & +1.72\\
\end{tabular}
\end{ruledtabular}
\end{table*}

\section{Thermal displacement correlator}

In second quantization, the displacement of basis atom $a$ in cell
$\mathbf R$ is
\begin{multline}
\hat u_a^\alpha(\mathbf R)=\frac{1}{\sqrt{N_c}}\sum_{\mathbf qs}
\sqrt{\frac{\hbar}{2M_a\omega_{\mathbf qs}}}\;e_a^\alpha(\mathbf qs)\\
\times\left(\hat a_{\mathbf qs}+\hat a^\dagger_{-\mathbf qs}\right)
e^{i\mathbf q\cdot\mathbf R}.
\end{multline}
The thermal occupations are
\begin{align}
\langle\hat a^\dagger_{\mathbf qs}\hat a_{\mathbf q's'}\rangle&=n_B\,
\delta_{\mathbf q\mathbf q'}\delta_{ss'},\\
\langle\hat a_{\mathbf qs}\hat a^\dagger_{\mathbf q's'}\rangle&=(n_B+1)\,
\delta_{\mathbf q\mathbf q'}\delta_{ss'},
\end{align}
the anomalous averages vanish, and $2n_B+1=\coth(\hbar\omega/2k_BT)$.
The canonical average $\langle\hat u_a^\alpha(\mathbf 0)\hat u_b^\beta(\mathbf R)\rangle$
then reduces to Eq.~(4) of the main text: only $\mathbf q'=-\mathbf q$
contributes, and $e^{\beta}_b(-\mathbf qs)=e^{\beta*}_b(\mathbf qs)$ gives
the factor
\begin{equation}
e^\alpha_a(\mathbf qs)\,e^{\beta*}_b(\mathbf qs)\,e^{-i\mathbf q\cdot\mathbf R}.
\end{equation}
Taking its trace with the completeness relation of the polarization vectors
gives, for a monatomic crystal,
\begin{equation}
U_{\rm iso}=\frac{\hbar}{6N_cM}\sum_{\mathbf qs}\frac{1}{\omega_{\mathbf qs}}
\coth\frac{\hbar\omega_{\mathbf qs}}{2k_BT}.
\end{equation}

In the commensurate supercell the same covariance is diagonal in the
normal-mode coordinates $Q_m$ with variances $\sigma_m^2$ given in Sec.~II~D
of the main text. A configuration of the form of Eq.~(7) has
$Q_m^2=\sigma_m^2$ for every mode, so that its single-configuration
second moment equals the canonical one mode by mode. The cross terms
\begin{equation}
Q_mQ_{m'}=s_ms_{m'}\,\sigma_m\sigma_{m'}
\end{equation}
average to zero only over the sign patterns, and their effect on the force fit is what the D-optimal sign
choice controls.

\begin{figure*}[t]
\centering
\includegraphics[width=\textwidth]{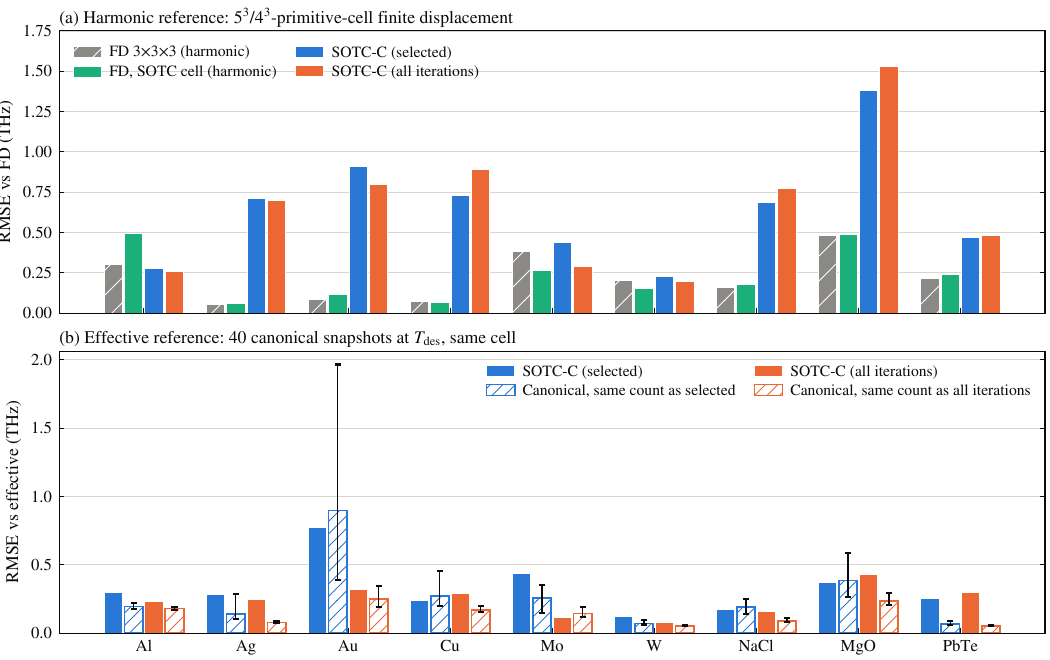}
\caption{(a) Mesh RMSE against the converged large-cell harmonic FD reference
for harmonic FD in $3\times3\times3$ cells and in the SOTC cells, and for the
selected and pooled SOTC-C models. (b) Mesh RMSE against the same-cell
effective FC2 for the selected and pooled SOTC-C models and for random
canonical subsets of the same size fitted with the same prior (median and
10--90\% range over 20 draws).}
\label{fig:accuracy}
\end{figure*}

\begin{figure*}[t]
\centering
\includegraphics[width=0.95\textwidth]{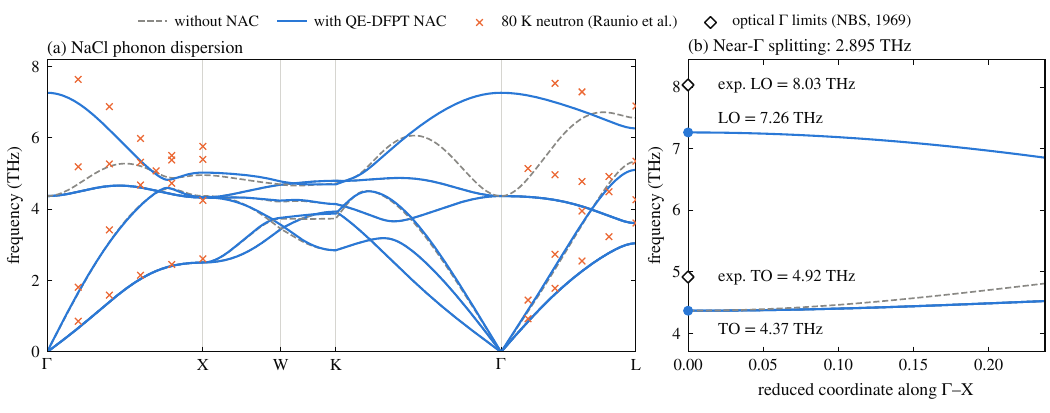}
\caption{NaCl harmonic FD FC2 ($3\times3\times3$ cell) without (gray dashed)
and with (blue) the non-analytic correction from QE-DFPT Born charges and
dielectric tensor. Crosses: 39 neutron points at 80~K along $\Gamma$--X
and $\Gamma$--L \cite{ref35}; open diamonds: infrared TO and LO limits
\cite{ref48}. The neutron data are temperature-mismatched to the
fixed-volume calculation and are shown without a global error metric.}
\label{fig:nacl}
\end{figure*}

\section{SOTC objective gradient}

For the shell objective of Eq.~(6) of the main text, write the cell
correlator as a sum over ordered atom pairs,
\begin{equation}
\bar C_{2,ab}^{\rm cell}(R;\mathbf u)=\sum_{i,j}P^{abR}_{ij}\,
\mathbf u_i\cdot\mathbf u_j ,
\end{equation}
where the symmetric weight matrix $P^{abR}_{ij}$ equals $1/(3N_R^{ab}N_a)$ (or half
of it for each order when $a\neq b$) if the pair belongs to shell $R$ of the
species channel and zero otherwise; the on-site term is the shell $R=0$. Then
\begin{equation}
\frac{\partial\mathcal Q}{\partial\mathbf u_i}=
4\sum_{a,b,R}w_{ab}(R)\,\Delta_{ab}(R)\sum_jP^{abR}_{ij}\,\mathbf u_j ,
\end{equation}
with $\Delta_{ab}(R)=\bar C_{2,ab}^{\rm cell}-\bar C_{2,ab}$. The
mode-amplitude gradient follows by the chain rule through
\begin{equation}
\mathbf u_i=\frac{1}{\sqrt{M_i}}\sum_m A_m\,\mathbf e_{im}.
\end{equation} The optimization is
carried out in mode amplitudes and phases with random restarts.

\section{Accuracy overview}

Figure~\ref{fig:accuracy} summarizes the reference accuracy of the main text:
the truncation error of the harmonic FD models in the $3\times3\times3$ and SOTC
cells and of the SOTC-C models against the converged references (a), and the
SOTC-C models against the same-cell effective FC2 compared with random canonical
subsets of the same size (b).

\section{NaCl optic modes and neutron data}

Figure~\ref{fig:nacl} shows the effect of the non-analytic correction on the
optic branches of NaCl and compares the corrected dispersion with neutron and
infrared data; the LO/TO ratio agrees with experiment within 2\%.

\section{Branch-resolved convergence and neutron model for Al}

\begin{figure*}[tp]
\centering
\includegraphics[width=\textwidth]{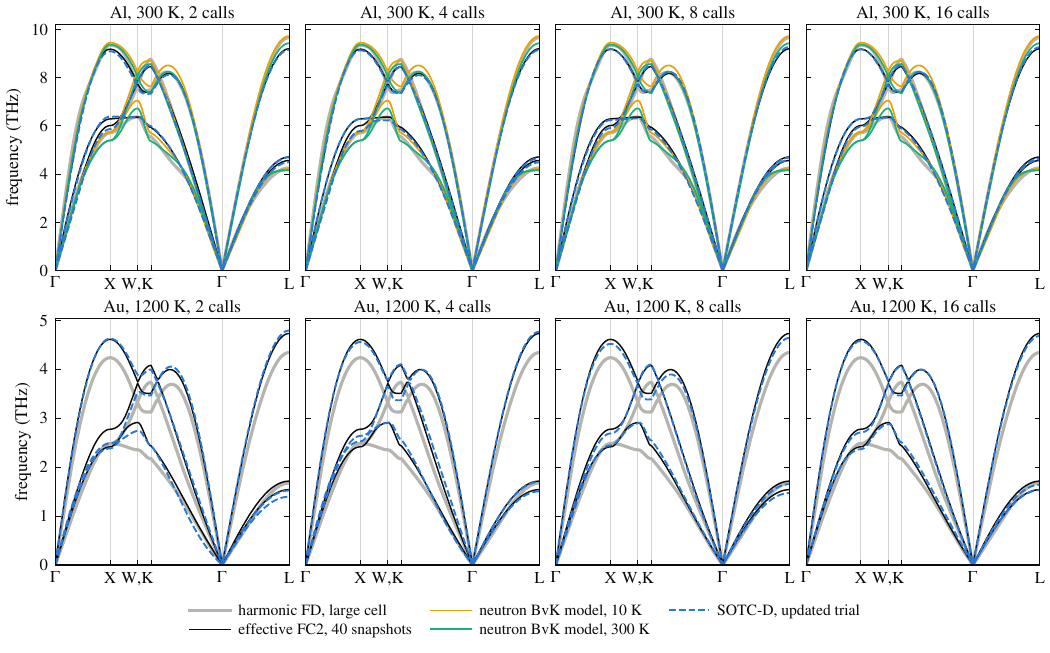}
\caption{Dispersions of SOTC-D from the updated trial (blue dashed) with 2,
4, 8, and 16 force calls, the converged large-cell harmonic FD reference
(gray), the 40-snapshot iteration-1 effective FC2 (black), and, for Al, the
Born--von K\'arm\'an models fitted to inelastic neutron data at 10~K (yellow)
and 300~K (green). Same-cell models are Fourier interpolated from the
32-atom SOTC cell.}
\label{fig:dispconv}
\end{figure*}

Figure~\ref{fig:dispconv} shows the dispersions of SOTC-D from the updated
trial with 2, 4, 8, and 16 force calls for Al (300~K) and Au (1200~K),
together with the converged large-cell harmonic FD reference and the
40-snapshot iteration-1 effective FC2 of the same cell. For Au, whose
single-configuration SOTC-C fits are unstable (Table~\ref{tab:iterations}),
the SOTC-D dispersions are stable at every budget and approach the effective
FC2 branch by branch; with two calls the largest deviations are in the
transverse branches along X--W--K and $\Gamma$--L. For Al the Born--von
K\'arm\'an models fitted to the inelastic neutron spectrum at 10 and 300~K
\cite{kresch2008} (eight shells) are also shown. The converged harmonic reference agrees with the 10~K
model within 1.1\% at X and L. The 300~K model is softer in the transverse
branches (TA(X) 5.39~THz) than both the converged harmonic reference
(5.72~THz) and the effective FC2 of the 32-atom cell (6.01 and 6.29~THz, the
degeneracy being lifted by the cell), most of whose stiffening is already
present in the harmonic FD model of that cell (5.87 and 6.09~THz). A
quantitative test of the finite-temperature renormalization against
experiment is therefore limited by the supercell truncation of the
32-atom cell and by the non-uniqueness of density-of-states-fitted
force-constant models for individual branches, especially away from the
high-symmetry points.

\section{Designs at four and sixteen force calls}

Tables~\ref{tab:matched_4_16}, \ref{tab:selfconsistent_crystal_4_16}, and
\ref{tab:selfconsistent_4_16} extend Tables~IV and V of the main text and
Table~\ref{tab:selfconsistent_crystal} to four and sixteen force calls, with
the same definitions.

\begin{table*}[t]
\caption{Matched comparison in the SOTC cell at equal numbers of displaced force calls, all designed from the same-cell harmonic FD trial and fitted in the supercell space-group basis. D: SOTC-D; C: random canonical subsets; P: random canonical $\pm\mathbf u$ pairs (medians over 20 draws); Z: special displacements with Walsh sign patterns (ZG); R: thermal rattle (independent Gaussian atoms with the canonical per-atom variance). Upper block: held-out force RMSE on 8 canonical test snapshots relative to that of the 40-snapshot effective FC2. Lower block: mesh RMSE (THz) against an effective FC2 fitted to $40-n$ canonical snapshots disjoint from the scored design. The best entry is bold; the last row counts the materials in which D is best. Entries $>$10~THz ($>$100 in force) are singular fits of designs that do not identify the model.}
\label{tab:matched_4_16}
\begin{ruledtabular}
\begin{tabular}{lcccccccccc}
 & \multicolumn{5}{c}{4 force calls} & \multicolumn{5}{c}{16 force calls}\\
\cmidrule(lr){2-6}\cmidrule(lr){7-11}
 & D & C & P & Z & R & D & C & P & Z & R\\
\hline
\multicolumn{11}{l}{Held-out force RMSE / floor}\\
Al & \textbf{1.005} & 1.053 & 1.015 & 1.025 & 1.050 & \textbf{0.997} & 1.007 & 1.001 & 1.020 & 1.019\\
Ag & \textbf{1.000} & 1.061 & 1.029 & 1.032 & 1.107 & \textbf{0.997} & 1.011 & 1.009 & 1.016 & 1.060\\
Au & \textbf{1.004} & 1.048 & 1.024 & 1.034 & 1.130 & \textbf{1.002} & 1.010 & 1.005 & 1.020 & 1.110\\
Cu & \textbf{1.038} & 1.154 & 1.186 & 1.567 & 1.366 & \textbf{1.006} & 1.032 & 1.013 & 1.026 & 1.073\\
Mo & \textbf{1.006} & 1.071 & 1.023 & 1.021 & 1.040 & \textbf{0.999} & 1.011 & 1.003 & 1.018 & 1.010\\
W & \textbf{1.007} & 1.057 & 1.017 & 1.026 & 1.024 & \textbf{0.997} & 1.010 & 1.002 & 1.010 & 1.005\\
NaCl & 1.005 & 1.005 & 1.005 & 1.002 & \textbf{0.996} & 1.002 & 1.001 & 1.000 & 0.999 & \textbf{0.982}\\
MgO & \textbf{1.012} & 1.035 & 1.031 & 1.030 & 1.135 & \textbf{0.998} & 1.007 & 1.005 & 1.005 & 1.058\\
PbTe & 1.027 & 1.018 & 1.018 & \textbf{1.011} & 1.045 & 1.012 & 1.003 & \textbf{1.001} & 1.016 & 1.141\\
Si & \textbf{0.999} & 1.005 & 1.002 & 1.001 & 1.131 & \textbf{0.998} & 1.002 & 1.000 & 1.000 & 1.116\\
Mg & \textbf{1.000} & 1.007 & 1.002 & 1.002 & 1.023 & 1.000 & 1.001 & 1.000 & \textbf{1.000} & 1.016\\
D best & \multicolumn{5}{c}{9/11} & \multicolumn{5}{c}{8/11}\\
\hline
\multicolumn{11}{l}{Mesh RMSE vs disjoint effective FC2 (THz)}\\
Al & \textbf{0.11} & 0.25 & 0.14 & 0.15 & 0.27 & \textbf{0.08} & 0.14 & 0.10 & 0.12 & 0.19\\
Ag & \textbf{0.11} & 0.32 & 0.21 & 0.24 & 0.41 & \textbf{0.10} & 0.18 & 0.14 & 0.18 & 0.30\\
Au & \textbf{0.17} & 0.27 & 0.20 & 0.30 & 0.40 & \textbf{0.11} & 0.17 & 0.13 & 0.19 & 0.42\\
Cu & \textbf{0.40} & 0.78 & 0.92 & 2.68 & 1.78 & 0.29 & 0.36 & \textbf{0.28} & 0.38 & 0.51\\
Mo & \textbf{0.08} & 0.17 & 0.10 & 0.11 & 0.14 & 0.06 & 0.11 & \textbf{0.05} & 0.11 & 0.10\\
W & \textbf{0.06} & 0.13 & 0.07 & 0.09 & 0.08 & \textbf{0.04} & 0.07 & 0.04 & 0.06 & 0.06\\
NaCl & 0.25 & 0.29 & \textbf{0.24} & 0.34 & 0.59 & 0.15 & 0.17 & \textbf{0.14} & 0.15 & 0.66\\
MgO & \textbf{0.40} & 0.57 & 0.52 & 0.47 & 1.05 & 0.34 & 0.39 & \textbf{0.30} & 0.36 & 0.74\\
PbTe & 0.16 & 0.13 & 0.13 & \textbf{0.12} & 0.27 & 0.10 & \textbf{0.08} & 0.08 & 0.09 & 0.37\\
Si & 0.04 & 0.12 & 0.07 & \textbf{0.03} & 0.62 & 0.04 & 0.07 & 0.04 & \textbf{0.04} & 0.56\\
Mg & \textbf{0.03} & 0.10 & 0.05 & 0.04 & 0.21 & 0.03 & 0.06 & \textbf{0.03} & 0.05 & 0.18\\
D best & \multicolumn{5}{c}{8/11} & \multicolumn{5}{c}{4/11}\\
\end{tabular}
\end{ruledtabular}
\end{table*}
\begin{table*}[t]
\caption{Updated trial (iteration 1, crystal basis). Mesh RMSE (THz) against the iteration-1 effective FC2. Matched trial: SOTC-D designed from the iteration-1 trial (D$_1$) versus the better of random canonical subsets and pairs drawn from the same trial (C$_1$), scored against the disjoint effective FC2. Equal cost: bootstrapped SOTC-D (B; first pair from the harmonic trial, later pairs from the model refitted to the first pair, no extra force calls) and one-shot SOTC-D (D$_0$) versus canonical pairs from the harmonic trial (C$_0$; a single sTDEP iteration), scored against the 40-snapshot iteration-1 effective FC2, which shares no configuration with them. B and D$_0$ coincide at 2 calls by construction. The best entry of each group is bold.}
\label{tab:selfconsistent_crystal_4_16}
\begin{ruledtabular}
\begin{tabular}{lcccccccccc}
 & \multicolumn{5}{c}{4 force calls} & \multicolumn{5}{c}{16 force calls}\\
\cmidrule(lr){2-6}\cmidrule(lr){7-11}
 & \multicolumn{2}{c}{matched trial} & \multicolumn{3}{c}{equal cost} & \multicolumn{2}{c}{matched trial} & \multicolumn{3}{c}{equal cost}\\
\cmidrule(lr){2-3}\cmidrule(lr){4-6}\cmidrule(lr){7-8}\cmidrule(lr){9-11}
 & D$_1$ & C$_1$ & B & D$_0$ & C$_0$ & D$_1$ & C$_1$ & B & D$_0$ & C$_0$\\
\hline
Al & 0.07 & \textbf{0.06} & \textbf{0.02} & 0.06 & 0.06 & \textbf{0.03} & 0.04 & \textbf{0.02} & 0.02 & 0.03\\
Ag & \textbf{0.04} & 0.08 & 0.08 & \textbf{0.06} & 0.08 & 0.05 & \textbf{0.05} & \textbf{0.04} & 0.06 & 0.10\\
Au & \textbf{0.03} & 0.07 & \textbf{0.07} & 0.16 & 0.19 & \textbf{0.04} & 0.05 & \textbf{0.01} & 0.17 & 0.21\\
Cu & \textbf{0.04} & 0.14 & 0.24 & 0.14 & \textbf{0.08} & \textbf{0.03} & 0.06 & 0.16 & \textbf{0.10} & 0.14\\
Mo & \textbf{0.02} & 0.04 & 0.02 & \textbf{0.02} & 0.06 & \textbf{0.02} & 0.03 & \textbf{0.02} & 0.03 & 0.05\\
W & \textbf{0.03} & 0.03 & 0.03 & \textbf{0.03} & 0.03 & \textbf{0.02} & 0.02 & 0.02 & \textbf{0.01} & 0.02\\
NaCl & 0.09 & \textbf{0.08} & \textbf{0.07} & 0.21 & 0.15 & 0.07 & \textbf{0.06} & \textbf{0.04} & 0.14 & 0.12\\
MgO & \textbf{0.13} & 0.18 & \textbf{0.27} & 0.37 & 0.31 & \textbf{0.10} & 0.12 & \textbf{0.10} & 0.32 & 0.25\\
PbTe & \textbf{0.02} & 0.05 & \textbf{0.07} & 0.17 & 0.10 & \textbf{0.03} & 0.03 & \textbf{0.04} & 0.11 & 0.08\\
Si & \textbf{0.05} & 0.07 & \textbf{0.06} & 0.09 & 0.09 & 0.06 & \textbf{0.04} & \textbf{0.04} & 0.09 & 0.06\\
Mg & \textbf{0.04} & 0.05 & 0.03 & \textbf{0.03} & 0.04 & 0.04 & \textbf{0.04} & \textbf{0.02} & 0.04 & 0.03\\
\hline
D$_1<$C$_1$ / B$<$C$_0$ & \multicolumn{2}{c}{9/11} & \multicolumn{3}{c}{9/11} & \multicolumn{2}{c}{7/11} & \multicolumn{3}{c}{10/11}\\
held-out force D$_1<$C$_1$ & \multicolumn{2}{c}{10/11} & \multicolumn{3}{c}{} & \multicolumn{2}{c}{6/11} & \multicolumn{3}{c}{}\\
\end{tabular}
\end{ruledtabular}
\end{table*}
\begin{table*}[t]
\caption{Updated trial (iteration 1, supercell basis). Mesh RMSE (THz) against the iteration-1 effective FC2. Matched trial: SOTC-D designed from the iteration-1 trial (D$_1$) versus the better of random canonical subsets and pairs drawn from the same trial (C$_1$), scored against the disjoint effective FC2. Equal cost: bootstrapped SOTC-D (B; first pair from the harmonic trial, later pairs from the model refitted to the first pair, no extra force calls) and one-shot SOTC-D (D$_0$) versus canonical pairs from the harmonic trial (C$_0$; a single sTDEP iteration), scored against the 40-snapshot iteration-1 effective FC2, which shares no configuration with them. B and D$_0$ coincide at 2 calls by construction. The best entry of each group is bold.}
\label{tab:selfconsistent_4_16}
\begin{ruledtabular}
\begin{tabular}{lcccccccccc}
 & \multicolumn{5}{c}{4 force calls} & \multicolumn{5}{c}{16 force calls}\\
\cmidrule(lr){2-6}\cmidrule(lr){7-11}
 & \multicolumn{2}{c}{matched trial} & \multicolumn{3}{c}{equal cost} & \multicolumn{2}{c}{matched trial} & \multicolumn{3}{c}{equal cost}\\
\cmidrule(lr){2-3}\cmidrule(lr){4-6}\cmidrule(lr){7-8}\cmidrule(lr){9-11}
 & D$_1$ & C$_1$ & B & D$_0$ & C$_0$ & D$_1$ & C$_1$ & B & D$_0$ & C$_0$\\
\hline
Al & \textbf{0.10} & 0.21 & \textbf{0.07} & 0.09 & 0.15 & \textbf{0.08} & 0.10 & \textbf{0.06} & 0.06 & 0.10\\
Ag & \textbf{0.15} & 0.15 & 0.12 & \textbf{0.12} & 0.17 & 0.10 & \textbf{0.10} & \textbf{0.08} & 0.09 & 0.13\\
Au & \textbf{0.07} & 0.15 & \textbf{0.14} & 0.20 & 0.24 & \textbf{0.07} & 0.09 & \textbf{0.06} & 0.19 & 0.24\\
Cu & \textbf{0.48} & 0.74 & 0.55 & \textbf{0.37} & 1.18 & \textbf{0.21} & 0.21 & 0.25 & \textbf{0.21} & 0.29\\
Mo & \textbf{0.07} & 0.10 & 0.06 & \textbf{0.06} & 0.10 & \textbf{0.06} & 0.06 & 0.05 & \textbf{0.05} & 0.07\\
W & \textbf{0.05} & 0.08 & \textbf{0.05} & 0.06 & 0.07 & \textbf{0.04} & 0.04 & 0.03 & \textbf{0.03} & 0.04\\
NaCl & \textbf{0.14} & 0.19 & \textbf{0.16} & 0.30 & 0.23 & \textbf{0.10} & 0.11 & \textbf{0.07} & 0.16 & 0.15\\
MgO & \textbf{0.23} & 0.38 & \textbf{0.33} & 0.44 & 0.44 & \textbf{0.21} & 0.22 & \textbf{0.16} & 0.36 & 0.30\\
PbTe & \textbf{0.05} & 0.11 & \textbf{0.09} & 0.18 & 0.14 & \textbf{0.05} & 0.06 & \textbf{0.06} & 0.11 & 0.10\\
Si & \textbf{0.05} & 0.07 & \textbf{0.06} & 0.09 & 0.09 & 0.06 & \textbf{0.04} & \textbf{0.04} & 0.09 & 0.06\\
Mg & \textbf{0.04} & 0.05 & 0.03 & \textbf{0.03} & 0.04 & 0.04 & \textbf{0.04} & \textbf{0.02} & 0.04 & 0.03\\
\hline
D$_1<$C$_1$ / B$<$C$_0$ & \multicolumn{2}{c}{11/11} & \multicolumn{3}{c}{11/11} & \multicolumn{2}{c}{8/11} & \multicolumn{3}{c}{11/11}\\
held-out force D$_1<$C$_1$ & \multicolumn{2}{c}{10/11} & \multicolumn{3}{c}{} & \multicolumn{2}{c}{6/11} & \multicolumn{3}{c}{}\\
\end{tabular}
\end{ruledtabular}
\end{table*}

\section{Nearest-neighbor force constants against neutron data}

Table~\ref{tab:bvk} compares the leading real-space force constants with
Born--von K\'arm\'an fits to inelastic neutron data \cite{kresch2009thesis}
where the measurement temperature matches the model: Al at 300~K and Cu at
1200~K (interpolated between 973 and 1336~K). For the other crystals the
neutron fits are at room temperature, whereas the models are at
1000--1500~K and the corresponding lattice parameters, so they are not
compared. The converged harmonic reference of Al reproduces the 300~K
constants within 4\%. In the 32-atom Al cell, truncation distorts
individual constants (the second-shell constant falls to 0.5~N/m). The
effective and SOTC-D models partly compensate, so shell constants of the
small cell should not be read as converged. For Cu, where truncation is small
($K_1$ 23.5~N/m in both cells), the harmonic model at the 1200~K lattice
parameter is 8\% softer than experiment in $K_1$. The effective and SOTC-D
models are 16--18\% stiffer (bootstrapped SOTC-D 30\%). The SCHA auxiliary
FC2 omits the cubic (bubble) contribution to the physical spectrum, which
softens it, so a stiffer auxiliary FC2 than the measured spectrum is the
expected sign.

\begin{table}[t]
\caption{Nearest-neighbor force constants (N/m): first- and second-shell longitudinal ($K_1$, $K_2$) and mean first-shell transverse ($K_{T1}$), averaged over all pairs of the shell. Experiment: Born--von K\'arm\'an fits to inelastic neutron data compiled in Ref.~\onlinecite{kresch2009thesis} (Cu interpolated linearly between 973 and 1336~K). All models at the production lattice parameters (300~K for Al, 1200~K for Cu).}
\label{tab:bvk}
\footnotesize
\begin{ruledtabular}
\begin{tabular}{lcccccc}
 & \multicolumn{3}{c}{Al, 300~K} & \multicolumn{3}{c}{Cu, 1200~K}\\
\cmidrule(lr){2-4}\cmidrule(lr){5-7}
 & $K_1$ & $K_2$ & $K_{T1}$ & $K_1$ & $K_2$ & $K_{T1}$\\
\hline
harmonic, large cell & 20.2 & 2.14 & -1.35 & 23.5 & 0.79 & -1.04\\
harmonic, SOTC cell & 17.6 & 0.49 & 0.19 & 23.5 & 0.40 & -1.03\\
effective FC2 & 18.8 & 0.86 & -0.51 & 30.1 & 0.04 & -1.57\\
SOTC-D (8, updated) & 18.8 & 0.69 & -0.57 & 29.6 & 0.99 & -1.19\\
bootstrapped (8) & 19.0 & 0.71 & -0.61 & 33.4 & 1.14 & -2.11\\
neutron BvK & 20.1 & 2.22 & -1.36 & 25.6 & 0.73 & -1.93\\
\end{tabular}
\end{ruledtabular}
\end{table}

\section{Thermodynamics of Al on the volume grid}

Table~\ref{tab:qha_thermo} complements Table~XIII of the main text. The
Gr\"uneisen formula $\alpha=\sum_k\gamma_kc_k/3B_TV$, with mode Gr\"uneisen
parameters from the converged harmonic FC2 on the volume grid, reproduces the
free-energy minimization to $0.1\times10^{-6}$~K$^{-1}$ at every temperature
when it is evaluated at $V(T)$ with $B_T$, which checks the numerics of the
minimization. Evaluated with the static $B_0$ and $V_0$ (the common
shortcut), it is 8\% low at 100~K and 13--44\% low at 300--900~K, because it
omits the softening of $B_T$ and the volume dependence of the mode
Gr\"uneisen parameters; the shortcut is therefore not a valid check of a QHA
calculation above about 100~K for Al. The converged QHA gives $C_P$ within
0.1--1.0~J~mol$^{-1}$K$^{-1}$ of the NIST-JANAF values between 300 and
900~K, a difference of the size of the electronic heat capacity
($\gamma_eT\approx0.4$--1.2~J~mol$^{-1}$K$^{-1}$), which is not included.
The self-consistent model gives the same $C_P$ and $B_T$ at 300~K; above
400~K its free-energy curvature, and hence $\alpha$, $B_T$, and $C_P$, are
not resolved with 32--128 force calls per volume (main text, Sec.~V\,K),
and its values are not listed.

\begin{table*}[t]
\caption{Al thermodynamics on the volume grid (PBE). $\alpha$ ($10^{-6}$~K$^{-1}$) of the converged 125-atom QHA from free-energy minimization and from the Gr\"uneisen formula $\alpha=\sum_k\gamma_kc_k/3BV$, evaluated with the static $B_0$ and $V_0$ (shortcut) and with $B_T$ and $V(T)$ of the minimization (exact). $B_T$ (GPa) and $C_P=C_V+TV\beta^2B_T$ (J~mol$^{-1}$K$^{-1}$, phonon part only) of the converged QHA and of the self-consistent SOTC-D (32 calls, independent-seed grid). Experiment: NIST-JANAF $C_P$ (Shomate fit, 298--933~K, including the electronic contribution) \cite{chase1998}.}
\label{tab:qha_thermo}
\begin{ruledtabular}
\begin{tabular}{lcccccccc}
 & \multicolumn{3}{c}{$\alpha$, converged QHA} & \multicolumn{2}{c}{$B_T$} & \multicolumn{3}{c}{$C_P$}\\
\cmidrule(lr){2-4}\cmidrule(lr){5-6}\cmidrule(lr){7-9}
$T$ (K) & min. & Gr\"uneisen, exact & Gr\"uneisen, shortcut & QHA & SC & QHA & SC & expt.\\
\hline
100 & 12.9 & 12.9 & 11.9 & 75.0 & 75.3 & 13.24 & 13.48 & --\\
300 & 24.1 & 24.1 & 20.9 & 69.6 & 69.5 & 24.14 & 24.01 & 24.25\\
400 & 26.2 & 26.3 & 21.7 & 66.6 & 68.9 & 25.56 & 25.21 & 25.78\\
500 & 28.2 & 28.2 & 22.1 & 63.3 & -- & 26.61 & -- & 26.84\\
700 & 32.9 & 32.9 & 22.4 & 55.8 & -- & 28.62 & -- & 29.10\\
900 & 40.2 & 40.1 & 22.6 & 46.6 & -- & 31.34 & -- & 32.31\\
\end{tabular}
\end{ruledtabular}
\end{table*}

\end{document}